\documentclass[prc,aps,floats,floatfix,twocolumn,showpacs,nofootinbib,%
superscriptaddress]{revtex4}

\usepackage{amsmath}
\usepackage{amssymb}
\usepackage{amsfonts}
\usepackage{graphicx}
\usepackage{color}

\makeatletter
\def\@dotsep{4.5}
\makeatother

\newcommand{\old}[1]{{\rule{0cm}{0cm}}}

\renewcommand{\vec}[1]{{\mathbf #1}}
\newcommand{\nuc}[2]{$^{#1}${#2}}
\newcommand{\etal}{\emph{et~al.}}
\newcommand{\ie}{i.e.}
\newcommand{\nn}{\nonumber}

\newcommand{\grad}{\boldsymbol{\nabla}}

\newcommand{\vnabla}{\boldsymbol{\mathbf\nabla}}
\newcommand{\vsigma}{\boldsymbol{\mathbf\sigma}}

\newcommand{\central}{c}
\renewcommand{\tensor}{t}

\newcommand{\CI}{\mathcal{I}}

\newcommand{\nrm}{\mathrm{n}}
\newcommand{\prm}{\mathrm{p}}

\begin{document}

\title{The tensor part of the Skyrme energy density functional.
       I. Spherical nuclei}


\author{T. Lesinski}
\email{lesinski@ipnl.in2p3.fr}
\affiliation{Universit{\'e} de Lyon, F-69003 Lyon, France;
             Institut de Physique Nucl{\'e}aire de Lyon,
             CNRS/IN2P3, Universit{\'e} Lyon 1,
             F-69622 Villeurbanne, France}

\author{M. Bender}
\email{bender@cenbg.in2p3.fr}
\affiliation{DSM/DAPNIA/SPhN, CEA Saclay,
             F-91191 Gif-sur-Yvette Cedex,
             France}
\affiliation{Universit{\'e} Bordeaux 1; CNRS/IN2P3;
             Centre d'{\'E}tudes Nucl{\'e}aires de Bordeaux Gradignan, UMR5797,
             Chemin du Solarium, BP120, F-33175 Gradignan, France}

\author{K. Bennaceur}
\affiliation{Universit{\'e} de Lyon, F-69003 Lyon, France;
             Institut de Physique Nucl{\'e}aire de Lyon,
             CNRS/IN2P3, Universit{\'e} Lyon 1,
             F-69622 Villeurbanne, France}
\affiliation{DSM/DAPNIA/SPhN, CEA Saclay,
             F-91191 Gif-sur-Yvette Cedex,
             France}

\author{T. Duguet}
\affiliation{National Superconducting Cyclotron Laboratory
             and Department of Physics and Astronomy,
             Michigan State University, East Lansing, MI 48824,
             USA}

\author{J. Meyer}
\affiliation{Universit{\'e} de Lyon, F-69003 Lyon, France;
             Institut de Physique Nucl{\'e}aire de Lyon,
             CNRS/IN2P3, Universit{\'e} Lyon 1,
             F-69622 Villeurbanne, France}


\begin{abstract}
We perform a systematic study of the impact of the $\vec{J}^2$ tensor term in
the Skyrme energy functional on properties of spherical nuclei. In the Skyrme
energy functional, the tensor terms originate both from zero-range central and
tensor forces. We build a set of 36 parameterizations which cover a wide range
of the parameter space of the isoscalar and isovector tensor term coupling
constants with a fit protocol very similar to that of the successful SLy
parameterizations. We analyze the impact of the tensor terms on a large variety
of observables in spherical mean-field calculations, such as the spin-orbit
splittings and single-particle spectra of doubly-magic nuclei, the evolution of
spin-orbit splittings along chains of semi-magic nuclei, mass residuals of
spherical nuclei, and known anomalies of radii. The major findings of our study
are (i) tensor terms should not be added perturbatively to existing
parameterizations, a complete refit of the entire parameter set is imperative.
(ii) The free variation of the tensor terms does not lower the $\chi^2$ within a
standard Skyrme energy functional. (iii) For certain regions of the parameter
space of their coupling constants, the tensor terms lead to instabilities of the
spherical shell structure, or even the coexistence of two configurations with
different spherical shell structure. (iv) The standard spin-orbit interaction
does not scale properly with the principal quantum number, such that
single-particle states with one or several nodes have too large spin-orbit
splittings, while those of nodeless intruder levels are tentatively too small.
Tensor terms with realistic coupling constants cannot cure this problem. (v)
Positive values of the coupling constants of proton-neutron and like-particle
tensor terms allow for a qualitative description of the evolution of spin-orbit
splittings in chains of Ca, Ni and Sn isotopes. (vi) For the same values of the
tensor term coupling constants, however, the overall agreement of the
single-particle spectra in doubly-magic nuclei is deteriorated, which can be
traced back to features of the single-particle spectra that are not related to
the tensor terms. We conclude that the currently used central and spin-orbit
parts of the Skyrme energy density functional are not flexible enough to allow
for the presence of large tensor terms.
\end{abstract}



\date{April 4, 2007}


\maketitle


\section{Introduction}
\label{sect:intro}


The strong nuclear spin-orbit interaction in nuclei is responsible for the
observed magic numbers in heavy nuclei~\cite{GM48a,Hax49a,Fee49a,GM49a}.
While a simple spin-orbit interaction allows for the qualitative description
of the global features of shell structure, the available data suggest that
single-particle energies evolve with neutron and proton number in a manner that
cannot be related to the geometrical growth of the single-particle potential
with $N$ and $Z$. Many anomalies of shell structure have been identified that
do not fit into simple experimental systematics, and that challenge any global
model of nuclear structure.

The evolution of shell structure with $N$ and $Z$ as a feature of
self-consistent mean-field models has been known for long. To quote the
pioneering study of shell structure in a self-consistent model performed by
Beiner \etal~\cite{Bei75a}, the ``most striking effect is the appearance of
\mbox{$N=16$}, 34 and 56 as neutron magic numbers for unstable nuclei, together
with a weakening of the shell closure at \mbox{$N=20$} and 28''. Various
mechanisms that modify the appearance of gaps in the single-particle spectra
have been discussed in detail in the literature. The two most prominent ones
that were worked out by Dobaczewski \etal\ in Ref.~\cite{Dob94a}, however, play
mainly a role for weakly-bound exotic nuclei far from stability, as
they are directly or indirectly related to the physics of loosely bound
single-particle states, namely that the enhancement of the diffuseness of
neutron density distribution reduces the spin-orbit coupling in neutron-rich
nuclei on the one hand, and the interaction between bound orbitals and the
continuum results in a quenching of shell effects in light and medium systems
on the other hand. The former effect was also extensively discussed in the
framework of relativistic models by Lalazissis \etal~\cite{Lal98a,Lal98b},
while the latter triggered a number of studies that discussed the potential
relevance of this so-called ``Boguliubov enhanced shell quenching'' to
explain the abundance pattern from the astrophysical $r$-process of
nucleosynthesis~\cite{Che95a,Dob95b,Pea96a,Pfe97a}.

These two effects take place in neutron-rich nuclei. In proton-rich nuclei, the
Coulomb barrier suppresses both the diffuseness of the proton density and the
coupling of bound proton states to the continuum. But the Coulomb interaction
itself can also modify the shell structure: for super-heavy nuclei, it begins
to destabilize the nucleus as a whole. Mean-field models predict that it
amplifies the shell oscillations of the densities for incomplete filled
oscillator shells, which leads to strong variations of the density profile that
feed back onto the single-particle spectra~\cite{Dec99a,Ben99b}.

Interestingly, most theoretical papers about the evolution of shell structure
from the last decade have speculated about new effects that mainly affect
neutron shells in nuclei far from stability in the anticipation of the
rare-isotope physics that might become accessible with the next generation of
experimental facilities. The known anomalies, some of which have been known for
a long time, and many more have been identified recently, concern also proton
shells and already appear sufficiently close to stability that ``exotic
phenomena can be ruled out for their explanation'' in most cases, to paraphrase
the authors of Ref.~\cite{Lan03a}. By contrast, this suggests that there exists
a mechanism that induces a strong evolution of single-particle spectra already
in stable nuclei that has been overlooked for long.

There is a prominent ingredient of the nucleon-nucleon interaction that has been
ignored for decades in virtually all global nuclear structure models for medium
and heavy nuclei, be it macroscopic-microscopic approaches or self-consistent
mean-field methods. It is only very recently, that the systematic discrepancies
between model predictions and experiment have triggered a renaissance of the
tensor force in the description of finite medium- and heavy-mass nuclei.

The tensor force is a crucial and necessary ingredient of the bare
nucleon-nucleon interaction~\cite{Wir95a,Mac01a}, and consequently is
contained in all \emph{ab-initio} approaches that are available for light,
mainly $p$-shell nuclei~\cite{Pie01a,Nav03a}. One of the first experimental
signatures of the tensor force was the small, but finite quadrupole moment of
the
deuteron. In a boson-exchange picture of the bare nucleon-nucleon interaction,
the tensor force originates from the exchange of pseudoscalar pions, which have
both central and tensor couplings, see for example section 2.3 in
Ref.~\cite{EGIII} or appendix~13A of Ref.~\cite{Nil95a}. In a nuclear many-body
system, the bare tensor force induces a strong correlation between the spatial
and spin orientations in the two-body density matrix. For two nucleons with
parallel spins, the tensor force energetically favors the configuration where
the distance vector is aligned with the spins, while for anti-parallel spins
the tensor force prefers when the distance vector is perpendicular to the
spins, see the discussion of Fig.~13 in Ref.~\cite{Nef03a} and of Fig.~3 in
Ref.~\cite{Rot04a}. The authors of these papers also demonstrate very nicely
the well-known fact~\cite{Bet68a,Neg70a} that in an approach that starts
from the bare nucleon-nucleon interaction, nuclei are not bound without taking
into account the two-body correlations induced by the tensor force.

The role of the tensor force, however, manifests itself differently in
self-consistent mean-field models, otherwise called energy density functional
(EDF) methods, the tool of choice for medium and heavy nuclei. The latter
methods use an independent-particle state as a reference state to express the
energy of the correlated nuclear ground state. Thus, correlations are not
explicitly present in the higher-order density matrices of the reference state,
but rather included under the form of a more elaborate functional of the (local
and nearly local parts of the) one-body density matrix of that reference state.
In such a scheme, most of the effect of the bare tensor force on the binding
energy is integrated out through the renormalization of the coupling constants
associated with a central effective vertex, in a similar fashion as the tensor
part of the bare interaction is renormalized into the central one when going
from the bare nucleon-nucleon force to a Brueckner $G$ matrix. The tensor terms
of the EDF relate to a \emph{residual} tensor vertex, that gives nothing but a
correction to the spin-orbit splittings, which for light $p$-shell nuclei might
be of the same order as the contribution from the genuine spin-orbit force. The
interplay of spin-orbit and tensor forces in the mean field of medium and heavy
nuclei was explored in Refs.~\cite{Sche76a,Goo78a,Zhe91a}, where the particular
role of spin-unsaturated shells was pointed out.

There are two widely used effective interactions for non-relativistic
self-consistent mean-field models~\cite{RMP}, the zero-range non-local Skyrme
interaction~\cite{Sky56a,Sky59a,Bel56a,Sky59b} on the one hand and the
finite-range Gogny force~\cite{Gog75a,Dec80a} on the other hand.

In fact, the effective zero-range non-local interaction proposed by Skyrme in
1956~\cite{Sky56a,Sky59a,Bel56a,Sky59b} already contained a zero-range tensor
force. The first applications of Skyrme's interaction in self-consistent
mean-field models that became available around 1970, however, neglected the
tensor force, and the simplified effective Skyrme interaction used in the
seminal paper by Vautherin and Brink~\cite{Vau72a} soon became the standard
Skyrme interaction that was used in most applications ever since. Until very
recently, there was only very little exploratory work on Skyrme's tensor force.
In their early study, Stancu, Brink and Flocard~\cite{Sta77a}, who added the
tensor force perturbatively to the SIII parameterization, pointed out that some
spin-orbit splittings in magic nuclei can be improved with a tensor force. A
complete fit including the terms from the tensor force that contribute in
spherical nuclei was attempted by Tondeur~\cite{Ton83a}, with the relevant
coupling constants of the spin-orbit and tensor terms adjusted to selected
spin-orbit splittings in \nuc{16}{O}, \nuc{48}{Ca} and \nuc{208}{Pb}. Another
complete fit of a generalized Skyrme interaction including a tensor force was
performed by Liu \etal~\cite{Liu91a}, but the authors did not investigate the
effect of the tensor force in detail, nor was the resulting parameterization
ever used in the literature thereafter.

Similarly, the seminal paper by Gogny~\cite{Gog75a} on the evaluation of matrix
elements of a finite-range force of Gaussian shape in an harmonic oscillator
basis contains the expressions for a finite-range tensor force, which, however,
was omitted in the parameterizations of Gogny's force adjusted by the
Bruy{\`e}res-le-Ch{\^a}tel group~\cite{Dec80a}. It were Onishi and
Negele~\cite{Oni78a} who first published an effective interaction that
combined a Gaussian two-body central force, a finite-range tensor force
with a zero-range spin-orbit force and a zero-range non-local three-body
force, which, however, also fell into oblivion.

The role of the tensor force is slightly different in Skyrme and Gogny
interactions. In the Gogny force, the contributions from the central and
tensor parts remain explicitly distinct, although, of course, this does not
prevent a certain entanglement of their physical effects. In the context of
Skyrme's functional, however, the contribution of a zero-range tensor force to
the spherical mean-field state of an even-even nucleus has exactly the same
form as a particular exchange term from the non-local part of the central
Skyrme force. When looking at spherical nuclei only, adding Skyrme's tensor
force simply allows one to decouple a term that is already provided by the
central force.
This indeed makes the effective-interaction-restricted functional
more flexible, as the additional degrees of freedom from the tensor force
remove an interdependence between the effective mass, the surface terms and the
``tensor terms''. However, one must always keep in mind that both
the central and tensor part of the effective vertex contribute to the so-called
$\vec{J}^{2}_{t}$ ``tensor'' terms of the functional.\footnote{As we will
outline below, and as was already pointed out in Ref.~\cite{Bei75a}, this
argument does not hold for deformed even-even nuclei or any situation where
intrinsic time-reversal is broken, for example odd nuclei or dynamics. There,
the tensor and non-local central parts of the effective Skyrme interaction
give contributions to the mean-fields and the binding energy with different
analytical expressions. This will be discussed in a companion
article~\cite{II}.}

In the context of relativistic mean-field models, the equivalent of the
non-relativistic tensor force appears as the exchange term of effective fields
with the quantum numbers of the pion, which by construction do not appear in
the standard relativistic Hartree models. Only relativistic Hartree-Fock models
contain this tensor force, with the first predictive parameterizations becoming
available just recently~\cite{Lon06a}.

We also mention that there is a large body of work on the tensor force in the
interacting shell model, see Ref.~\cite{Fay97a} for a review, that concentrates
on a completely different aspect of the tensor force, namely its unique
contribution to excitations with unnatural parity.

The recent interest in the effect of the tensor force in the context of
self-consistent mean field models was triggered by the observed evolution of
single-particle levels of one nucleon species in dependence of the number of
the other nucleon species. Otsuka \etal~\cite{Ots05a} proposed that at least
part of the effect is caused by the proton-neutron tensor force from pion
exchange. Many groups attempt now to explain known, but so far unresolved,
anomalies of shell structure in terms of a tensor force. A particularly popular
playground is the relative shift of the proton $1g_{7/2}$ and $1h_{11/2}$
levels in tin isotopes, which is interpreted as the reduction of the spin-orbit
splittings of both levels with their respective partners with increasing
neutron number~\cite{Schi04aE}.

Otsuka \etal~\cite{Ots06a} added a Gaussian tensor force, adjusted on the
long-range part of a one-pion+$\rho$ exchange potential, to a standard Gogny
force. After a consistent readjustment of the parameters of its central and
spin-orbit parts, they were able to explain coherently the anomalous relative
evolution of some single-particle levels without, however, being able to
describe their absolute distance in energy. Dobaczewski~\cite{Dobaczewskitalk}
has pointed out that a perturbatively added tensor interaction with suitably
chosen coupling constants in the Skyrme energy density functional does not only
modify the evolution of shell structure, but does also improve the description
of nuclear masses around magic nuclei. Brown \etal~\cite{Bro06a} have fitted a
Skyrme interaction with added zero-range tensor force with emphasis on the
reproduction of single-particle spectra. While the authors appreciate the
qualitatively correctly described evolution of relative level distances, they
point out that the combination of zero-range spin-orbit and tensor forces does
not and can not correctly describe the $\ell$-dependence of spin-orbit
splittings. Col{\`o} \etal~\cite{Col07a}, and Brink \etal~\cite{Bri07a}
have added Skyrme's tensor force perturbatively to the existing standard
parameterization SLy5~\cite{Cha97a,Cha98a}, and to the SIII~\cite{Bei75a} one,
respectively. They have investigated some single-particle energy differences:
the $1h_{11/2}$ and $1g_{7/2}$ proton states in tin isotopes as well as
$1i_{13/2}$ and $1h_{9/2}$ neutron states in $N=82$ isotones and propose
similar parameters as in Ref.~\cite{Bro06a}. The effect of the tensor force on
the centroid of the GT giant resonance is also estimated by Col{\`o}
\etal\ using a sum-rule approach and found to be substantial. Long
\etal~\cite{Lon06b}, demonstrate that the tensor force that emerges naturally
in relativistic Hartree-Fock also improves the relative shifts of the
proton $1g_{7/2}$ and $1h_{11/2}$ levels in tin isotopes.

The work on the tensor force published so far aims at an optimal single
parameterization, that establishes a best fit to either the underlying bare
tensor force~\cite{Ots06a,Bro06a} or empirical
data~\cite{Ton83a,Dobaczewskitalk,Col07a}. The published results, as well as
our first exploratory studies, however, suggest that adding a tensor force to
the existing mean-field models gives only a local improvement of the relative
change of certain single-particle energies, but not necessarily a global
improvement of single-particle spectra or other observables. In the framework
of the Skyrme interaction, that we will employ throughout this work, there is
also the already mentioned ambiguity that the contribution from the tensor
force to spherical nuclei has the same structure as a term from the central
force. In view of this situation, we will pursue a different strategy and
investigate the effect of the tensor terms on a multitude of observables in
nuclei though a set of Skyrme interactions with systematically varied
coupling constants of the tensor terms.

The present study was motivated by the finding that the performance of the
existing Skyrme-type effective interactions for masses and spectroscopic
properties is limited by systematic deficiencies of the single-particle
spectra~\cite{BBH06a,Ben03c,Ben06a,Cha06a} that seem to be impossible to remove
within the standard Skyrme interaction. The details of single-particle spectra
were so far somewhat outside the focus of self-consistent mean-field methods,
on the one hand as they do \emph{not} correspond directly to empirical
single-particle energies (we will come back to that below), and on the other
hand because many of the observables that are usually calculated with
self-consistent mean-field methods are not very sensitive to the exact
placement of single-particle levels. By
contrast, there is an enormous body of work that examines the infinite and
semi-infinite nuclear matter properties of the effective interactions that are
the analog of liquid-drop and droplet parameters in great detail. The reason
is, of course, that the global trends over the whole chart of nuclei have to be
understood before one can look into details. The last few years have seen
an increasing demand on predictive power. Moreover, beyond-mean-field
approaches of the projected generator coordinate method (GCM), or
Bohr-Hamiltonian type, have become widely used tools to analyze and predict
spectroscopic properties in medium and heavy nuclei, employing either Gogny or
Skyrme interactions. The underlying single-particle spectra thus now deserve
more attention, as many of the spectroscopic properties of interest turn out to
be extremely sensitive to even subtle details of the single-particle
spectra. As the tensor force is the most obvious missing piece in all
standard mean-field interactions, it is the natural starting point for the
systematic investigation of possible generalizations with the ultimate goal to
improve the predictive power of the interactions for spectroscopy.

In the present paper, we will outline the formalism of a Skyrme interaction
with added tensor force, describe the fit of the parameterizations, analyze the
role of the tensor terms for single-particle spectra, masses and radii of
spherical even-even nuclei. A second paper~\cite{II} studies the surface and
deformation
properties of these Skyrme interactions for even-even nuclei, and future work
will examine the stability of nuclear matter and the role of the time-odd terms
from the tensor force in odd and rotating nuclei. Only deformed nuclei and, in
particular, observables sensitive to the time-odd contributions, will possibly
allow to distinguish clearly between the non-local central and tensor
parts of the Skyrme force.


\section{The Skyrme interaction with tensor terms}
\label{sect:skyrme}


\subsection{The energy density functional}


The usual \emph{ansatz} for the Skyrme effective
interaction~\cite{Cha97a,Cha98a} leads to an energy density functional which
can
be written as the sum of a kinetic term, the Skyrme potential energy functional
that models the effective strong interaction in the particle-hole channel, a
pairing energy functional corresponding to a density-dependent contact pairing
interaction, the Coulomb energy functional (calculated using the Slater
approximation~\cite{Sla51}) and correction terms to approximately remove the
excitation energy from spurious motion caused by broken symmetries
\begin{equation}
\label{eq:efu:complete}
\mathcal{E}
=   \mathcal{E}_{\text{kin}}
  + \mathcal{E}_{\text{Skyrme}}
  + \mathcal{E}_{\text{pairing}}
  + \mathcal{E}_{\text{Coulomb}}
  + \mathcal{E}_{\text{corr}}\,.
\end{equation}


\subsection{The Skyrme energy density functional}


Throughout this work, we will use an effective Skyrme energy functional that
corresponds to an anti\-sym\-me\-trized density-dependent two-body
vertex in the
particle-hole channel of the strong interaction, that can be decomposed into a
central, spin-orbit and tensor contribution
\begin{equation}
v^{\text{Skyrme}}
=   v^{\text{\central}}
  + v^{\text{\tensor}}
  + v^{\text{LS}}\,.
\end{equation}
Other choices for the writing of the Skyrme energy functional are
possible and have been made in the literature, which might affect the form of
the effective interaction, its interpretation and the results obtained from it.
We will come back to that in section~\ref{sect:vertex:edf:dft} below.

The Skyrme energy density functional
is a functional of local densities and currents
\begin{equation}
\mathcal{E}_{\text{Skyrme}}
= \int \! d^3 r \; \mathcal{H}^{\text{Skyrme}} (\vec{r})\,,
\end{equation}
which has many technical advantages compared to finite-range forces such
as the Gogny force. All exchange terms have the same structure as the direct
terms, which greatly reduces the number of necessary integrations during a
calculation.


\subsubsection{Local densities and currents}


Throughout this paper we will assume that we have pure proton and neutron
states. The formal framework of the general case including proton-neutron mixing
is discussed in Ref.~\cite{Per04a}. Without making reference to any
single-particle basis, we start from the density matrices of protons and
neutrons in coordinate space~\cite{Dob00a}
\begin{eqnarray}
\rho_q (\vec{r} \sigma ,\vec{r}' \sigma')
& = & \langle \hat{a}^\dagger_{r' \sigma' q} \hat{a}_{r \sigma q} \rangle
      \nn \\
& = &   \tfrac{1}{2} \, \rho_q (\vec{r},\vec{r}') \delta_{\sigma \sigma'}
      + \tfrac{1}{2} \, \vec{s}_q (\vec{r},\vec{r}')
                        \cdot \langle \sigma' | \hat{\vsigma} | \sigma \rangle
      \nn \\
\end{eqnarray}
where
\begin{eqnarray}
\rho_q (\vec{r}, \vec{r}')
& = & \sum_{\sigma} \rho_q (\vec{r} \sigma ,\vec{r}' \sigma)
      \nn \\
\vec{s}_q (\vec{r},\vec{r}')
& = & \sum_{\sigma \sigma'} \rho_q (\vec{r} \sigma ,\vec{r}' \sigma') \;
      \langle \sigma' | \hat{\vsigma} | \sigma \rangle\,.
\end{eqnarray}
The Skyrme energy functional up to second order in derivatives that we
will introduce below can be expressed in terms of seven local densities
and currents~\cite{Per04a} that are defined as
\begin{eqnarray}
\label{eq:locdensities:rho}
\rho_q (\vec{r})
& = & \rho_q (\vec{r},\vec{r}') \big|_{\vec{r} = \vec{r}'}
      \nn \\
\label{eq:locdensities:s}
\vec{s}_q (\vec{r})
& = & \vec{s}_q (\vec{r},\vec{r}') \big|_{\vec{r} = \vec{r}'}
      \nn \\
\label{eq:locdensities:tau}
\tau_q (\vec{r})
& = & \vnabla \cdot \vnabla' \; \rho_q (\vec{r},\vec{r}')
      \big|_{\vec{r} = \vec{r}'}
      \nn \\
\label{eq:locdensities:T}
T_{q, \mu}(\vec{r})
& = & \vnabla \cdot \vnabla' \; s_{q,\mu} (\vec{r},\vec{r}')
      \big|_{\vec{r} = \vec{r}'}
      \nn \\
\label{eq:locdensities:j}
\vec{j}_q (\vec{r})
& = & - \tfrac{i}{2} (\vnabla - \vnabla') \;
      \rho_q (\vec{r},\vec{r}') \big|_{\vec{r} = \vec{r}'}
      \nn \\
\label{eq:locdensities:J}
J_{q,\mu \nu}(\vec{r})
& = & - \tfrac{i}{2} (\nabla_\mu - \nabla_\mu^\prime) \;
      s_{q, \nu} (\vec{r},\vec{r}') \big|_{\vec{r} = \vec{r}'}
      \nn \\
F_{q, \mu} (\vec{r})
& = & \tfrac{1}{2}
      \sum_{\nu = x}^{z}
      \big(   \nabla_\mu \nabla_\nu^\prime
            + \nabla_\mu^\prime \nabla_\nu
      \big) \, s_{q, \nu} (\vec{r}, \vec{r}') \big|_{\vec{r} = \vec{r}'}
\end{eqnarray}
which are
the density $\rho_q (\vec{r})$,
the kinetic density $\tau_q (\vec{r})$,
the current (vector) density $\vec{j}_q (\vec{r})$,
the spin (pseudovector) density $\vec{s}_q (\vec{r})$,
the spin kinetic (pseudovector) density $\vec{T}_q (\vec{r})$,
the spin-current (pseudotensor) density  $J_{q,\mu \nu}(\vec{r})$, and
the tensor-kinetic (pseudovector) density $\vec{F}_{q} (\vec{r})$.
$\rho_q (\vec{r})$, $\tau_q (\vec{r})$ and $J_{q, \mu \nu}(\vec{r})$
are time-even, while $\vec{s}_q (\vec{r})$, $\vec{T}_q (\vec{r})$,
$\vec{j}_q (\vec{r})$ and $\vec{F}_{q} (\vec{r})$ are time-odd. For
a detailed discussion of their symmetries see Ref.~\cite{Dob00a}. There
are other local densities up to second order in derivatives that can
be constructed, but when constructing an energy functional they either
cannot be combined with others to terms with proper symmetries or they
lead to terms that are not independent from the others~\cite{Dob96a}.

The cartesian spin-current pseudotensor density $J_{\mu\nu}$ can be
decomposed into pseudoscalar, (anti-symmetric) vector and
(symmetric) traceless pseudotensor parts, all of which have well-defined
transformation properties under rotations
\begin{equation}
J_{\mu\nu} (\vec{r})
=   \tfrac{1}{3} \delta_{\mu \nu} \, J^{(0)} (\vec{r})
  + \tfrac{1}{2} \sum_{\kappa = x}^{z} \epsilon_{\mu\nu\kappa} \,
                 J^{(1)}_{\kappa} (\vec{r})
  + J^{(2)}_{\mu\nu} (\vec{r})\,,
\end{equation}
where $\delta_{\mu\nu}$ is the Kronecker symbol and $\epsilon_{\mu\nu\kappa}$
the Levi-Civita tensor. The pseudoscalar, vector and pseudotensor
parts expressed in terms of the cartesian tensor are given by
\begin{eqnarray}
\label{eq:J:recoupled}
J^{(0)}(\vec{r})
& = & \sum_{\mu = x}^{z} J_{\mu\mu}(\vec{r})\,,
       \\
J^{(1)}_{\kappa} (\vec{r})
& = & \sum_{\mu,\nu = x}^{z} \epsilon_{\kappa \mu \nu} \, J_{\mu\nu}
      (\vec{r})\,,      \nn \\
J^{(2)}_{\mu \nu} (\vec{r})
& = &   \tfrac{1}{2} [ J_{\mu \nu}(\vec{r}) + J_{\nu \mu}(\vec{r}) ]
      - \tfrac{1}{3} \delta_{\mu\nu} \sum_{\kappa = x}^{z}
        J_{\kappa \kappa} (\vec{r})\,.\nn
\end{eqnarray}
The vector spin current density
$\vec{J}^{(1)}(\vec{r}) \equiv \vec{J} (\vec{r})$ is often called
spin-orbit current, as it enters the spin-orbit energy density.
\footnote{Some authors call $\vec{J} (\vec{r})$ \emph{spin density},
which is ambiguous and confusing when discussing the complete energy
density functional including terms that contain the time-odd
$\vec{s}(\vec{r})$.}

For the formal discussion of the physical content of the Skyrme energy
functional it is of advantage to recouple the proton and
neutron densities to isoscalar and isovector densities, for example
\begin{eqnarray}
\rho_{0} (\vec{r})
& = & \rho_\nrm (\vec{r}) + \rho_\prm (\vec{r})\,,
      \nn \\
\rho_{1} (\vec{r})
& = & \rho_\nrm (\vec{r}) - \rho_\prm (\vec{r})
\end{eqnarray}
and similar for all others. As we assume pure proton and neutron states,
only the $T_z = 0$ component of the isovector density is non-zero,
which we exploit to drop the index $T_z$ from the isovector
densities $\rho_{1 T_z} (\vec{r})$ etc.


\subsubsection{Skyrmes's central force}


We will use the standard density-dependent central Skyrme force
\begin{eqnarray}
\label{eq:Skyrme:central}
v^{\text{\central}} (\vec{R},\vec{r})
& = &  t_0 \, ( 1 + x_0 \hat{P}_\sigma ) \; \delta (\vec{r})
      \nn \\
& + & \tfrac{1}{6} \, t_{3} \, ( 1 + x_{3} \hat{P}_\sigma ) \,
        \rho^{\alpha} (\vec{R}) \; \delta (\vec{r})
      \nn \\
& + & \tfrac{1}{2} \, t_1 \, ( 1 + x_1 \hat{P}_\sigma )
        \big[   \hat{\vec{k}}^{\prime 2} \; \delta (\vec{r})
              + \delta (\vec{r}) \; \hat{\vec{k}}^2
        \big]
      \nn \\
& + & t_2 \ ( 1 + x_2 \hat{P}_\sigma ) \,
      \hat{\vec{k}}^{\prime} \cdot \delta (\vec{r}) \; \hat{\vec{k}}
\end{eqnarray}
where we use the shorthand notation
\begin{eqnarray}
\label{eq:r:R}
\vec{r}
& = & \vec{r}_1 - \vec{r}_2\,,
      \nn \\
\vec{R}
& = & \tfrac{1}{2} ( \vec{r}_1 + \vec{r}_2 )\,,
\end{eqnarray}
while $\hat{\vec{k}}$ is the usual operator for relative momenta
\begin{equation}
\label{eq:k:kp}
\hat{\vec{k}}
 =  - \tfrac{i}{2} ( \vnabla_1 - \vnabla_2 )
\end{equation}
and $\hat{\vec{k}}^{\prime}$ its complex conjugated acting on the left.
Finally, $\hat{P}_\sigma$ is the spin exchange operator that controls the
relative strength of the $S=0$ and $S=1$ channels for a given term
in the two-body interaction
\begin{equation}
\hat{P}_\sigma
= \tfrac{1}{2} \, ( 1 + \hat{\vsigma}_1 \cdot \hat{\vsigma}_2 )\,.
\end{equation}
As said above, we restrict ourselves to a parameterization of the Skyrme
energy functional as obtained from the average value of an effective two-body
vertex in the reference Slater determinant. We decompose the isoscalar and
isovector parts of the resulting energy density functional
$\mathcal{H}^{\text{\central}}$ into a part
$\mathcal{H}_t^{\text{\central,even}}$
that is composed entirely of time-even densities and currents, and a part
$\mathcal{H}_t^{\central,\text{odd}}$ that contains terms which are bilinear in
time-odd densities and currents and vanishes in intrinsically time-reversal
invariant systems
\begin{eqnarray}
\label{eq:skyrme:energy}
\mathcal{H}^{\text{\central}} (\vec{r})
& = & \sum_{t=0,1}
      \big[   \mathcal{H}_t^{\text{\central,even}} (\vec{r})
            + \mathcal{H}_t^{\text{\central,odd}}  (\vec{r})
      \big]\,.
\end{eqnarray}
Both $\mathcal{H}_t^{\text{\central,even}}$ and
$\mathcal{H}_t^{\text{\central,odd}}$ are of course constructed
such that they are time-even; they are given by~\cite{Eng75a,Per04a}
\begin{eqnarray}
\label{eq:ef:central}
\mathcal{H}_t^{\text{\central,even}}
& = &   A^\rho_t [\rho_0 ] \, \rho_t^2
      + A^{\Delta \rho}_t  \, \rho_t \Delta \rho_t
      + A^\tau_t           \, \rho_t \tau_t
      \nn \\
&   & \quad
      - A^{T}_t \sum_{\mu, \nu = x}^{z} J_{t, \mu \nu} J_{t, \mu \nu}\,,
      \nn \\
\mathcal{H}_t^{\text{\central,odd}}
& = &   A^s_t [\rho_0 ] \, \vec{s}_t^2
      - A^\tau_t        \, \vec{j}_t^2
      \nn \\
&   & \quad
      + A^{\Delta s}_t  \, \vec{s}_t \cdot \Delta \vec{s}_t
      + A^{T}_t         \, \vec{s}_t \cdot \vec{T}_t
\,,
\end{eqnarray}
where $A^\rho_t [\rho_0 ]$ and $A^s_t [\rho_0 ]$ are density dependent
coupling constants that depend on the total (isoscalar) density. The
detailed relations between the coupling constants of the functional and
the central Skyrme force are given in appendix~\ref{app:sect:cpl}.
The notation reflects that two pairs of terms in
$\mathcal{H}_t^{\text{\central,even}}$ and
$\mathcal{H}_t^{\text{\central,odd}}$ are connected by the requirement of
local gauge invariance of the Skyrme energy functional~\cite{Dob95a}.


\subsubsection{A zero-range spin-orbit force}


The spin-orbit force used with most standard Skyrme interactions
\begin{equation}
\label{eq:Skyrme:LS}
v^{\text{LS}} (\vec{r})
= i W_0 \, ( \hat{\vsigma}_1 + \hat{\vsigma}_2 ) \cdot
  \hat{\vec{k}}^{\prime} \times \delta (\vec{r}) \; \hat{\vec{k}}
\end{equation}
is a special case of the one proposed by Bell and Skyrme~\cite{Bel56a,Sky59b}.
Again, the corresponding energy  functional~\cite{Eng75a,Per04a} can be
separated into a time-even and a time-odd term
\begin{eqnarray}
\mathcal{H}^{\text{LS}} (\vec{r})
& = & \sum_{t=0,1}
      \big[   \mathcal{H}_t^{\text{LS,even}} (\vec{r})
            + \mathcal{H}_t^{\text{LS,odd}}  (\vec{r})
      \big]
\end{eqnarray}
where
\begin{eqnarray}
\label{eq:ef:ls}
\mathcal{H}_t^{\text{LS,even}}
& = & A^{\nabla \cdot J}_t \; \rho_t \nabla \cdot \vec{J}_t
      \nn \\
\mathcal{H}_t^{\text{LS,odd}}
& = & A^{\nabla \cdot J}_t \; \vec{s}_t \cdot \nabla \times \vec{j}_t
\end{eqnarray}
which share the same coupling constant as again both terms are linked by the
local gauge invariance of the energy functional. The relation between
the $A^{\nabla \cdot J}_t$ and the one coupling constant of the two-body
spin-orbit force $W_0$ is given in appendix~\ref{app:sect:cpl}.


\subsubsection{Skyrme's tensor force}
\label{sect:tensor:force}


By convention, the tensor operator in the tensor force is
constructed using the unit vectors in the direction of the
relative coordinate $\vec{e}_r = \vec{r} / |\vec{r}|$
and subtracting $\hat\vsigma_1 \cdot \hat\vsigma_2$
\begin{equation}
\hat{S}_{12}
= 3 (\hat\vsigma_1 \cdot \vec{e}_r ) (\hat\vsigma_2 \cdot \vec{e}_r )
  - \hat\vsigma_1 \cdot \hat\vsigma_2
\,,
\end{equation}
such that its mean value vanishes for a relative $S$ state, which decouples the
central and tensor channels of the interaction. The operator $\hat{S}_{12}$
commutes with the total spin $[\hat{S}_{12}, \hat{\vec{S}}^2] = 0$, therefore it
does not mix partial waves with different spin, i.e.~spin singlet and spin
triplet states. In particular, it does not act in spin singlet states at all, as
$\hat{S}_{12} \hat{P}_{S=0} = 0$ (see section 13.6 of Ref.~\cite{Nil95a}). As a
consequence, there is no point in multiplying a tensor force with an exchange
operator $(1 + x_t \hat{P}_{\sigma})$ as done for the central force, as this
will only lead to an overall rescaling of its strength.

The derivation of the general energy functional from a zero-range two-body
tensor force is discussed in detail in Refs.~\cite{Flothesis,Per04a}. We repeat
here the details relevant for our discussion, starting from the two zero-range
tensor forces proposed by Skyrme~\cite{Sky56a,Sky59a}
\begin{widetext}
\begin{eqnarray}
\label{eq:Skyrme:tensor}
v^{\text{\tensor}} (\vec{r})
& = & \tfrac{1}{2} \, t_e \;
    \Big\{
    \big[ 3 \,( \vsigma_1 \cdot \vec{k}' ) \, ( \vsigma_2 \cdot \vec{k}' )
          - ( \vsigma_1 \cdot \vsigma_2 ) \, \vec{k}^{\prime 2} \,
    \big] \; \delta (\vec{r})
   +  \delta (\vec{r}) \;
      \big[ 3 \, ( \vsigma_1 \cdot \vec{k} ) \, ( \vsigma_2 \cdot \vec{k} )
            -  ( \vsigma_1 \cdot \vsigma_2) \, \vec{k}^{2}
      \Big]
    \Big\}
   \nn \\
&  & + t_o \,
     \Big[
       3 \, ( \vsigma_1 \cdot \vec{k}' ) \, \delta (\vec{r}) \,
            ( \vsigma_2 \cdot \vec{k} )
      - ( \vsigma_1 \cdot \vsigma_2 ) \, \vec{k}' \cdot \,
        \delta (\vec{r}) \, \vec{k}
     \Big]\,
\end{eqnarray}
\end{widetext}
where $\vec{r}$, $\hat{\vec{k}}$ and $\hat{\vec{k}}^{\prime}$
are defined as above, Eqs.~(\ref{eq:r:R}) and~(\ref{eq:k:kp}).
The corresponding energy density functional can again be
decomposed in a time-even and a time-odd part
\begin{eqnarray}
\label{eq:EF:tensor}
\mathcal{H}^{\text{\tensor}} (\vec{r})
& = & \sum_{t=0,1}
      \big[   \mathcal{H}_t^{\text{\tensor,even}} (\vec{r})
            + \mathcal{H}_t^{\text{\tensor,odd}}  (\vec{r})
      \big]
\end{eqnarray}
with~\cite{Per04a}
\begin{eqnarray}
\label{eq:ef:tensor}
\mathcal{H}_t^{\text{\tensor,even}}
& = & - B^{T}_t \sum_{\mu, \nu = x}^{z} J_{t, \mu \nu} J_{t, \mu \nu}
      -\tfrac{1}{2} \, B^{F}_t \, \Big( \sum_{\mu = x}^{z} J_{t,\mu \mu} \Big)^2
      \nn \\
&   & - \tfrac{1}{2} \, B^{F}_t \sum_{\mu, \nu = x}^{z} J_{t, \mu \nu}
                                                        J_{t, \nu \mu}
       \nn \\
\mathcal{H}_t^{\text{\tensor,odd}}
& = &   B^{T}_t \, \vec{s}_t \cdot \vec{T}_t
      + B^{F}_t \, \vec{s}_t \cdot \vec{F}_t
      \phantom{\sum}
      \nn \\
&   & + B^{\Delta s}_t \, \vec{s}_t \cdot \Delta \vec{s}_t
      + B^{\nabla s}_t \, (\nabla \cdot \vec{s}_t)^2\,,
\end{eqnarray}
where we already used the local gauge invariance of the energy
functional~\cite{Per04a} for the expressions of the coupling constants.
The actual expressions for the coupling constants expressed in terms
of the two coupling constants $t_e$ and $t_o$ of the tensor forces
are given in appendix~\ref{app:sect:cpl}.

The ``even'' term proportional to $t_e$ in the two-body tensor force
(\ref{eq:Skyrme:tensor}) mixes relative $S$ and $D$ waves, while the
``odd'' term proportional to $t_o$ mixes relative $P$ and $F$ waves.
Thus, due to the fact that both act in spin-triplet states only,
antisymmetrization implies that the former acts in isospin-singlet
states (and hence contributes to the neutron-proton interaction
only) and the latter in isospin-triplet states (contributing both to
the like-particle and neutron-proton interactions).
The central and spin-orbit interactions as we use them, however, do
not contain $D$ or $F$ wave interactions. From this point of view,
one might suspect a mismatch when combining the various interaction
terms. From the point of view of the energy functional (\ref{eq:ef:tensor}),
however, all contributions from the zero-range tensor force are of the
same second order in derivatives as the contributions from the non-local
part of the central Skyrme force (\ref{eq:ef:central}) and from
the spin-orbit force (\ref{eq:ef:ls}).

In the time-even part of the energy functional
$\mathcal{H}_t^{\text{\tensor,even}}$, there appear three different
combinations of the cartesian components of the spin current tensor.
The term proportional to $B^{T}_t$ contains the symmetric combination
$J_{\mu \nu} J_{\mu \nu}$ as it already appeared in the
energy functional from the central Skyrme interaction
(\ref{eq:ef:central}), while the term proportional to $B^{F}_t$
contains two different terms, namely the antisymmetric combination
$J_{\mu \nu} J_{\nu \mu}$ and the square of the trace of $J_{\nu \mu}$.


\subsubsection{Combining central and tensor interactions}


The Skyrme energy functional representing central, tensor, and spin-orbit
interactions is given by
\begin{widetext}
\begin{eqnarray}
\label{eq:EF:full}
\mathcal{E}_{\text{Skyrme}}
& = &   \mathcal{E}_{\text{\central}}
      + \mathcal{E}_{\text{LS}}
      + \mathcal{E}_{\text{\tensor}}
      \nn \\
& = & \int \! d^3r \sum_{t=0,1}
      \bigg\{  C^\rho_t [\rho_0] \, \rho_t^2
             + C^s_t    [\rho_0] \, \vec{s}_t^2
             + C^{\Delta \rho}_t  \rho_t \Delta \rho_t
             + C^{\nabla s}_t     (\nabla \cdot \vec{s}_t)^2
             + C^{\Delta s}_t     \vec{s}_t \cdot \Delta \vec{s}_t
             + C^\tau_t           ( \rho_t \tau_t - \vec{j}_t^2 )
      \nn \\
&   &
             + C^{T}_t \Big(  \vec{s}_t \cdot \vec{T}_t
                 - \sum_{\mu, \nu = x}^{z} J_{t, \mu \nu} J_{t, \mu \nu} \Big)
             + C^{F}_t \Big[  \vec{s}_t \cdot \vec{F}_t
                  - \tfrac{1}{2} \Big( \sum_{\mu = x}^{z} J_{t,\mu \mu}
                                 \Big)^2
                 - \tfrac{1}{2}
                   \sum_{\mu, \nu = x}^{z} J_{t, \mu \nu} J_{t, \nu \mu}
                \Big]
      \nn \\
&   &
             + C^{\nabla \cdot J}_t ( \rho_t \nabla \cdot \vec{J}_t
                                    + \vec{s}_t \cdot \nabla \times \vec{j}_t)
      \bigg\}
\,.
\end{eqnarray}
\end{widetext}
This functional contains all possible bilinear terms up to second order in the
derivatives that can be constructed from local densities and that are
invariant under spatial and time inversion, rotations, and local gauge
transformations~\cite{Per04a}.

Some of the coupling constants are completely defined by the standard central
Skyrme force, \ie~$C^{\rho}_t =  A^{\rho}_t$, $C^{s}_t =  A^{s}_t$,
$C^{\tau}_t =  A^{\tau}_t$, and $C^{\Delta \rho}_t =  A^{\Delta \rho}_t$,
two by the spin-orbit force, $C^{\nabla J}_t = A^{\nabla J}_t$, others by the
tensor force, $C^{F}_t = B^{F}_t$ and $C^{\nabla s}_t = B^{\nabla s}_t$, while
some are the sum of coupling constants from both central and
tensor forces, $C^{T}_t = A^{T}_t + B^{T}_t$, and
$C^{\Delta s}_t = A^{\Delta s}_t + B^{\Delta s}_t$.

The three terms bilinear in $J_{\mu \nu}$ can be recoupled into
terms bilinear in its pseudoscalar, vector, and pseudotensor
components $J^{(0)}$, $J^{(1)}$, and $J^{(2)}$, Eq.~(\ref{eq:J:recoupled}),
which is prefered by some authors~\cite{Per04a}
\begin{widetext}
\begin{eqnarray}
\label{eq:EJ2:recoupled}
\sum_{\mu, \nu = x}^{z} J_{t, \mu \nu} J_{t, \mu \nu}
& = &   \tfrac{1}{3} \, \big( J_t^{(0)} \big)^2
      + \tfrac{1}{2} \, \vec{J}_t^2
      + \sum_{\mu, \nu = x}^{z} J_{t, \mu \nu}^{(2)} J_{t, \mu \nu}^{(2)}
      \\
\tfrac{1}{2}
\Big[ \Big( \sum_{\mu = x}^{z} J_{t,\mu \mu} \Big)^2
     + \sum_{\mu, \nu = x}^{z} J_{t, \mu \nu} J_{t, \nu \mu}
\Big]
& = &   \tfrac{2}{3} \, \big( J_t^{(0)} \big)^2
      - \tfrac{1}{4} \, \vec{J}_t^2
      + \tfrac{1}{2}
        \sum_{\mu, \nu = x}^{z} J_{t, \mu \nu}^{(2)} J_{t, \mu \nu}^{(2)}
\,.
\end{eqnarray}
\end{widetext}
After combining (\ref{eq:EF:full}) with the kinetic, Coulomb, pairing and other
contributions from (\ref{eq:efu:complete}), the mean-field equations are
obtained by standard functional derivative techniques from the total energy
functional~\cite{RMP,Per04a}.

The complete Skyrme energy functional (\ref{eq:EF:full}) has quite complicated
a structure, and in the most general case leads to seven distinct mean fields
in the single-particle Hamiltonian~\cite{Per04a}. As already mentioned, we
want to divide the examination of those terms that contain two derivatives and
two Pauli matrices in the complete functional, \ie~those terms from the
central Skyrme force that are often neglected and all the terms from the
tensor Skyrme force, into three distinct steps: First, in the present paper,
we enforce spherical symmetry which removes all time-odd densities and all
but one out of the nine components of the spin current tensor $J_{\mu \nu}$
as will be outlined in the following section. A subsequent paper~\cite{II}
will discuss deformed even-even nuclei where the complete spin current tensor
$J_{\mu \nu}$ is present, and future work will address the time-odd part
of the energy functional (\ref{eq:EF:full}).


\subsection{The Skyrme energy functional in spherical symmetry}
\label{sect:skyrmefu:sphere}


For the rest of this paper, we will concentrate on spherical nuclei, enforcing
spherical symmetry of the $N$-body wave functions. As a consequence, the
canonical
single-particle wave functions $\Psi_i$~\cite{Dob84a} can be labeled
by $j_i$, $\ell_i$ and $m_i$.
The index $n_i$ labels the different states with same $j_i$ and
$\ell_i$. The functions $\Psi_i$ separates into a radial part $\psi$
and an angular and spin part, represented by a tensor spherical harmonic
$\Omega_{j \ell m}$
\begin{equation}
\Psi_{n j \ell m} (\vec{r})
= \tfrac{1}{r} \psi_{n j \ell} (r) \;
  \Omega_{j \ell m} (\theta, \phi)\,.
\end{equation}
Spherical symmetry also enforces that all magnetic substates of
$\Psi_{n j \ell m}$ have the same occupation probability
$v^2_{n j \ell m} \equiv v^2_{n j \ell}$ for all
$- j \leq m \leq j$. For a static spherical state, all time-odd densities are
zero $\vec{s}_q(\vec{r}) = \vec{T}_q(\vec{r}) = \vec{j}_q(\vec{r}) =
\vec{F}_q(\vec{r}) = {\mathbf 0}$, as are the corresponding mean fields in the
single-particle Hamiltonian.

Enforcing spherical symmetry also greatly simplifies the spin-current tensor,
both the pseudoscalar and pseudotensor parts of $J_{\mu \nu}$ vanish. From the
vector spin-orbit current, only the radial component is non-zero, which is
given by~\cite{Vau72a}
\begin{eqnarray}
\label{eq:j:radial}
J_q (r)
& = & \frac{1}{4 \pi r^3} \! \sum_{n,j,\ell}
       ( 2 j + 1 ) \, v_{n j \ell}^2
      \nn \\
&   & \times
      \Big[  j ( j + 1)
           - \ell ( \ell + 1)
           - \tfrac{3}{4} \Big] \, \psi^2_{n j \ell} (r)
\end{eqnarray}
so that there is only one out of the nine components of the spin-current tensor
density that contributes in spherical nuclei. Unlike the total density $\rho$
and the kinetic density $\tau$, that are bulk properties of the nucleus and
grow with the size of the nucleus, the spin-orbit current is a shell effect
that shows strong fluctuations. Assume the two shells with same $n$ and $\ell$
which are split by the spin-orbit interaction, one coupled with the spin to
$j = \ell + \tfrac{1}{2}$, the other to $j = \ell - \tfrac{1}{2}$. It is easy
to verify that their contributions to $J_q (r)$ are equal but of opposite signs
such that they cancel when (i) both shells are completely filled and (ii) their
radial wave functions are identical $\psi_{n,\ell+1/2,\ell}
=\psi_{n,\ell-1/2, \ell}$.
Although the latter condition is never exactly fulfilled, this demonstrates
that
the spin-orbit current is not a bulk property, but a shell effect that strongly
fluctuates with $N$ and $Z$. It nearly vanishes in so-called
\emph{spin-saturated} nuclei, where all spin-orbit partners are either
completely occupied or empty, and it might be quite large when only the
$j=\ell+1/2$ level out of one or even several pairs of spin-orbit partners is
filled.

Altogether, the Skyrme part of the energy density functional in spherical
nuclei is reduced to
\begin{eqnarray}
\label{eq:EF:sphere}
\mathcal{H}^{\text{Skyrme}}
& = & \sum_{t=0,1}
      \Big\{  C^\rho_t [\rho_0] \, \rho_t^2
            + C^{\Delta \rho}_t  \, \rho_t \Delta \rho_t
            + C^\tau_t \, \rho_t \tau_t
      \nn \\
&   &       + \tfrac{1}{2} \, C^{J}_t \, \vec{J}_t^2
            + C^{\nabla \cdot J}_t \, \rho_t \nabla \cdot \vec{J}_t
      \Big\}
\,,
\end{eqnarray}
where we have introduced an effective coupling constant $C_t^{J}$ of the
$\vec{J}_t^2$ tensor terms at sphericity, such that the corresponding
contribution to the energy functional is given by
\begin{equation}
\label{eq:E:tensor:sphere}
\mathcal{H}^{\text{\tensor}}
= \sum_{t=0,1} \tfrac{1}{2} \, C_t^{J} \, \vec{J}_t^2
= \sum_{t=0,1}
    \left( -\tfrac{1}{2} C^{T}_t + \tfrac{1}{4} C^{F}_t \right) \,
    \vec{J}_t^2 \,.
\end{equation}
The effective coupling constants can be separated back into contributions from
the non-local central and tensor forces
\begin{equation}
C_t^{J} = A_t^{J} + B_t^{J}
\end{equation}
which are given by
\begin{eqnarray}
A_0^{J}
& = &       \tfrac{1}{8} \, t_1 \, \big( \tfrac{1}{2} - x_1 \big)
          - \tfrac{1}{8} \, t_2 \, \big( \tfrac{1}{2} + x_2 \big)
      \nn \\
A_1^{J}
& = & \tfrac{1}{16} \, t_1 - \tfrac{1}{16} \, t_2
      \nn \\
B_0^{J}
& = & \tfrac{5}{16} \,  ( t_e + 3 t_o )
  =   \tfrac{5}{48} \, (T + 3U)
      \nn \\
B_1^{J}
& = & \tfrac{5}{16} \, ( t_o - t_e )
  =   \tfrac{5}{48} \, (U - T)
\,,
\end{eqnarray}
where we also give the expressions using the notation $T = 3 t_e$
and $U = 3 t_o$ employed in~\cite{Flothesis,Sta77a,Col07a}.

For the following discussion it will be also illuminating to recouple this
expression to a representation that uses proton and neutron densities,
where we use the notation introduced in Ref.~\cite{Sta77a}
\begin{eqnarray}
\label{eq:Htensor:pn}
\mathcal{H}^{\text{\tensor}}
& = &   \tfrac{1}{2} \, \alpha \, ( \vec{J}_n^2 + \vec{J}_p^2 )
      + \beta \, \vec{J}_n \cdot \vec{J}_p
\,,
\end{eqnarray}
with
\begin{alignat}{3}
\label{eq:cjtot}
\alpha
& = C_0^{J} + C_1^{J}\,,
    & \quad
\beta
& =  C_0^{J} - C_1^{J}\,,
    \nn \\
C_0^{J}
& =  \tfrac{1}{2} \, (\alpha + \beta )\,,
     & \quad
C_1^{J}
& =  \tfrac{1}{2} \, (\alpha - \beta )
\,.
\end{alignat}
The proton-neutron coupling constants $\alpha = \alpha_C + \alpha_T$ and
$\beta = \beta_C + \beta_T$ can again be separated into contributions
from central and tensor forces
\begin{eqnarray}
\alpha_C
& = &   \tfrac{1}{8} \, ( t_1 - t_2 )
      - \tfrac{1}{8} \, ( t_1 x_1 + t_2 x_2 )\,,
      \nn \\
\beta_C
& = & - \tfrac{1}{8} \, ( t_1 x_1 + t_2 x_2 )\,,
      \nn \\
\alpha_T
& = & \tfrac{5}{4} \, t_o
  =   \tfrac{5}{12} \, U\,,
      \nn \\
\beta_T
& = & \tfrac{5}{8} \, ( t_e + t_o )
  =   \tfrac{5}{24} \, (T + U)\,.
\end{eqnarray}
As could be expected, the isospin-singlet tensor force contributes
only to the proton-neutron term, while the isospin-triplet
tensor force contributes to both.

The spin-orbit potential of the neutrons is given by
\begin{eqnarray}
\label{eq:wpot:tot}
W_n (r)
& = & \frac{\delta \mathcal{E}}
           {\delta \vec{J}_n (r)}\cdot \mathbf e_r
      \nn \\
& = & \frac{W_0}{2} \, \big( 2\nabla \rho_n + \nabla \rho_p )
      + \alpha\, J_n + \beta \,J_p
\,.
\end{eqnarray}
The expression
for the protons is obtained exchanging the indices for protons and neutrons.
In spherical symmetry, the tensor force gives a contribution to
the spin-orbit potential, but does not alter the structure of the
spin-orbit terms in the single-particle Hamiltonian as such.
This will be different in the case of deformed mean fields~\cite{Per04a,II}.

The dependence of the spin-orbit potential $W_q (r)$ on the spin-orbit
current $J_q(r)$ through the tensor terms is the source of a potential
instability. When the spin-orbit splitting becomes larger than the
splitting of the centroids of single-particle states with different
orbital angular momentum $\ell$, the reordering of levels might increase
the number of spin-unsaturated levels, which increases the spin-orbit
current $J_n$ and feeds back on the spin-orbit potential by increasing
it even further, which ultimately leads to an unphysical shell structure.
An example will be given in appendix~\ref{sect:app:instability}.


\subsection{A brief history of tensor terms in the central Skyrme
               energy functional}
\label{sect:vertex:edf:dft}


For the interpretation of the parameterizations we will describe below it
is important to point out that within our choice of the effective
Skyrme interaction as an antisymmetrized vertex the two coupling
constants of the contribution from the central force to $\mathcal{H}^{T}$,
Eq.~(\ref{eq:E:tensor:sphere}), either represented through $A_0^{J}$, $A_1^{J}$
or through $\alpha_C$, $\beta_C$, are not independent from the coupling
constants $A^\tau_0$, $A^\tau_1$, $A^{\Delta \rho}_0$, and
$A^{\Delta \rho}_1$, that appear in Eq.~(\ref{eq:EF:sphere}).
Through the expressions given in appendix~\ref{app:sect:cpl},
all six of them are determined by the four coupling constants
$t_1$, $x_1$, $t_2$, and $x_2$ from the central Skyrme force,
Eq.~(\ref{eq:Skyrme:central}). As a consequence, a tensor force
is absolutely necessary to decouple the values of the $C_t^{J}$
from those of the $C^\tau_t$ and $C^{\Delta \rho}_t$,
which determine the isoscalar and isovector effective masses
and give the dominant contribution to the surface and surface
asymmetry coefficients, respectively.

This interpretation of the Skyrme interaction is, however, far from being common
practice and a source of confusion and potential inconsistencies in the
literature. Many authors have used parameterizations of the central and
spin-orbit Skyrme energy functional with coupling constants that in one way or
the other do not exactly correspond to the functional obtained from
Eqns.~(\ref{eq:Skyrme:central}) and (\ref{eq:Skyrme:LS}), which, depending on
the point of view, can be seen as an approximation to or a generalization of the
original Skyrme interaction. As the most popular modification concerns the
tensor terms, a few comments on the subject are in order. Again, the practice
goes back to the seminal paper by Vautherin and Brink \cite{Vau72a}, who state
that ``the contribution of this term to [the spin-orbit potential] is quite
small. Since it is difficult to include such a term in the case of deformed
nuclei, it has been neglected''. This choice was further motivated by the
interpretation of the effective Skyrme interaction as a density-matrix expansion
(DME)~\cite{Neg70a,Neg72a,Neg75a,Cam78a}. All early parameterizations as SI and
SII~\cite{Vau72a}, SIII-SVI~\cite{Bei75a}, SkM~\cite{Kri80a} and
SkM$^*$~\cite{Bar82a} followed this example and did not contain the $\vec{J}^2$
terms. Beiner~\etal~\cite{Bei75a} weakened the case for $\vec{J}^2$ terms
further by pointing out that they might lead to unphysical single-particle
spectra. During the 1980s and later, however, it became more popular to include
them, for example in SkP~\cite{Dob84a}, the parameterizations T1-T9 by
Tondeur~\etal~\cite{Ton84a}, E$_{\sigma}$ and Z$_{\sigma}$ by Friedrich and
Reinhard~\cite{Fri86a}. Some of the recent parameterizations come in pairs,
where variants without and with $\vec{J}^2$ terms are fitted within the same fit
protocol, for example (SLy4, SLy5) and (SLy6, SLy7) in Ref.~\cite{Cha98a}, or
(SkO, SkO') in Ref.~\cite{Rei99a}.

Interestingly, all but one parameterization of the central Skyrme
interaction found in the literature set the coupling constants of the
$\vec{J}^2$ terms either to their Skyrme force value (\ref{eq:cpl:SF})
or strictly to zero. The exception is Ref.~\cite{Ton83a} by Tondeur, where
an independent fit of the coupling constants of the $\vec{J}^2$ terms
was attempted, making explicit reference to a DME interpretation
of the energy functional.

Setting the coupling constants of a term to zero when one does not know
how to adjust its parameters is of course an acceptable practise when
permitted by the chosen framework. For Skyrme interactions fitted
without the $\vec{J}^2$ terms, the situation becomes confusing when one
looks at deformed nuclei and any situation that breaks time-reversal
invariance. First of all, Galilean invariance of the energy functional
dictates that the coupling constant of the $\vec{s} \cdot \vec{T}$ terms
is also set to zero, as already indicated by the presentation of the
energy functional in Eq.~(\ref{eq:EF:full}). Second,
using a DME interpretation of the Skyrme energy functional
in one place, but the interrelations from the two-body Skyrme force in
all others is not entirely satisfactory. Many authors who drop the
$\vec{J}^2$ terms rarely show scruples to keep most of the time-odd
terms in the Skyrme energy functional (\ref{eq:EF:full}) with coupling
constants $A_t^s$ and $A_t^{\Delta s}$
from (\ref{eq:cpl:SF}), although they are not at all constrained
in the common fit protocols employing properties of even-even nuclei and
spin-saturated nuclear matter. For a list of exceptions see Sect.~II.A.2.d
of Ref.~\cite{RMP}. An alternative is to set up a hierarchy of terms, as
it was attempted by Bonche, Flocard and Heenen in their mean-field and
beyond codes, which set $A_t^{\Delta s} = 0$ in addition to the coupling
constant of the $\vec{J}^2$ terms, as all three terms have in common that
they couple two Pauli matrices with two derivatives in different manners,
see the footnote on page 129 of \cite{Bon87a}.

There are also inconsistent applications of parameterizations without
$\vec{J}^2 - \vec{s} \cdot \vec{T}$ terms to be found in the literature.
For example, almost all applications of Skyrme interactions to the Landau
parameters $g_\ell$ and $g_\ell'$ and the properties of polarized
nuclear matter, include the contribution from the $\vec{s} \cdot \vec{T}$
terms, although it should be dropped for parameterizations fitted without
$\vec{J}^2$ terms. Similarly, most RPA and QRPA codes include them for
simplicity, see the discussion in Refs.~\cite{Eng99a,Ben02a,Ter05a}.

As it is relevant for the subject of the present paper, we also mention
another generalization of the Skyrme interaction that invokes the
interpretation of the Skyrme energy functional in a DME framework.
The spin-orbit force (\ref{eq:Skyrme:LS}) fixes the isospin mix of the
corresponding terms in the Skyrme energy functional (\ref{eq:EF:full})
such that $A_0^{\nabla J} = 3 A_1^{\nabla J}$ (\ref{eq:cpl:LSF}).
There are a few parameterizations as MSkA~\cite{Sha95a}, SkI3 and
SkI4~\cite{Rei95a}, SkO and SkO'~\cite{Rei99a}
and SLy10~\cite{Cha98a} that liberate the isospin
degree of freedom in the spin-orbit functional. A DME interpretation
of the energy functional is mandatory for this generalization. It
is motivated by the better performance of standard relativistic
mean-field models for the kink of the charge radii in Pb isotopes.
Note that the standard RMF models are effective Hartree theories
without exchange terms, and that the standard Lagrangians have very
limited isovector degrees of freedom~\cite{RMP}, both of which supress a strong
isospin dependence of the spin-orbit interaction. It is interesting
to note that the existing fits of Skyrme energy functionals with
generalized spin-orbit interaction do not improve spin-orbit
splittings~\cite{Ben99b}.


\section{The fits}
\label{sect:fit}


\subsection{General remarks}


In order to study the effect of the $\vec{J}^2$ terms, we have built a set of
36 effective interactions that systematically cover the region of coupling
constants $C^{J}_0$ and $C^{J}_1$ that give a reasonable description of finite
nuclei in connection with the standard central and spin-orbit Skyrme forces. At
variance with the perturbative approach used in Refs.~\cite{Sta77a,Col07a},
each of these parameterizations has been fitted separately, following a
procedure nearly identical to that used for the construction of the SLy
parameterizations~\cite{Cha97a,Cha98a}, so that we can keep the connection
between the new fits with parameterizations that have been applied to a large
variety of observables and phenomena. The Saclay-Lyon fit protocol focuses on
the simultaneous reproduction of nuclear bulk properties such as binding
energies and radii of finite nuclei and the empirical characteristics of
infinite nuclear matter (\ie~symmetric and pure neutron matter).
The latter establishes an important, though highly
idealized, limiting case as it permits to confront the energy functional with
calculations from first principles using the bare nucleon-nucleon
force \cite{akmal98a}.

\begin{figure}[t]
  \includegraphics[width=\columnwidth]{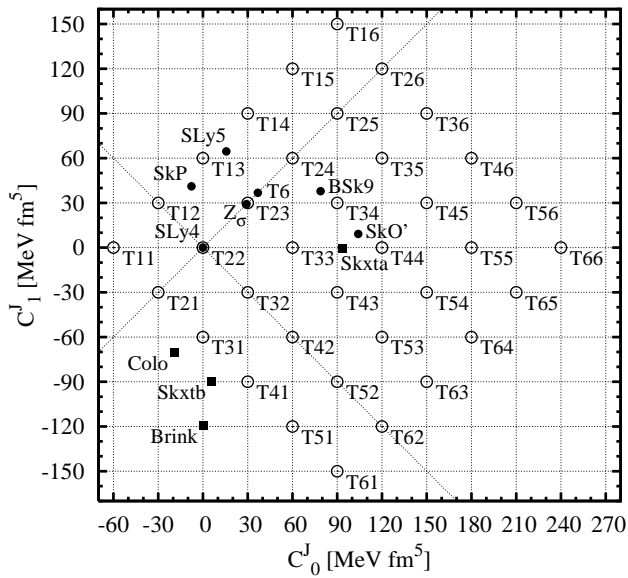}
  \caption{
    Values of $C^{J}_0$ and $C^{J}_1$ for our set of parameterizations
    (circles). Diagonal lines indicate $\alpha = C^{J}_0 + C^{J}_1 = 0$
    (pure neutron-proton coupling) and $\beta = C^{J}_0 - C^{J}_1 = 0$ (pure
    like-particle coupling). Values for classical parameter sets are also
    indicated (dots), with SLy4 representing all parameterizations
    for which $\vec J^2$ terms have been omitted in the fit. Recent
    parameterizations with tensor terms are indicated by squares.
  }
  \label{fig:cjplane-tot}
\end{figure}

The region of effective coupling constants $(C^{J}_0,C^{J}_1)$ of the
$\vec{J}^2$ terms acting in spherical nuclei, as defined in
Eq.~(\ref{eq:EF:sphere}), that we will explore, is shown in
Fig.~\ref{fig:cjplane-tot}. The parameterizations are labeled T$IJ$, where
indices $I$ and $J$ refer to the proton-neutron ($\beta$) and like-particle
($\alpha$) coupling constants in
Eq.~(\ref{eq:Htensor:pn}) such that
\begin{alignat}{3}
\alpha & =  60 \, (J-2) & \; \text{MeV} \, \text{fm}^5, \nn \\
 \beta & =  60 \, (I-2) & \; \text{MeV} \, \text{fm}^5 .
\label{eq:alphabeta_ij}
\end{alignat}
The corresponding values of $C^{J}_t$ can be obtained through
Eq.~(\ref{eq:cjtot}) or from Fig.~\ref{fig:cjplane-tot}. On the one hand, we
cover the positions of the most popular existing Skyrme interactions that take
the $\vec{J}^2$ terms from the central force into account, which are
SLy5~\cite{Cha98a}, SkP~\cite{Dob84a}, Z$_\sigma$~\cite{Fri86a},
T6~\cite{Ton84a}, SkO'~\cite{Rei99a} and BSk9~\cite{Gor05a}. On the other hand,
among recent parameterizations including a tensor term,
\ie~Skxta~\cite{Bro06a}, Skxtb~\cite{Bro06a,BrownPrivComm} as well as those
published by Col{\`o} \etal~\cite{Col07a} and Brink and Stancu~\cite{Bri07a},
most fall in a region of negative $C^J_1$ and vanishing $C^J_0$, that is to the
lower left of Fig.~\ref{fig:cjplane-tot}. Parameterizations of this region,
which also includes a part of the triangle advocated in the perturbative study
of Stancu \etal~\cite{Sta77a}, gave unsatisfactory results for many
observables. Moreover, when attempting to fit parameterizations with large
negative coupling constants, we sometimes obtained unrealistic single-particle
spectra or even ran into the instabilities already mentioned and outlined in
appendix~\ref{app:instab}. Parameterizations further to the lower and upper
right also have unrealistic deformations properties. The contribution from the
$\vec{J}^2$ terms vanishes for T22, which will serve as the reference point.
For the parameterizations T2$J$, only the proton-proton and neutron-neutron
terms in $\mathcal{H}^{\text{\tensor}}$ are non-zero ($\beta=0$), while for
the parameterizations T$I$2, only the proton-neutron term in
$\mathcal{H}^{\text{\tensor}}$ contributes ($\alpha=0$). Note that the earlier
parameterizations T6 and Z$_\sigma$ have a pure like-particle $\vec{J}^2$ terms
as a consequence of the constraint $x_1 = x_2 = 0$ employed for both (and
most other early parameterizations of Skyrme's interaction).


\subsection{The fit protocol and procedure}


The list of observables used to construct the cost function $\chi^2$ minimized
during the fit (see Eq.~(4.1) in Ref.~\cite{Cha97a}) reads as follows:
binding energies and charge radii of \nuc{40}{Ca}, \nuc{48}{Ca}, \nuc{56}{Ni},
\nuc{90}{Zr}, \nuc{132}{Sn} and \nuc{208}{Pb}; the binding energy of
\nuc{100}{Sn}; the spin-orbit splitting of the neutron $3p$ state in
\nuc{208}{Pb}; the empirical energy per particle and density at the saturation
point of symmetric nuclear matter; and finally, the equation of state of
neutron matter as predicted by Wiringa \etal~\cite{Wir95a}.

Furthermore, some properties of infinite nuclear matter are constrained
through analytic relations between coupling constants
in the same manner as they were in Refs.~\cite{Cha97a,Cha98a}:
the incompressibility modulus $K_\infty$ is kept at 230~MeV, while the
volume symmetry energy coefficient $a_\tau$ is set to 32~MeV. The
isovector effective mass, expressed through the Thomas-Reiche-Kuhn
sum rule enhancement factor $\kappa_v$, is taken such that
$\kappa_v = 0.25$.

When using a single density-dependent term in the central Skyrme
force~(\ref{eq:Skyrme:central}), the isoscalar effective mass $m^\ast_0$
cannot be chosen independently from the incompressibility modulus for a given
exponent $\alpha$ of $\rho_0$. We follow here the prescription used for the
SLy parameterizations~\cite{Cha97a,Cha98a} and use $\alpha = 1/6$, which leads
to an isoscalar effective mass close to 0.7 in units of the bare nucleon mass
for all T$IJ$ parameterizations. This value allows for a correct description
of dynamical properties, as for example the energy of the giant quadrupole
resonance~\cite{Liu76a}. Using such a protocol we cannot reproduce the
isovector effective mass consistent with recent \emph{ab-initio}
predictions~\cite{Les06a}. Regarding the present exploratory study of the
tensor terms this is not a critical limitation, in particular as the influence
of this quantity on static properties of finite nuclei turns out to be small.

There are three modifications of the fit protocol compared
to~\cite{Cha97a,Cha98a}. The obvious one is that the values for $C^{J}_0$ and
$C^{J}_1$ are fixed beforehand as the parameters that will later on label and
classify the fits. The second is that we have added the binding energies of
\nuc{90}{Zr} and \nuc{100}{Sn} to the set of data. Indeed, we observed that the
latter nucleus is usually significantly overbound when not included in the fit.
The third is that we have dropped the constraint $x_2 = -1$ that was imposed on
the SLy parameterizations~\cite{Cha97a,Cha98a} to ensure the stability of
infinite homogeneous neutron matter against a transition into a ferromagnetic
state.
On the one hand, this stability criterion is completely determined by the
coupling constants of the time-odd terms in the energy
functional~\cite{Ben02a},
that we do not want to constrain here, accepting that the parameterizations
might be of limited use beyond the present study. On the other hand, the tensor
force brings many new contributions to the energy per particle of polarized
nuclear matter that lead to a much more complex stability criterion.
We postpone
the entire discussion concerning the stability in polarized systems in the
presence of a tensor force to future work that will also address finite-size
instabilities~\cite{Les06a}. It also has to be stressed that the actual
stability criterion, as all properties of the time-odd part of the Skyrme
energy
functional, depends on the choices made for the interpretation of its coupling
constants, \ie~antisymmetrized vertex or density functional~\cite{Ben02a}.

\begin{figure}[t]
  \includegraphics[width=0.8\columnwidth]{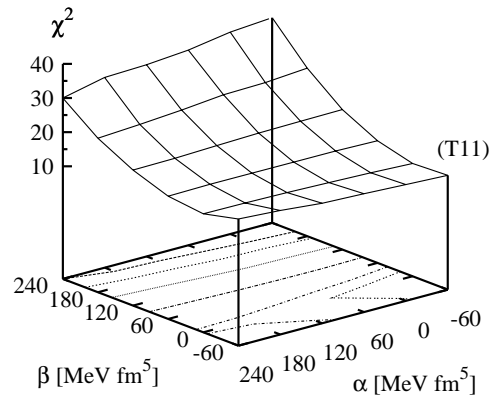}
  \caption{
    Values of the cost function $\chi^2$ as defined in the fit procedure, for
    the set of parameterizations T$IJ$. The label ``T11'' indicates the
    position of this parameterization in the ($\alpha$,$\beta$)-plane as
    obtained from Eqs.~(\ref{eq:alphabeta_ij}). Contour lines
    are drawn at $\chi^2 = 11$, 12, 15, 20, 25, and 30. The minimum value is
    found for T21 ($\chi^2 = 10.05$), the maximum for T61
    ($\chi^2 = 37.11$).
  }
  \label{fig:chi2}
\end{figure}

The properties of the finite nuclei entering the fit are computed using
a Slater determinant without taking pairing into account.
The cost function $\chi^2$ was minimized using a simulated
annealing algorithm. The annealing schedule was an exponential one, with a
characteristic time of 200 iterations (also referred to as ``simulated
quenching'') Thus, assuming a reasonably smooth cost function, we strive to
obtain satisfactory convergence to its absolute minimum in a single run,
allowing a systematic and straightforward production of a large series of
forces. The coupling constants for all 36 parameterizations can be found in
the \emph{Physical Review} archive \cite{EPAPS}.

Figure \ref{fig:chi2} displays the value of $\chi^2$ after minimization as a
function of the recoupled coupling constants $\alpha$ and $\beta$. The first
striking feature is the existence of a ``valley'' at $\beta = 0$, i.e.~a
pure like-particle tensor term $\sim (\vec{J}^2_\nrm + \vec{J}^2_\prm)$. The
abrupt rise of $\chi^2$ around this value can be attributed to the term
depending on nuclear binding energies, as sharp variations of energy residuals
can be seen between neighboring magic nuclei with functionals of the T6$J$
series ($\beta = 240$). For example, \nuc{48}{Ca} and \nuc{90}{Zr} tend to be
significantly overbound in this case. We will come back later to discussing the
implications for the quality of the functionals.


\subsection{General properties of the fits}


\begin{figure}[t]
  \includegraphics[width=\columnwidth]{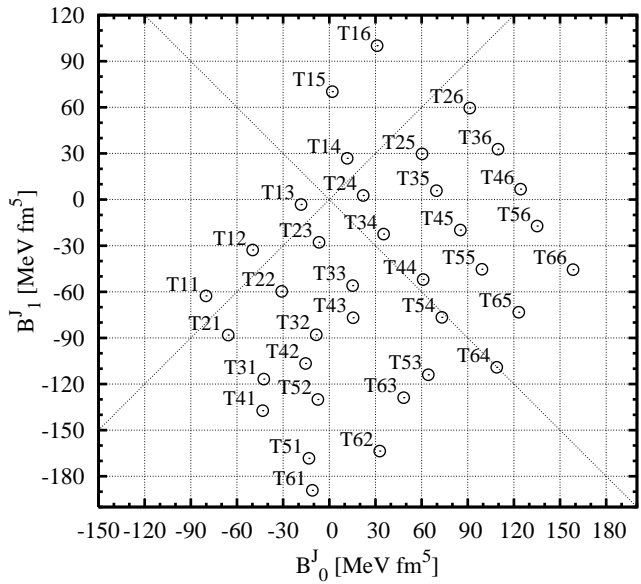}
  \caption{
    The contributions from the tensor force
    $B^{J}_0$ and $B^{J}_1$ to the effective coupling constants of the
    $\vec{J}^2$ term at sphericity.
    Diagonal lines as in Fig.~\ref{fig:cjplane-tot}. The diagonal where
    $B^J_0 + B^J_1 = \alpha_T = 0$ (pure proton-neutron contribution)
    additionally corresponds to an isospin-singlet force with
    $t_o \equiv U = 0$.
  }
  \label{fig:cjplane-tens}
\end{figure}

The coupling constants of the energy functional for spherical nuclei
(\ref{eq:EF:sphere}) obtained for T22 are very similar to those of SLy4, except
for a slight readjustment coming from the inclusion of the binding energies of
\nuc{90}{Zr} and \nuc{100}{Sn} in the fit as well as the abandoned constraint on
$x_2$. With its value of $-0.945$, the $x_2$ obtained for T22 still stays close
to the value $-1$ enforced for SLy4, which confirms that this is not too severe
a constraint for parameterizations \emph{without} effective $\vec{J}^2$ terms at
sphericity. Increasing the effective
tensor term coupling constants $C^{J}_t$, however, the values for $x_2$ start to
deviate strongly from the region around $-1$, which is to a large extent due to
the feedback from the contribution of the $\vec{J}^2$ terms to the surface and
surface symmetry energy coefficients in the presence of constraints on
isoscalar and isovector effective masses, all of which also depend on $x_2$. A
more detailed discussion of the contribution of the $\vec{J}^2$ terms to the
surface energy coefficients will be given elsewhere~\cite{II}.

From the constrained coupling constants $C^{J}_0$ and $C_1^{J}$, the respective
contributions $B_0^{J}$ and $B_1^{J}$ from the tensor force can be deduced
afterwards using the expressions given in Sect.~\ref{sect:skyrmefu:sphere}.
Their values, shown in Fig.~\ref{fig:cjplane-tens}, are less regularly
distributed, which is a consequence of the the non-linear interdependence of all
coupling constants. Still, a general trend can be observed, such that all
parameterizations are shifted towards the ``south-west'' compared to
Fig.~\ref{fig:cjplane-tot}. In turn, this indicates that the contribution from
the central Skyrme force always stays in the small region outlined by SkP, SLy5,
Z$_\sigma$, {\em etc} in Fig.~\ref{fig:cjplane-tot}, with values that range
between $28$ and $104$~MeV\,fm$^5$ for $A^{J}_0$ and between $38$ to
$62$~MeV\,fm$^5$ for $A^{J}_1$, respectively. This justifies a posteriori to use
the tensor force as a motivation to decouple the $\vec{J}^{2}_{t}$ terms from
the central part of the effective Skyrme vertex. We note in passing that all our
parameterizations T$I$4 correspond to an almost pure proton-neutron or
isospin-singlet tensor force, \ie~the term $\propto t_e$ in
Eq.~(\ref{eq:Skyrme:tensor}), as they are all located close to the $\alpha_T =
0$ line.

\begin{figure}[t!]
  \includegraphics[width=0.8\columnwidth]{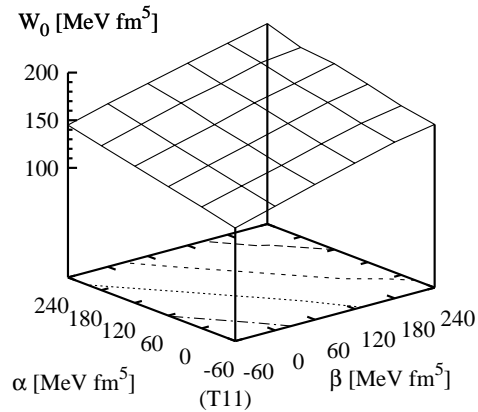}
  \caption{
    Value of spin-orbit coupling constant
    $W_0$ for each of the parameterizations T$IJ$,
    \emph{vs.} indices $I$ and $J$ (The ``(T11)'' label indicates the
    position of this parameterization in the $(\alpha, \beta)$-plane).
    The contour lines differ by 20~MeV\,fm$^5$.
    The values plotted here range from
    103.7~MeV\,fm$^5$~(T11) to 195.3~MeV\,fm$^5$~(T66).
  }
  \label{fig:w0}
\end{figure}

We also find a particularly strong and systematic variation
of the coupling constant $W_0$ of the spin-orbit force, which varies from
$W_0 = 103.7$~MeV\,fm$^5$ for T11 to $W_0 = 195.3$~MeV\,fm$^5$ for T66, see
Fig.~\ref{fig:w0}. This variation is of course correlated to the strength
of the tensor force. As already shown, the tensor force has the tendency
to reduce the spin-orbit splittings in spin-unsaturated nuclei. To maintain
a given spin-orbit splitting in such a nucleus, the spin-orbit coupling
constant $W_0$ has to be increased.


\section{Results and discussion}
\label{sect:results}


The calculations presented below include open-shell nuclei treated in the
Hartree-Fock-Bogoliubov (HFB) framework. In the particle-particle channel, we
use a zero-range interaction with a mixed surface/volume form factor (called
DFTM pairing in Ref.~\cite{Dug07}). The HFB equations were regularized with a
cutoff at 60~MeV in the quasiparticle equivalent spectrum~\cite{Ben05a}. The
pairing strength was adjusted in \nuc{120}{Sn} with the particle-hole
mean field calculated using the parameter set T33. The resulting strength
was kept at the same value for all parameterizations, which is justified by
the fact that the effective mass parameters are the same. Moreover,
we thus avoid including, in the adjustment of the pairing strength, local
effects linked with changes in details of the single-particle spectrum.


\subsection{Spin-orbit currents and potentials}


\begin{figure}[t!]
  \includegraphics[width=0.8\columnwidth]{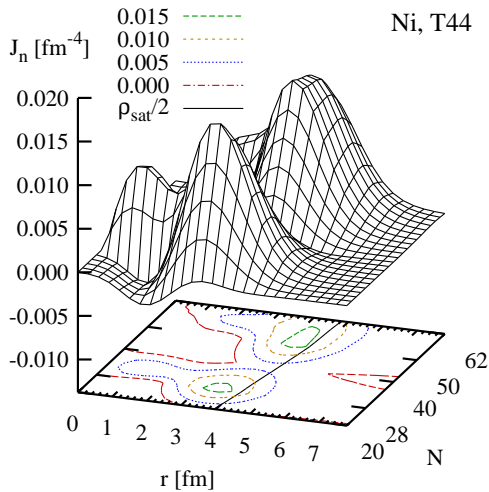}
  \caption{
    (Color online)
    Radial component of the neutron spin-orbit current for
    the chain of Ni isotopes, plotted against radius and neutron number $N$.
    The solid line on the base plot indicates the radius where the
    total density has half its saturation value.
  }
  \label{fig:Ni-Jn}
\end{figure}

As a first step in the analysis of the role of the tensor terms and their
interplay with the spin-orbit interaction in spherical nuclei, we analyze the
spin-orbit current density and its relative contribution to the
spin-orbit potential. We choose the chain of nickel isotopes, $Z=28$, as it
covers the largest number of spherical neutron shells and subshells
($N=20$, 28, 40 and 50) of any isotopic chain, two of which are
spin-saturated ($N=20$ and 40), while the other two are not.
Figure~\ref{fig:Ni-Jn} displays the radial component of the neutron
spin-orbit current $\vec J_\nrm$ for isotopes from the proton to the
neutron drip-lines. The calculations are performed with T44, but the
spin-orbit current is fairly independent from the parameterization. Starting
from $N=20$, which corresponds to a completely filled and spin-saturated
$sd$-shell, the next magic number at $N=28$ is reached by filling the
$1f_{7/2}$ shell, which leads to the steeply rising bump in the plot of
$\vec J_\nrm$ in the foreground, peaked around $r \simeq 3.5$~fm. Then,
from $N=28$ to $N=40$ the
rest of the $fp$ shell is filled, which first produces the small bump at small
radii that corresponds to the filling of the $2p_{3/2}$ shell, but ultimately
leads to a vanishing spin-orbit current when the $1f$ and $2p$ levels are
completely filled for the $N=40$ isostope, visible as the deep valley in
Fig.~\ref{fig:Ni-Jn}. Adding more neutrons, the filling of the $1g_{9/2}$ shell
leads again to a strong neutron spin-orbit current at $N=50$. For the remaining
isotopes up to the neutron drip line, the evolution of $\vec J_\nrm$ is slower
with the filling of the $2d$ and $3s$ orbitals.

A few further comments are in order. First, the spin-orbit current clearly
reflects the spatial probability distribution of the single-particle wave
function in pairs of unsaturated spin-orbit partners. Within a given shell, the
high-$\ell$ states contribute at the surface, represented by the solid line on
the base of Fig.~\ref{fig:Ni-Jn}, while low-$\ell$ states contribute at the
interior. The peak from the high-$\ell$ orbitals, however, is always located on
the inside of the nuclear surface, as defined by the radius of half saturation
density. Second, within a given shell, the largest contributions to the
spin-orbit current density obviously come from the levels with largest
$\ell$, as they have the largest degeneracy factors in (\ref{eq:j:radial}),
and because they do not have nodes, which leads to a single, sharply peaked
contribution. Third, the spin-orbit current is not exactly zero for nominally
``spin-saturated'' nuclei, exemplified by the $N=20$ and $N=40$ isotopes in
Fig.~\ref{fig:Ni-Jn}, as the radial single-particle wave functions are not
exactly identical for all pairs of spin-orbit partners, which is a necessary
requirement to obtain $\vec J_\nrm = 0$ at all radii ({\em Cf.}~the example of
the $\nu$ $2d$ states in \nuc{132}{Sn} in Fig.~\ref{fig:Sn132-field} below).
Fourth, pairing and other correlations will always smooth the fluctuations of
the spin-orbit current with nucleon numbers, as levels in the vicinity of the
Fermi energy will never be completely filled or empty.

\begin{figure}[t!]
  \includegraphics[width=0.8\columnwidth]{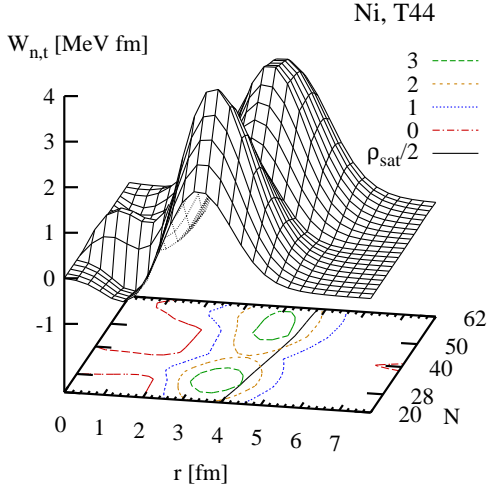}
  \caption{
    (Color online)
    Contribution from the tensor terms to the neutron spin-orbit
    potential for the chain of Ni isotopes as obtained with the
    parameterization T44. The solid line on the base plot indicates the
    radius where the isoscalar density $\rho_0$ crosses half its saturation
    value.
  }
  \label{fig:Ni-Wn-t}
\end{figure}

\begin{figure}[t!]
  \includegraphics[width=0.8\columnwidth]{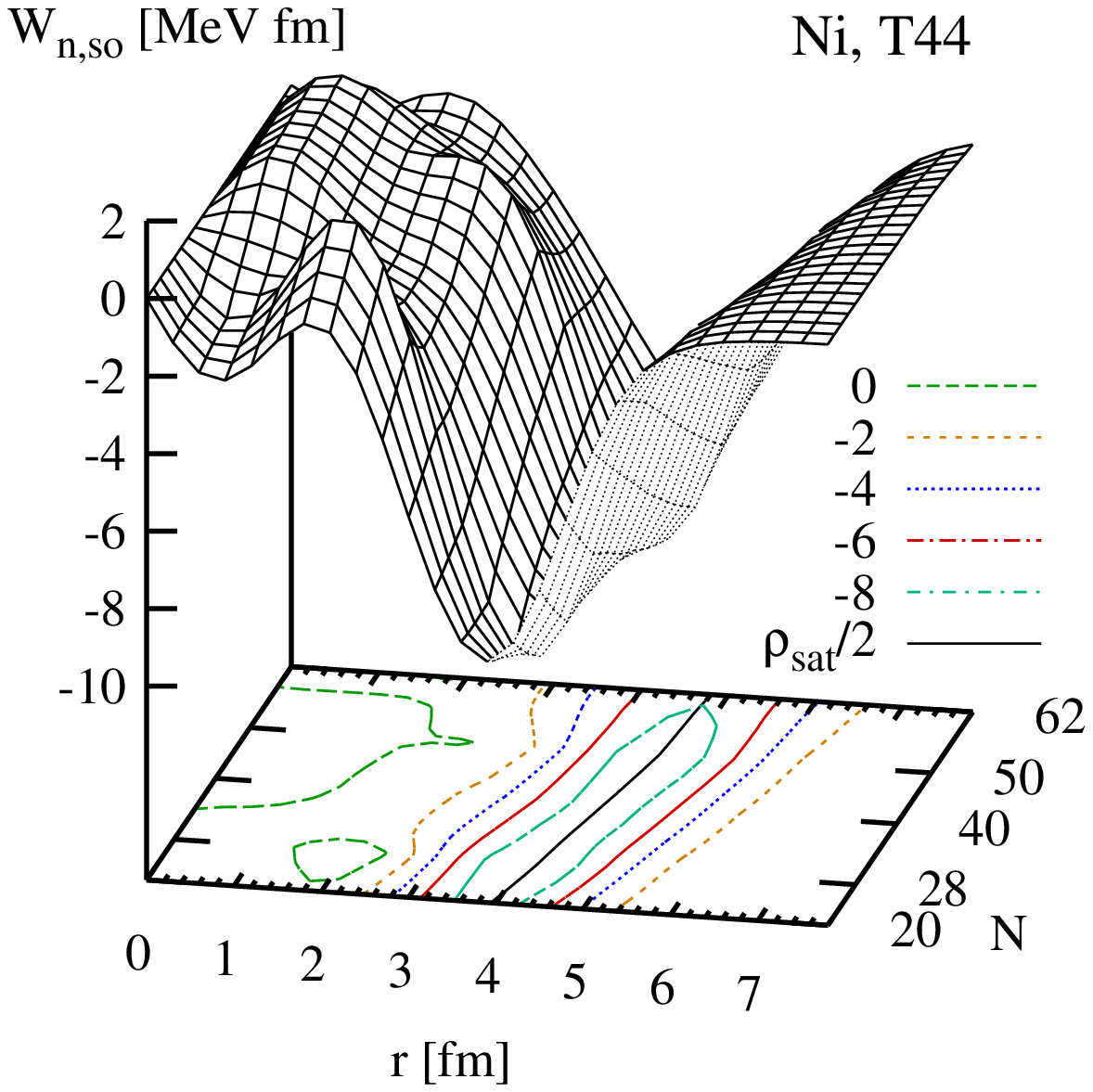}
  \caption{
    (Color online)
    Contribution from the spin-orbit force to the neutron spin-orbit
    potential for the chain of Ni isotopes as obtained with the
    parameterization T44. The solid line on the base plot indicates the
    radius where the isoscalar density $\rho_0$ crosses half its saturation
    value.
  }
  \label{fig:Ni-Wn-so}
\end{figure}

\begin{figure}[t!]
  \includegraphics[width=0.8\columnwidth]{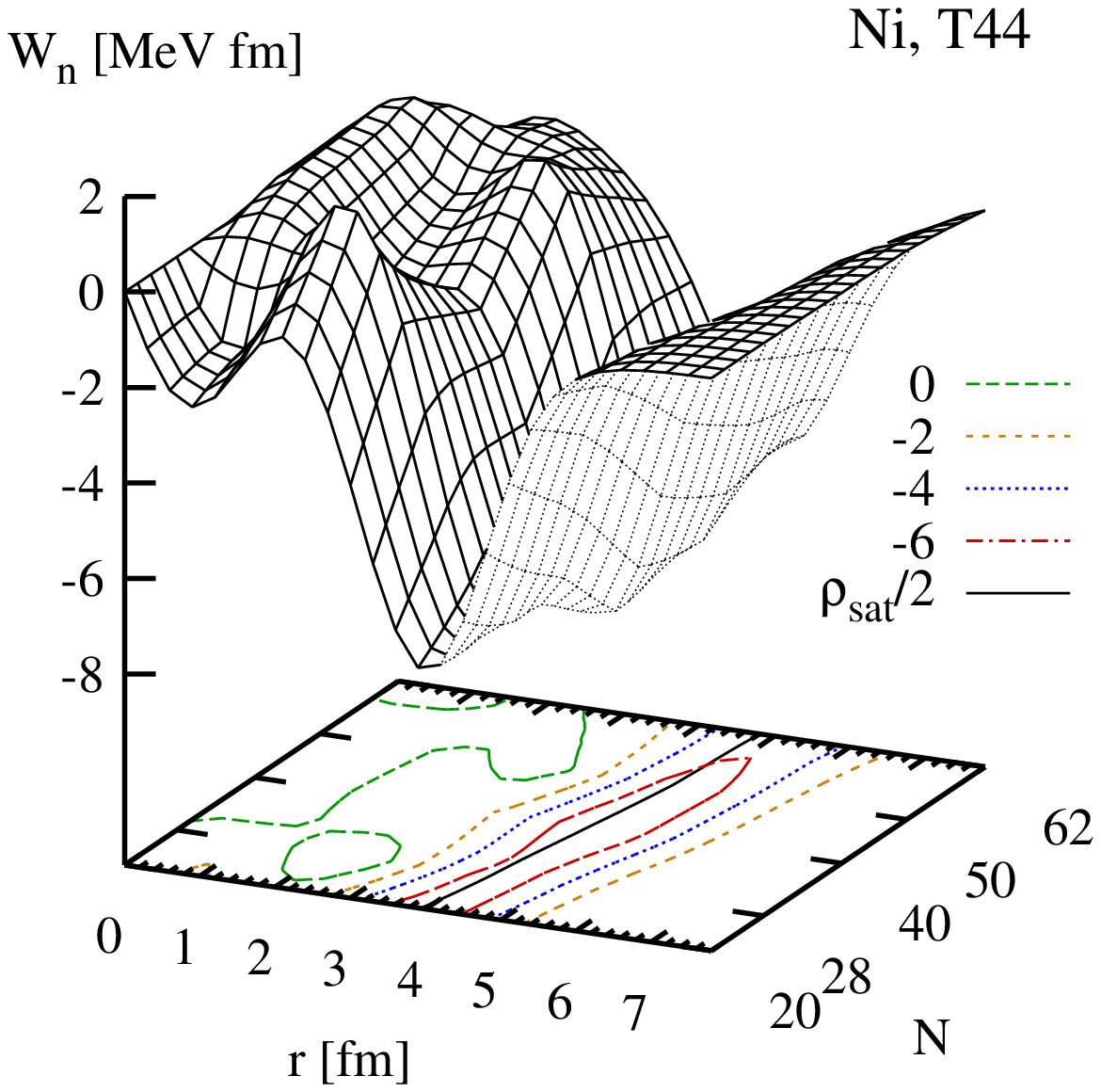}
  \caption{
    (Color online)
    Total neutron spin-orbit potential for the chain of Ni isotopes
    as obtained with the parameterization T44.
    The solid line on the base plot indicates the radius where the isoscalar
    density $\rho_0$ crosses half its saturation value.
  }
  \label{fig:Ni-Wn-total}
\end{figure}

\begin{figure}[t!]
  \includegraphics[width=0.8\columnwidth]{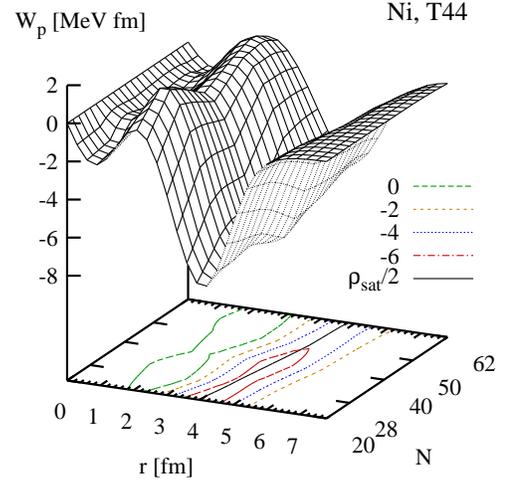}
  \caption{
    (Color online)
    Total proton spin-orbit potential for the chain of Ni isotopes
    as obtained with the parameterization T44. The solid line on the base
    plot indicates the radius where the isoscalar density $\rho_0$ crosses
    half its saturation value.
  }
  \label{fig:Ni-Wp-total}
\end{figure}

Next, we compare the contributions from the tensor terms and from the spin-orbit
force to the spin-orbit potentials of protons and neutrons,
Eq.~(\ref{eq:wpot:tot}). The contributions from the tensor force to the
spin-orbit potential are proportional to the spin-orbit currents of protons and
neutrons. For the Ni isotopes, the proton spin-orbit current is very similar to
that of the neutrons at $N=28$ displayed in Fig.~\ref{fig:Ni-Jn}. For the
parameterization T44 we use here as an example, we have contributions from both
proton and neutron spin-orbit currents, which come with equal weights. Their
combined contribution to the spin-orbit potential of the neutron $W_n$ might be
as large as 4 MeV, see Fig.~\ref{fig:Ni-Wn-t}. This is more than a third of the
maximum contribution from  the spin-orbit force to $W_n$, see
Fig.~\ref{fig:Ni-Wn-so}. The latter is proportional to a combination of the
gradients of the proton and neutron densities, $2 \nabla \rho_n (r) + \nabla
\rho_p (r)$, see Eq.~(\ref{eq:wpot:tot}). As a consequence, it has a smooth
behavior as a function of particle number, with slowly and monotonically varying
width, depth and position. Only limited local variations can be seen on the
interior due to small variations of the density profile originating from the
successive filling of different orbits. Furthermore, one can easily verify that
the contribution from the spin-orbit force is peaked at the surface of the
nucleus (the solid line on the base plot). The strongest variation of the depth
of this potential occurs just before the neutron drip line at $N=62$, where is
becomes wider and shallower due to the development of a diffuse neutron skin,
which reduces the gradient of the neutron density~\cite{Dob94a,Lal98a,Lal98b}.

\begin{figure}[t!]
  \includegraphics[width=\columnwidth]{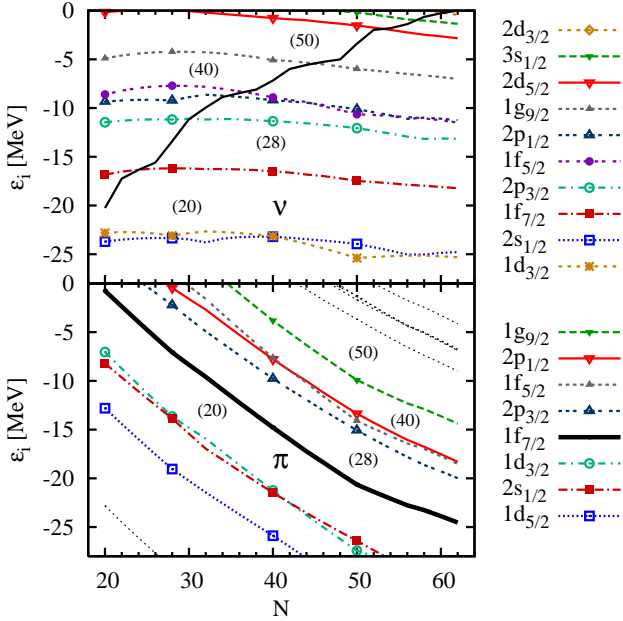}
  \caption{
    (Color online)
    Single-particle spectra of neutrons (upper panel) and
    protons (lower panel) for the chain of Ni isotopes, as obtained with
    the parameterization T22 with vanishing combined $\vec{J}^2$ terms.
    The thick solid line in the upper panel denotes the Fermi energy
    for neutrons.
  }
  \label{fig:esp:ni:t22}
\end{figure}

Adding the contributions from the proton and neutron tensor terms to that from
the spin-orbit force, the total neutron spin-orbit potential for neutrons in Ni
isotopes is shown in Fig.~\ref{fig:Ni-Wn-total}. For the parameterization T44
used here (and most others in the sample of parameterizations used in this
study) the dominating contributions from the spin-orbit and tensor forces to
the spin-orbit potential are of opposite sign. For Ni isotopes, $\vec{J}_p$ is
always quite large, while $\vec{J}_n$ varies as shown in Fig.~\ref{fig:Ni-Jn}.
Notably, both are peaked inside of the surface. When examining the combined
contribution from the spin-orbit and tensor forces to the spin-orbit potential
(\ref{eq:wpot:tot}), one must keep in mind that they are peaked at different
radii. Moreover, the variation of tensor-term coupling constants among a set of
parameterizations implies a rearrangement of the spin-orbit term strength, as
will be discussed later. As a consequence, taking into account the tensor
force modifies the width and localization of the spin-orbit potential
$W_q(r)$ much more than it modifies its depth through the variation of the
spin-orbit currents.

\begin{figure}[t!]
  \includegraphics[width=\columnwidth]{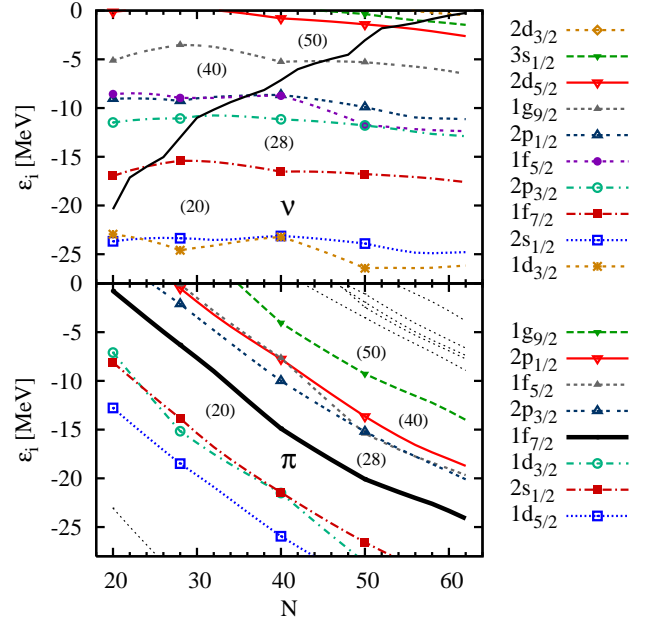}
  \caption{
    (Color online)
    The same as Fig.~\protect\ref{fig:esp:ni:t22},
    obtained with T44 with proton-neutron and like-particle tensor
    terms of equal strength.
  }
  \label{fig:esp:ni}
\end{figure}

Our observations also confirm the finding of Otsuka \etal~\cite{Ots06a}
that the spin-orbit splittings might be more strongly modified by the
tensor force than they are by neutron skins in neutron-rich
nuclei through the reduction of the gradient of the density.

Figure~\ref{fig:Ni-Wp-total} shows the spin-orbit potential of the protons for
the chain of Ni isotopes. Here, the contribution from the spin-orbit force
has a larger contribution coming from the gradient of the proton density
that just grows with the mass number, without being subject to varying
shell fluctuations. The same holds for the proton contribution from the
tensor terms. Only the neutron contribution from the tensor terms varies
rapidly, proportional to $\vec{J}_n$ displayed in Fig.~\ref{fig:Ni-Jn},
which has a very limited effect on the total spin-orbit potential, though.

With that,  we can
examine how the tensor terms affect the evolution of single-particle spectra.
To that end, Fig.~\ref{fig:esp:ni:t22} shows the single-particle energies of
protons and neutrons along the chain of Ni isotopes for the parameterization T22
with vanishing combined tensor terms, which will serve as a reference,
while Fig.~\ref{fig:esp:ni} shows the same for the parameterization T44 with
proton-neutron and like-particle tensor terms of equal strength. For the
latter, the variation of the neutron spin-orbit current with $N$ influences
both neutron
and proton single-particle spectra. The effect of the tensor terms is subtle,
but clearly visible: for T22, the major change of the single-particle energies
is their compression with increasing mass number, while for T44 the level
distances oscillate on top of this background correlated to the neutron shell
and sub-shell closures at $N=20$, 28, 40 and 50. As shown above, the neutron
spin-orbit current vanishes for $N=20$, where it consequently has no effect on
the spin-orbit potentials and splittings. By contrast, the neutron spin-orbit
current is large for $N=28$ and 50, where its contribution to the spin-orbit
potential reduces the splittings from the spin-orbit force.

\begin{figure}[t!]
  \includegraphics[width=0.8\columnwidth]{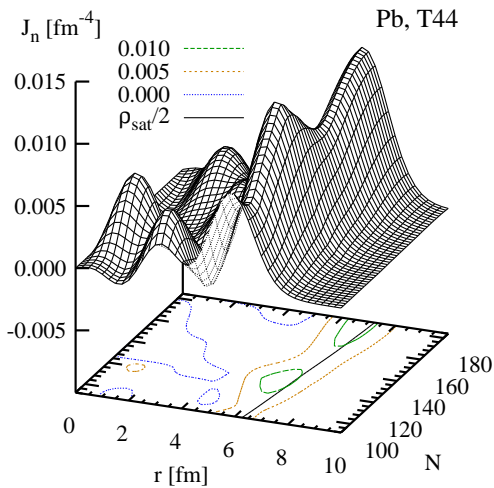}
  \caption{
    (Color online)
      Radial component of the Neutron spin-orbit current for
      the chain of Pb isotopes plotted in the same manner as in
      Fig.~\protect\ref{fig:Ni-Jn}.
  }
  \label{fig:Pb-Jn}
\end{figure}

The strong variation of the spin-orbit current with nucleon numbers is typical
for light nuclei up to about mass 100. For heavier nuclei, its variation
becomes much smaller. This is exemplified in Fig.~\ref{fig:Pb-Jn} for the
neutron spin-orbit current in the chain of Pb isotopes. There remain the fast
fluctuations at small radii which we already saw for the Ni isotopes and that
reflect the subsequent filling of low-$\ell$ levels with many nodes, but which
have a very limited impact on the spin-orbit splittings when fed into the
spin-orbit potential. The dominating peak of the spin-orbit current, just
beneath the surface shows only small fluctuations, as the overlapping
spin-orbit splittings of levels with different $\ell$ never give rise
to a spin-saturated configuration in heavy nuclei.

Note that both the spin-orbit current $\vec{J}$ and the spin-orbit potential
are exactly zero at $r=0$ as they are vectors with negative parity.


\subsection{Single-particle energies}
\label{sect:spe}


As a next step, we analyze the modifications that the presence of $\vec{J}^2$
terms brings to single-particle energies in detail. Before we do so, a few
general comments on the definition and interpretation of single-particle
energies are in order. From an experimental point of view, empirical
single-particle energies in a doubly-magic nucleus are determined as the
separation energies between the even-even doubly magic nucleus and low-lying
states in the adjacent odd-$A$ nuclei, i.e.~they are differences of
binding energies. In nuclear models, however, it is customary to discuss shell
structure and single-particle energies in terms of the spectrum of eigenvalues
$\epsilon_i$ of the Hartree-Fock mean-field Hamiltonian (in even-even nuclei),
as we have done already in Figs.~\ref{fig:esp:ni:t22} and~\ref{fig:esp:ni}:
\begin{equation}
\label{eq:spe:def}
\hat{h} \, \Phi_i
= \epsilon_i \, \Phi_i\,.
\end{equation}
In the nuclear EDF approach without pairing, the reference state is directly
constructed as a Slater determinant of eigenstates of $\hat{h}$; hence, the
corresponding eigenvalues are directly connected to the fundamental building
blocks of the theory and reflect the mean field in the nucleus. The density of
single-particle levels around the Fermi surface drives the magnitude of pairing
correlations, the relative distance of single-particle levels at sphericity and
their quantum numbers determine to a large extent the detailed structure of the
deformation energy landscape which in turn, determines the collective
spectroscopy. The spectroscopic properties of even-even nuclei, in particular
when they exhibit shape coexistence, provide valuable benchmarks for the
underlying single-particle spectrum~\cite{Ben06a}. The link between the spectrum
of single-particle energies on the one hand and the collective excitation
spectrum on the other hand, however, always remains indirect.

On the other hand, ``single-particle'' states near the Fermi level of a magic
nucleus can be observed by adding or removing a particle in one of these states,
and thus correspond to the ground and excited states of the neighboring odd-mass
nuclei. Assuming an infinitely stiff magic core, which is neither subject to any
rearrangement or polarization, nor to any collective excitations following the
addition (or removal) of a nucleon, the separation energies with the states in
the odd-mass neighbors are equal to the single-particle energies as defined
through (\ref{eq:spe:def}). This highly idealized situation is modified by
static~\cite{Rut98a} and dynamic~\cite{Ber80a,Lit06a} correlations, often called
``core polarization'' (see chapter~7 of Ref.~\cite{Ber72aB}) and
``particle-vibration coupling'' (see section~9.3.3 of Ref.~\cite{Rin80aB}) in
the literature, that alter the separation energies. The main effect of the
correlations is that they compress the spectrum, pulling down the levels from
above the Fermi energy and pushing up those from below. The gross features,
i.e.~the ordering and relative placement of single-particle states, however, are
more weakly affected by correlations. The particle-vibration coupling, however,
is also responsible for the fractionization of the single-particle strength.
When the latter is too large, the naive comparison between the calculated
$\epsilon_i$ given by Eq.~(\ref{eq:spe:def}) and the energy of the lowest
experimental state with the same quantum numbers is not even qualitatively
meaningful anymore~\cite{Bro06a}.

We mention that a part of the static correlations originate from
the non-vanishing time-odd densities in the mean-field ground-state of
an odd-$A$ nucleus, that also cannot be truly spherical, so that the
complete energy functional from Eq.~(\ref{eq:EF:full}) should be considered
in a fully self-consistent calculation of the separation energies.

The effective single-particle energies that are used to characterize the
underlying shell structure in the interacting shell model~\cite{Cau05a} have
a slightly different meaning. Their definition usually renormalizes
polarization and particle-vibration coupling effects around a doubly-magic
nucleus whereas their evolution is discussed in terms of monopole
shifts~\cite{Duf99a}. A collection of effective single-particle energies and
their evolution was collected by Grawe~\cite{Gra05a,Gra06a}. Note that the
SkX parameterization of the Skyrme energy functional by Brown and its
variants~\cite{Bro98a,Bro06a} were constructed aiming at a description of
effective single-particle energies along these lines.

It should be kept in mind that the obvious, coarse discrepancies between the
calculated spectra of $\epsilon_\mu$ and the empirical single-particle energies
are often larger than the uncertainties coming from the missing correlations,
as long as one observes some elementary precautions. We took care to ensure
that the states used in the analysis below were one-quasiparticle states weakly
coupled to core phonons. First, we checked that the even-even nucleus of
interest could be described as spherical, indicated by a sufficiently
high-lying 2$^+$ state. Second, we avoided all levels which were obviously
correlated with the energies of 2$^+$ states in the adjacent semi-magic series,
as this indicates strong coupling with core excitations. Finally, we carefully
examined states, lying above the 2$^+$ energy and/or twice the pairing gap of
adjacent semi-magic nuclei, in order to eliminate those more accurately
described as an elementary core excitation coupled to one or more
quasiparticles, which generally appear as a multiplet of states. We did not
attempt to use energy centroids calculated with use of spectroscopic factors,
as these are not systematically available. Indeed, our requirement is that if
some collectivity is present, it should be similar among all nuclei considered,
in order to be easily subtracted out. Empirical single-particle levels shown
below are determined from the lowest states having given quantum numbers in an
odd-mass nucleus.


\subsubsection{Spin-orbit splittings}
\label{sect:splittings}


The primary effect one expects from a tensor term is that it affects
spin-orbit splittings by altering the strength of the spin-orbit field in
spin-unsaturated nuclei, according to Eq.~(\ref{eq:wpot:tot}). One should
remember, though, that the spin-orbit coupling itself is readjusted for each
pair of coupling constants $C^J_0$, and $C^J_1$.
The effect of this readjustment is generally opposite to that of the variation
of the isoscalar tensor term coupling constant. It should thus be stressed
that the effects described result from the balance between the variation
of tensor and spin-orbit terms, which for most of our parameterizations pull
into opposite directions.

\begin{figure}[t!]
  \includegraphics[width=\columnwidth]{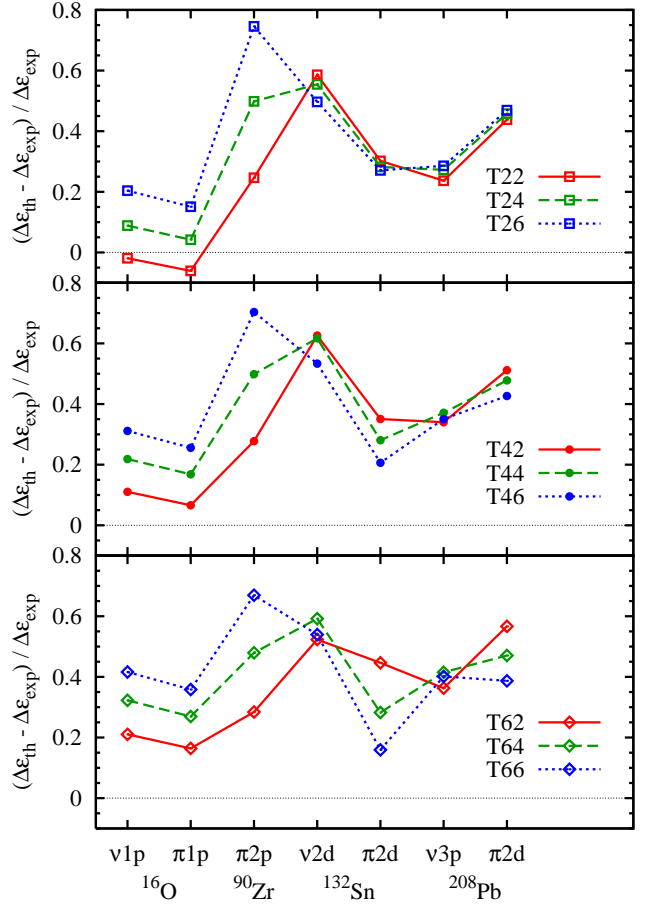}
  \caption{
    (Color online)
    Relative error of the  spin-orbit splittings in
    doubly-magic nuclei for $\ell \leq 2$ levels.
  }
  \label{fig:splittings}
\end{figure}

Common wisdom states that the energy spacing between levels that are both above
or both below the magic gap are not much affected by correlations, even when
their absolute energy changes; hence it is common practice to confront only
the spin-orbit splittings between pairs of particle or hole states with
calculated single-particle energies from the spherical mean field.
Figure~\ref{fig:splittings} shows the relative error of single-particle
splitting of such levels for doubly-magic nuclei throughout the chart of
nuclei. The calculated values are typically 20~to~$60\,\%$ larger than the
experimental ones, with the exception of \nuc{16}{O}, where the splittings of
the neutron and proton $1p$ states are acceptably reproduced at least
for the parameterizations T22, T24 and T42, \ie~those with the weakest
tensor terms in the sample.

It is noteworthy that the calculated splittings depend much more sensitively
on the tensor terms for light nuclei with spin-saturated shells (protons and
neutrons in \nuc{16}{O}, protons in \nuc{90}{Zr}) than for the heavy
doubly-magic \nuc{132}{Sn} and \nuc{208}{Pb}, which are quite robust
against a variation of the tensor terms. The reason will become clear below.


\subsubsection{Connection between tensor and spin-orbit terms}
\label{sect:J2-so-rearrangement}


The finding that our parameterizations systematically overestimate the
spin-orbit  splittings deserves an explanation. It was earlier already
noted that all standard Skyrme interactions, including the SLy
parameterizations that share our fit protocol, have an unresolved trend
that overestimates the spin-orbit splittings in heavy
nuclei~\cite{Ben99b,RMP,Lop00a}. Adding the tensor terms, however,
further deteriorates the overall description of spin-orbit splittings, instead
of improving it. It is particularly disturbing that the spin-orbit splitting of
the $3p$ level in \nuc{208}{Pb} that was used to constrain $W_0$ in the fit is
overestimated by 30 to 40$\%$, which is larger than the relative tolerance of
20$\%$ included in the fit protocol. In fact, it turns out that the coupling
constant $W_0$ of the spin-orbit force is more tightly constrained by the
binding energies of light nuclei than by this or any other spin-orbit
splitting. In the HF approach used during the fit, the structure of
\nuc{40}{Ca}, \nuc{48}{Ca}, and \nuc{56}{Ni} differs by the occupation of
the neutron and proton $1f_{7/2}$ levels. First, we have to note that the
terms in the energy functional that contain the spin-orbit current play
an important role for the energy difference between \nuc{40}{Ca} and
\nuc{56}{Ni}. The combined contribution from the tensor and spin-orbit
terms varies from a near-zero value in the spin-saturated \nuc{40}{Ca}
to about $-60$ MeV in \nuc{56}{Ni} for all our parameterizations, which
is a large fraction of the $-142$ MeV difference in total binding energy
between both nuclei. The $Z=40$ subshell and $Z=50$ shell are another
example of abrupt variation of the spin-orbit current with the
filling of the $1g_{9/2}$ level, which strongly affects the relative binding
energy of $N=50$ isotones \nuc{90}{Zr} and \nuc{100}{Sn}. Second, the fit
to phenomenological data can take advantage of the large relative variation of
these terms to mock up missing physics in the energy functional that should
contribute to the energy difference, but that is absent in it.
The consequence will be a spurious increase of the spin-orbit and
tensor term coupling constants. The resulting energy functional will correctly
describe the mass difference, but not the physics of the spin-orbit and tensor
terms.

In order to test the above interpretation, we performed a refit of selected
T${IJ}$ parameterizations without taking into account the masses of
\nuc{40}{Ca}, \nuc{48}{Ca}, \nuc{56}{Ni} and \nuc{90}{Zr} in the fit procedure.
In the resulting parameterizations, the spin-orbit coefficient $W_0$ is
typically $20\,\%$ lower than in the original ones. As a consequence, the
empirical value for the spin-orbit splitting of the neutron $3p$ level in
\nuc{208}{Pb} is met well within tolerance, at the price of binding energy
residuals in light nuclei being unacceptably large, \ie~\nuc{56}{Ni}
being underbound by 5~MeV while \nuc{40}{Ca} and \nuc{90}{Zr} are overbound by
up to 10~MeV. While the global trend of the spin-orbit splittings shown in
Fig.~\ref{fig:splittings} is enormously improved with these fits, in particular
for heavy nuclei, the overall agreement of the single-particle spectra with
experiment is not, so that we had to discard these parameterizations. This
finding hints at a deeply rooted deficiency of the Skyrme energy functional.
The spin-orbit and, when present, tensor terms indeed do simulate missing
physics of
the energy functional at the price of unrealistic spin-orbit splittings. This
also hints why perturbative studies, as those performed in~\cite{Sta77a,Col07a}
give much more promising results than what we will find below with our complete
refits. We will discuss mass residuals in more detail in
Sect.~\ref{sect:masses:chains} below.

\begin{figure}[t!]
  \includegraphics[width=0.8\columnwidth]{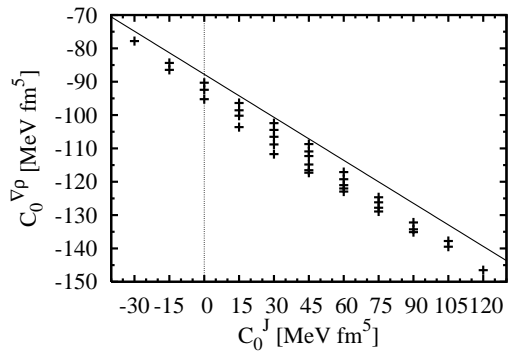}
  \caption{
    Correlation between the values of spin-orbit coupling constant
    $C_0^{\nabla J}$ and the isoscalar spherical effective
    spin-current coupling constant $C^{J}_0$. Dots: values for the
    actual parameterizations T$IJ$,
    solid line: trend estimated through Eq.~(\ref{eq:J2so}) (see text).
  }
  \label{fig:J2so}
\end{figure}

During the fit, the masses of light nuclei do not only compromise
the spin-orbit splittings, they also establish a correlation between
$W_0$ and $C^J_0$ in all
our parameterizations. The combined spin-orbit and spin-current  energy of a
given spherical nucleus $(N,Z)$ is given by (keeping only the isoscalar part
since we shall focus on the $N=Z$ nuclei \nuc{40}{Ca} and \nuc{56}{Ni})
\begin{equation}
\label{eq:Espin}
E^{\text{spin}}_0 (N,Z)
=   C_0^{\nabla J} \, \CI_0^{\nabla J} (N,Z)
  + C_0^{J} \, \CI_0^{J} (N,Z)
\end{equation}
with
\begin{eqnarray}
\label{eq:Jint1}
\CI_0^{\nabla J}(N,Z)
& = & \int \! \mathrm{d}^3 r \; \rho_0 \grad \cdot \vec J_0
     \\
\label{eq:Jint2}
\CI_0^{J}(N,Z)
& = & \int \! \mathrm{d}^3 r \; \vec{J}_0^2\,.
\end{eqnarray}
The difference of $E^{\text{spin}}_0$ between \nuc{56}{Ni} and \nuc{40}{Ca}
\begin{equation}
\label{eq:DeltaEspin}
  E^{\text{spin}}_0 \left( \text{\nuc{56}{Ni}} \right)
- E^{\text{spin}}_0 \left( \text{\nuc{40}{Ca}} \right)
= \Delta E^{\text{spin}}
\end{equation}
turns out to be fairly independent from the parameterization. Averaged over
all 36 parameterizations T$IJ$ used here, $\Delta E^{\text{spin}}$ has a value
of $-58.991$ MeV with a standard deviation as small as 3.202~MeV, or $5.4\%$.

The integrals in Eqs.~(\ref{eq:Jint1},\ref{eq:Jint2}) are fairly independent
from the actual parameterization. For a rough estimate, we can replace
them in Eq.~(\ref{eq:Espin}) by their average values. Plugged into
Eq.~(\ref{eq:DeltaEspin}) this yields
\begin{eqnarray}
\label{eq:J2so}
C_0^{\nabla J}
& = & \frac{\Delta E^{\text{spin}}
      - C_0^{J}\langle\CI_0^{J}\left(\text{\nuc{56}{Ni}}\right)
      - \CI_0^{J}\left(^{40}\mathrm{Ca}\right)\rangle}
           {\langle  \CI_0^{\nabla J}\left(^{56}\mathrm{Ni}\right)
                   - \CI_0^{\nabla J}\left(^{40}\mathrm{Ca}\right) \rangle}
\,.
\end{eqnarray}
Figure~\ref{fig:J2so} compares the values of $C_0^{\nabla J}$ as obtained
through (\ref{eq:J2so}) with the values for the actual parameterizations.
The estimate works very well, which demonstrates that $C_0^{\nabla J} =
-\tfrac{3}{4}W_0$ and $C^{J}_0$ are
indeed correlated and cannot be varied independently within a high quality fit
of the energy functional (\ref{eq:EF:sphere}). As the combined strength of the
spin-orbit and tensor terms in the energy functional is mainly determined by
the mass difference of the two $N=Z$ nuclei \nuc{40}{Ca} and \nuc{56}{Ni}, the
spin-orbit coupling constant $W_0$ depends more or less linearly on the
isoscalar tensor coupling constant $C^{J}_0$, while for all practical purposes
it is independent from the isovector one, see also Fig.~\ref{fig:w0} above.


\subsubsection{Splitting of high-$\ell$ states and the role of the radial
               form factor}


\begin{figure}[t!]
  \includegraphics[width=\columnwidth]{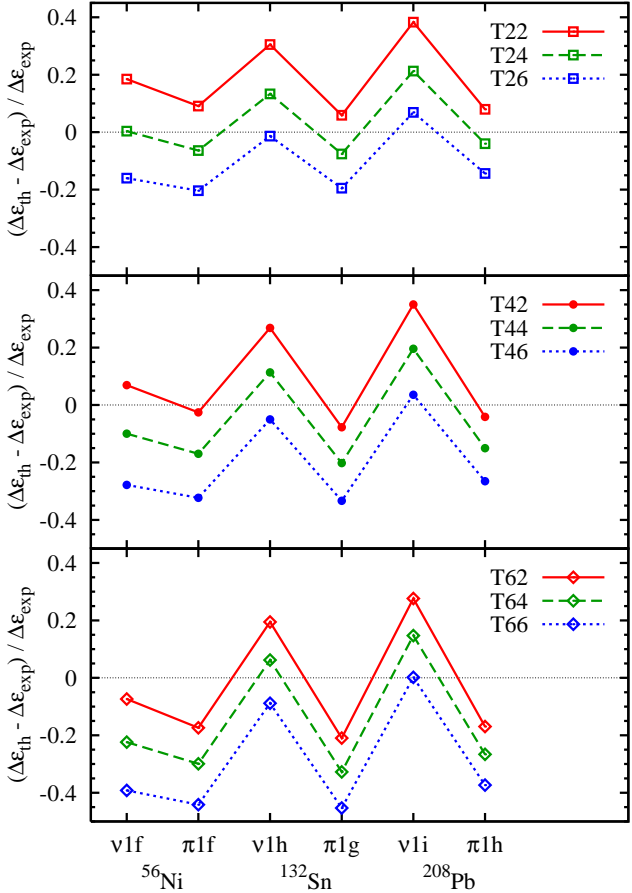}
  \caption{
    (Color online)
    Spin-orbit splittings of high-$\ell$ levels in magic nuclei across
    the Fermi energy. The calculated values are less robust against correlation
    effects than those shown in Fig.~\ref{fig:splittings} and have to be
    interpreted with caution (see text).
  }
  \label{fig:splihell}
\end{figure}

As stated above, it is common practice to confront only the spin-orbit
splittings between pairs of particle or hole states with calculated
single-particle energies from the spherical mean field. The spin-orbit
splitting of intruder states is rarely examined. Figure~\ref{fig:splihell}
displays the
relative deviation of the spin-orbit splittings of the intruder states with
$\ell \geq 3$ that span across major shell closures and are thus given by
the energy difference of a particle and a hole state. These splittings are not
``safe'', \ie~they can be expected to be strongly decreased by
polarization and correlation effects~\cite{Rut98a,Ber80a, Lit06a}.
To leave room
for this effect, a mean-field calculation should overestimate the empirical
spin-orbit splittings. We observe, however, that mean-field calculations done
here give values that are quite close to the experimental ones, or even smaller
for parameterizations with large positive isoscalar tensor coupling (cf.\ the
evolution from T22 to T66).

This means that the spin-orbit splittings are not too large in general,
as might be concluded from Fig.~\ref{fig:splittings}, but that there
is a wrong trend of the splittings with $\ell$ with the strength of
the spin-orbit potential establishing a compromise between the in-shell
splittings of small $\ell$ orbits
that are too large and the across-shell splittings of the intruders that are
tentatively too small. In fact, the levels in Fig.~\ref{fig:splihell} obviously
have in common that their radial wave functions do not have nodes, while the
levels in Fig.~\ref{fig:splittings} have one or two nodes, with the notable
exception of the $1p$ levels in \nuc{16}{O}, for which we also find smaller
deviations of the spin-orbit splittings than for the other levels in
Fig.~\ref{fig:splittings}.

Underestimating the spin-orbit splittings of intruder levels has immediate and
obvious consequences for the performance of an effective interaction, as this
closes the magic gaps in the single-particle spectra and compromises the
predictions for doubly-magic nuclei, as we will demonstrate in detail below. By
contrast, the spin-orbit splittings of the low-$\ell$ states within the major
shells have no obvious direct impact on bulk properties. Their deviation from
empirical data is less dramatic, as the typical bulk observables discussed
with mean-field approaches are not very sensitive to them. It is
only in applications
to spectroscopy that their deficiencies become evident. It is noteworthy that
the parameterization T22 \emph{without} effective tensor terms at sphericity
provides a reasonable compromise between the tentatively underestimated
splittings of the intruder levels shown in Fig.~\ref{fig:splihell} and the
tentatively overestimated splittings of the levels within major shells shown in
Fig.~\ref{fig:splittings} above, while for parameterizations with tensor terms
this balance is lost.

There clearly is a proton-neutron staggering in Figs.~\ref{fig:splittings}
and~\ref{fig:splihell}, such that calculated proton splittings are relatively
smaller than the neutron ones. The effect appears both when comparing proton
and neutron levels with different $\ell$ in the same nucleus, and when
comparing proton and neutron levels with the same $\ell$ in the same or
different nuclei (see the $1h$ levels in \nuc{132}{Sn} and \nuc{208}{Pb}).
The staggering for the intruder levels is even amplified for parameterizations
with large proton-neutron tensor term, as T62, T64 or T66. The effect is
particularly prominent for the heavy \nuc{132}{Sn} and \nuc{208}{Pb} with a
large proton-to-neutron ratio $N/Z$, which might hint at unresolved isospin
dependence of the spin-orbit interaction, although alternative explanations
that involve how single-particle states in different shells should interact
through tensor and spin-orbit forces are possible as well, see also the next
paragraph.

Note that also the spin-orbit splittings of the low-$\ell$ levels shown in
Fig.~\ref{fig:splittings} exhibit a staggering, which is of smaller amplitude,
though. It has been pointed out by
Skalski~\cite{Ska01a}, that an exact treatment of the Coulomb exchange term
(compared to the Slater approximation used here and nearly all existing
literature) does indeed slightly increase the spin-orbit splittings of protons
across major shells. This effect might give a clue to the staggering observed
for the $N=Z$ nucleus \nuc{56}{Ni}, but the magnitude of the effect reported
in~\cite{Ska01a} is too small to explain the large staggering we find for the
heavier $N \neq Z$ nuclei.

\begin{figure}[t!]
  \includegraphics[width=\columnwidth]{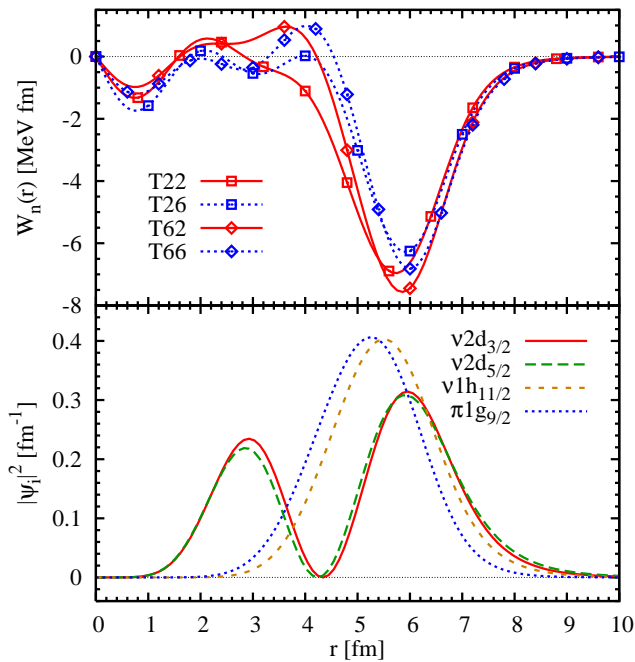}
  \caption{
    (Color online)
    Neutron spin-orbit potential (top) and the radial wave function
    of selected orbitals (bottom) in \nuc{132}{Sn}.
  }
  \label{fig:Sn132-field}
\end{figure}

Next, we use the example of \nuc{132}{Sn} to demonstrate why the spin-orbit
splittings of nodeless high-$\ell$ states are more sensitive to the tensor terms
than low-$\ell$ states with one or several nodes, see
Fig.~\ref{fig:Sn132-field}. The lower panel shows the neutron spin-orbit
potential in \nuc{132}{Sn} for four different parameterizations, while the upper
panel shows selected radial single-particle wave functions. The $\nu$
$1h_{11/2}$ and $\pi$ $1g_{9/2}$ levels give the main contribution to the
neutron and proton spin-orbit currents in this nucleus, and consequently to the
tensor contribution to the spin-orbit potential. Indeed, the largest differences
between the spin-orbit potentials from the chosen parameterizations are caused
by the varying contribution from the tensor terms and appear for the region
between 3 and 6~fm, where the wave functions of the $1g$ and $1h$ states are
peaked. This region corresponds to the inner flank of the spin-orbit potential
well, while the outer flank is much less affected. While the $1g$ and $1h$ wave
functions are peaked at the inner flank, the $2d$ orbitals have their node in
this region. Consequently, the splittings of the $1g$ and $1h$ levels are
strongly modified by the tensor terms, while those of the $2d$ orbitals are
quite insensitive.

As a rule of thumb, the tensor contribution to the spin-orbit potential in
doubly-magic nuclei comes mainly from the nodeless intruder states, which, when
present, in turn mainly affect their own spin-orbit splittings, leaving the
splittings of the low-$\ell$ states with one or more nodes nearly unchanged for
reasons of geometrical overlap.

We note in passing that the slightly different radial wave functions of
the $2d$ orbitals demonstrate nicely that their contribution to the
spin-orbit current, Eq.~(\ref{eq:j:radial}), cannot completely cancel.

In fact, when regarding more specifically the evolution of the spin-orbit
potential between the parameterizations T22 and T66, it is striking that for T66
it is essentially narrowed and its minimum slightly pushed towards larger radii,
while its depth remains unaltered. Recalling that T66 shows a pathological
behavior of too weak spin-orbit splitting of the intruder states, it appears
that a correct $\ell$-dependence of spin-orbit splittings might require to
modify the radial dependence of the spin-orbit potential such that it becomes
wider towards smaller radii. This uncalled-for modification of the shape of the
spin-orbit field has previously been put forward by Brown \etal~\cite{Bro06a} as
an argument for a negative like-particle $\vec{J}^2$ coupling constant $\alpha$.
However, as will be discussed in paragraph~\ref{sect:spectro-nn}
below, the evolution of single-particle levels along isotopic chains calls
for $\alpha > 0$, see also \cite{Bro06a}.
Additionally, as we will show
in appendix~\ref{sect:app:instability}, large negative values of $\alpha$
pose the risk of instabilities towards the transition to states with unphysical
shell structure.


\subsubsection{Single-particle spectra of doubly-magic nuclei}
\label{sect:doublymagic}


\begin{figure}[t!]
  \includegraphics[width=\columnwidth]{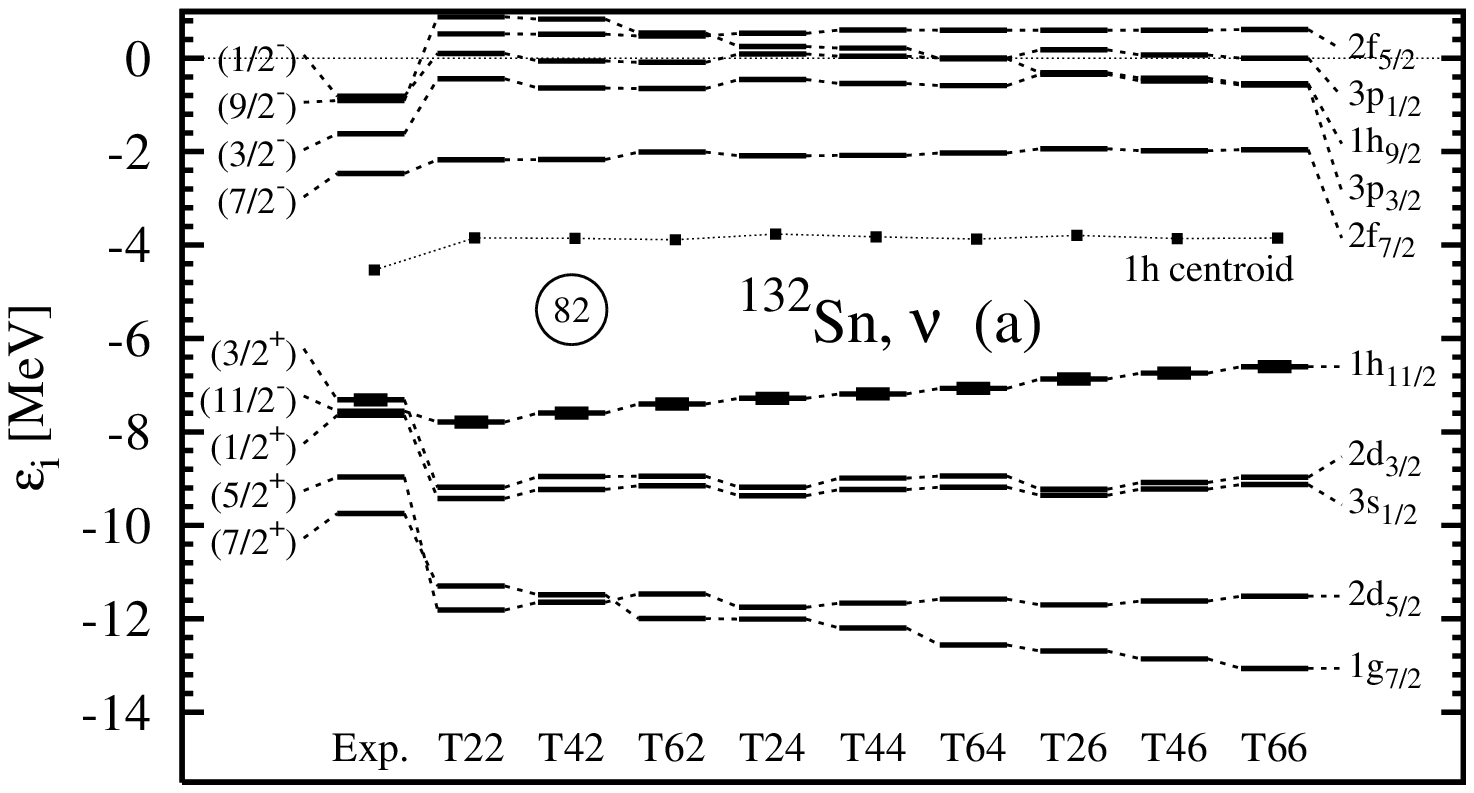}
  \\ \vspace{-12pt}
  \includegraphics[width=\columnwidth]{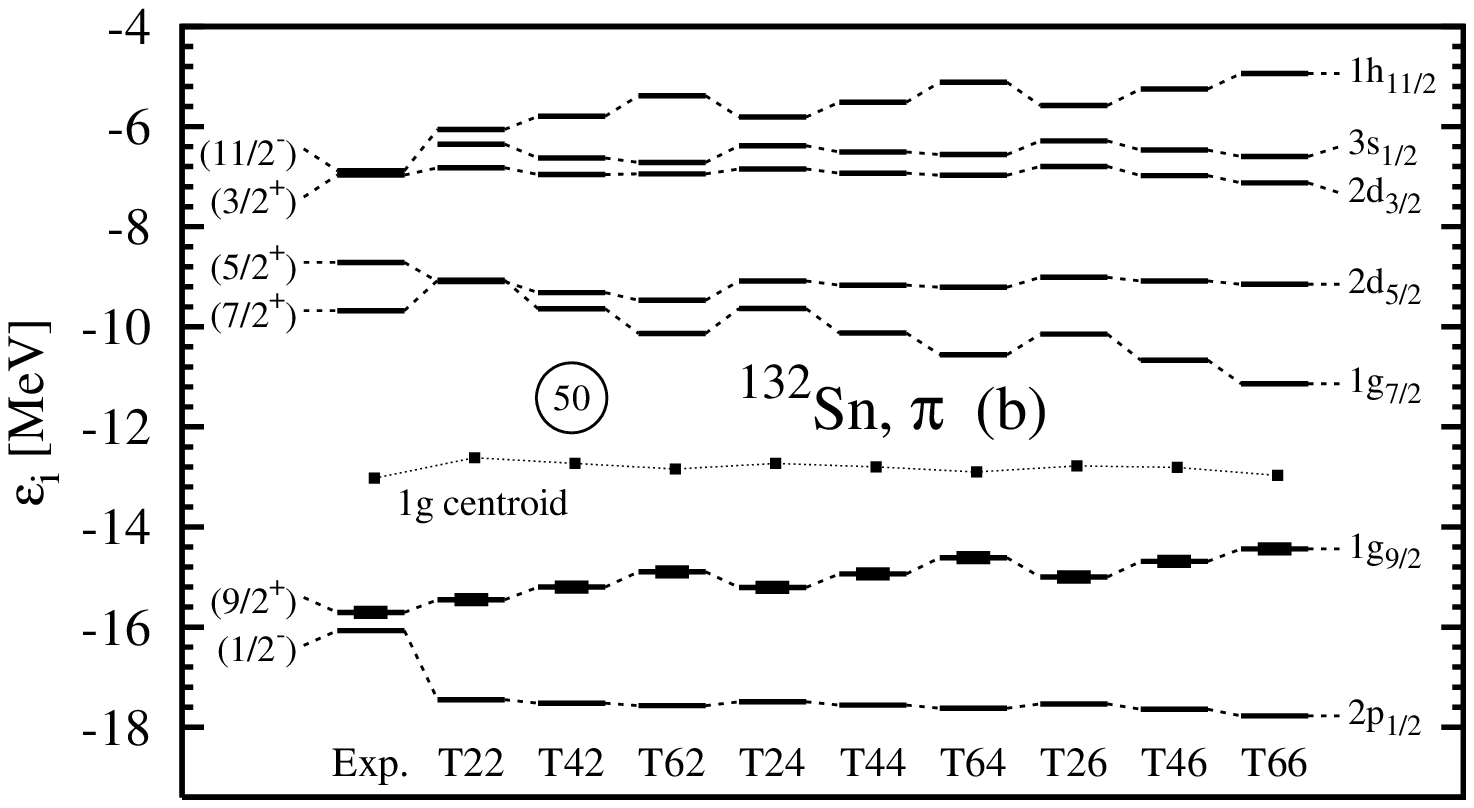}
  \caption{
    Single-particle energies in \nuc{132}{Sn} for a subset of our
    parameterizations. We also show the centroid of
    the intruder levels, defined through Eq.~(\protect\ref{eq:centroid})
    Top panel: neutron levels, bottom panel: proton levels. A thick mark
    indicates the Fermi level.
  }
  \label{fig:spe-Sn132}
\end{figure}

\begin{figure}[th!]
  \includegraphics[width=\columnwidth]{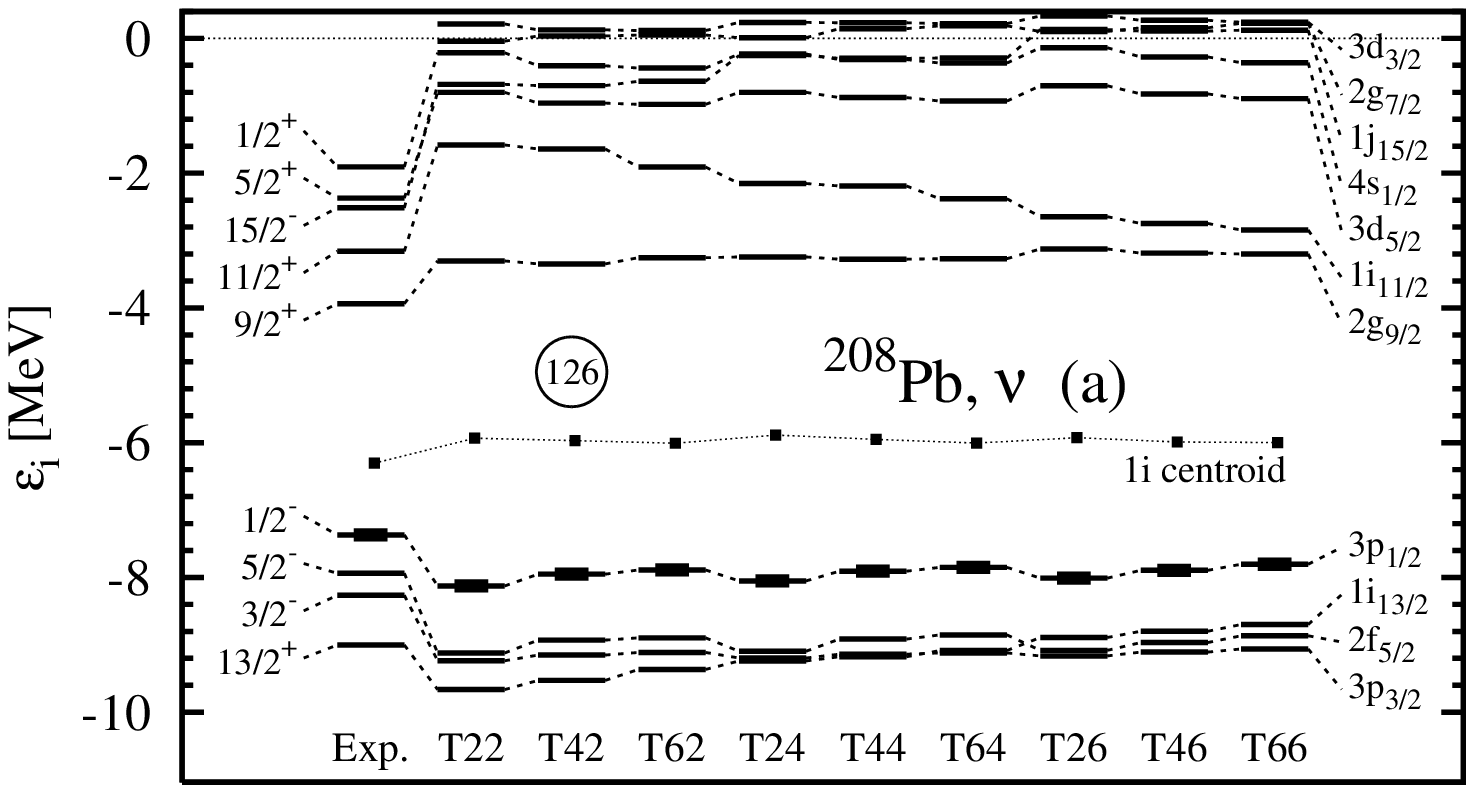}
  \\ \vspace{-12pt}
  \includegraphics[width=\columnwidth]{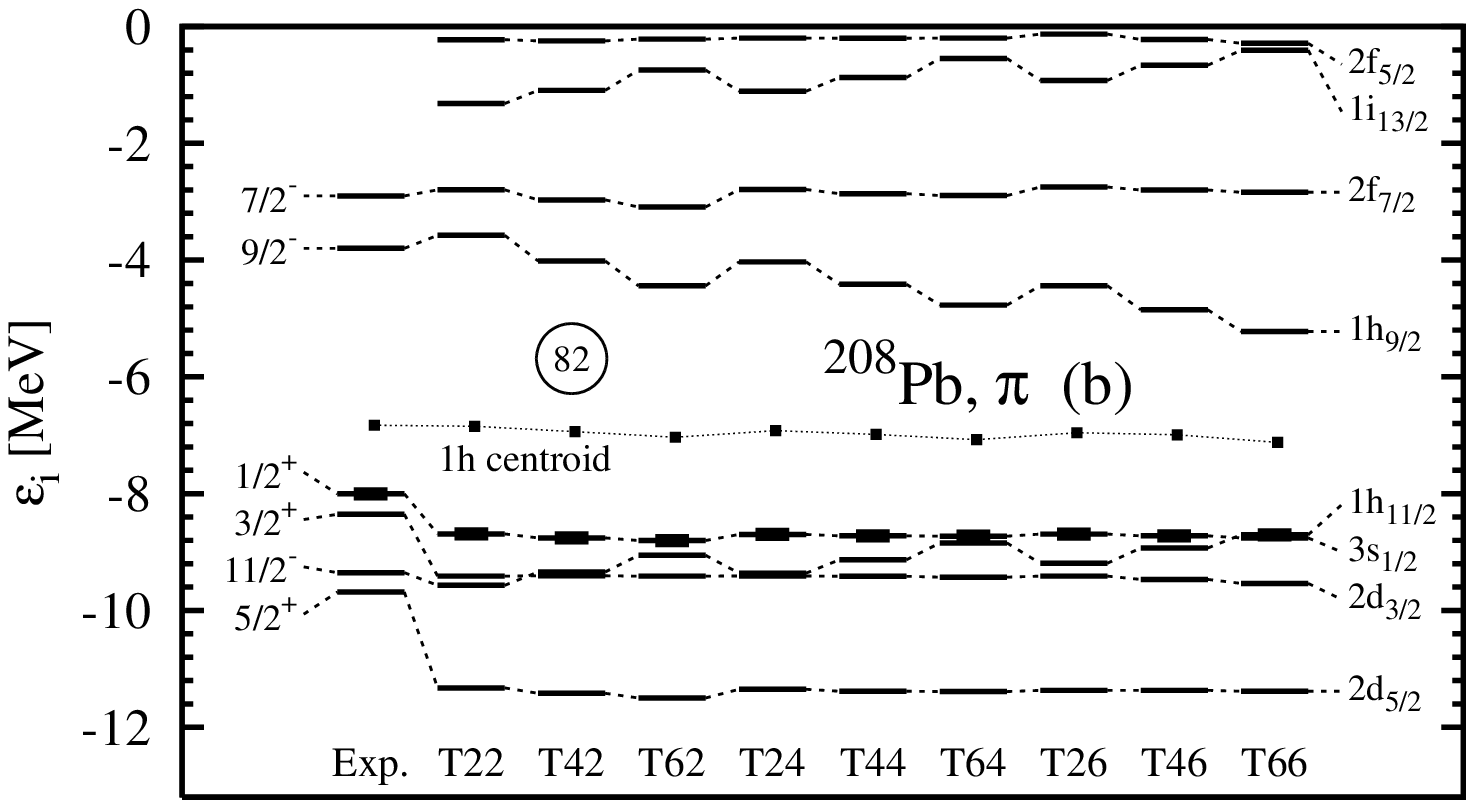}
  \caption{
    Same as Fig.~\ref{fig:spe-Sn132} for \nuc{208}{Pb}.
  }
  \label{fig:spe-Pb208}
\end{figure}

After we have examined the predictions for spin-orbit splittings, we will now
turn to the overall quality of the single-particle spectra of doubly-magic
nuclei. Figure~\ref{fig:spe-Sn132} shows the single-particle spectrum of
\nuc{132}{Sn}. It is evident that as a consequence of the underestimated
spin-orbit splittings of the intruder levels that we discussed in the last
section, the spectrum is deteriorated for large positive isoscalar tensor term
coupling constants $C^{J}_0$ (see T66), as, for example, a decrease of the
spin-orbit splitting of the neutron $1h$ shell pushes the $1h_{11/2}$ further
up, closing the $N=82$ gap. As a consequence, the presence of the tensor terms
cannot remove the problem shared by all standard mean-field methods that always
wrongly put the neutron $1h_{11/2}$ level above the $2d_{3/2}$ and $3s_{1/2}$
levels~\cite{RMP}, which compromises the description of the entire mass region.
For the same reason, the proton spectrum of \nuc{132}{Sn} also excludes
interactions with large positive $C^{J}_0$, which reduces the $Z=50$ gap between
the $1g$ levels to unacceptable small values.

Figure~\ref{fig:spe-Sn132} also shows the energy centroids of the
$\nu$ $1h$ and $\pi$ $1g$ levels, defined as
\begin{equation}
\label{eq:centroid}
\varepsilon_{qn\ell}^\text{cent}
=   \frac{\ell+1}{2\ell + 1} \, \varepsilon_{qn\ell,j=\ell+1/2}
  + \frac{\ell}{2\ell + 1}   \, \varepsilon_{qn\ell,j=\ell-1/2}\,.
\end{equation}
The position of the centroid is fairly independent from the parameterization.
Assuming that the calculated energy of the centroid of an intruder state is more
robust against corrections from core polarization and particle-vibration
coupling that its spin-orbit splitting, we see that the $\nu\,1h$ centroid is
clearly too high in energy by about 1~MeV. In combination with its
tentatively too small spin-orbit splitting, see Fig.~\ref{fig:splihell}, this
offers an explanation for the notorious wrong  positioning of the
$\nu$~$1h_{11/2}$, $2d_{3/2}$ and $3s_{1/2}$ levels in
\nuc{132}{Sn}~\cite{RMP}. The near-degeneracy of the $\nu$~$2d_{3/2}$ and
$3s_{1/2}$ levels is always well reproduced, while the $1h_{11/2}$ comes out
much too high. As the $1h_{11/2}$ is the last occupied neutron level,
self-consistency puts it close to the Fermi energy, which, in turn, pushes
the $2d_{3/2}$ and $3s_{1/2}$ levels down in the spectrum.

The overall situation is similar for \nuc{208}{Pb}, see
Fig.~\ref{fig:spe-Pb208}. Again, the high-$\ell$ intruder states move too close
to the $Z=82$ and $N=126$ gaps for large positive $C^{J}_0$. The effect is
less obvious than for \nuc{132}{Sn} as the intruders and their spin-orbit
partners are further away from the gaps. Still, the level ordering and the
size of the $Z=82$ gap become unacceptable for parameterizations with large
tensor coupling constants. For strong tensor term coupling constants (both
like-particle and proton-neutron), a $Z=92$ gap opens in the single-particle
spectrum of the protons that is also frequently predicted by relativistic
mean-field models~\cite{Rut98a,Ben99b} but absent in experiment~\cite{Hau01aE}.

\begin{figure}[t!]
  \includegraphics[width=\columnwidth]{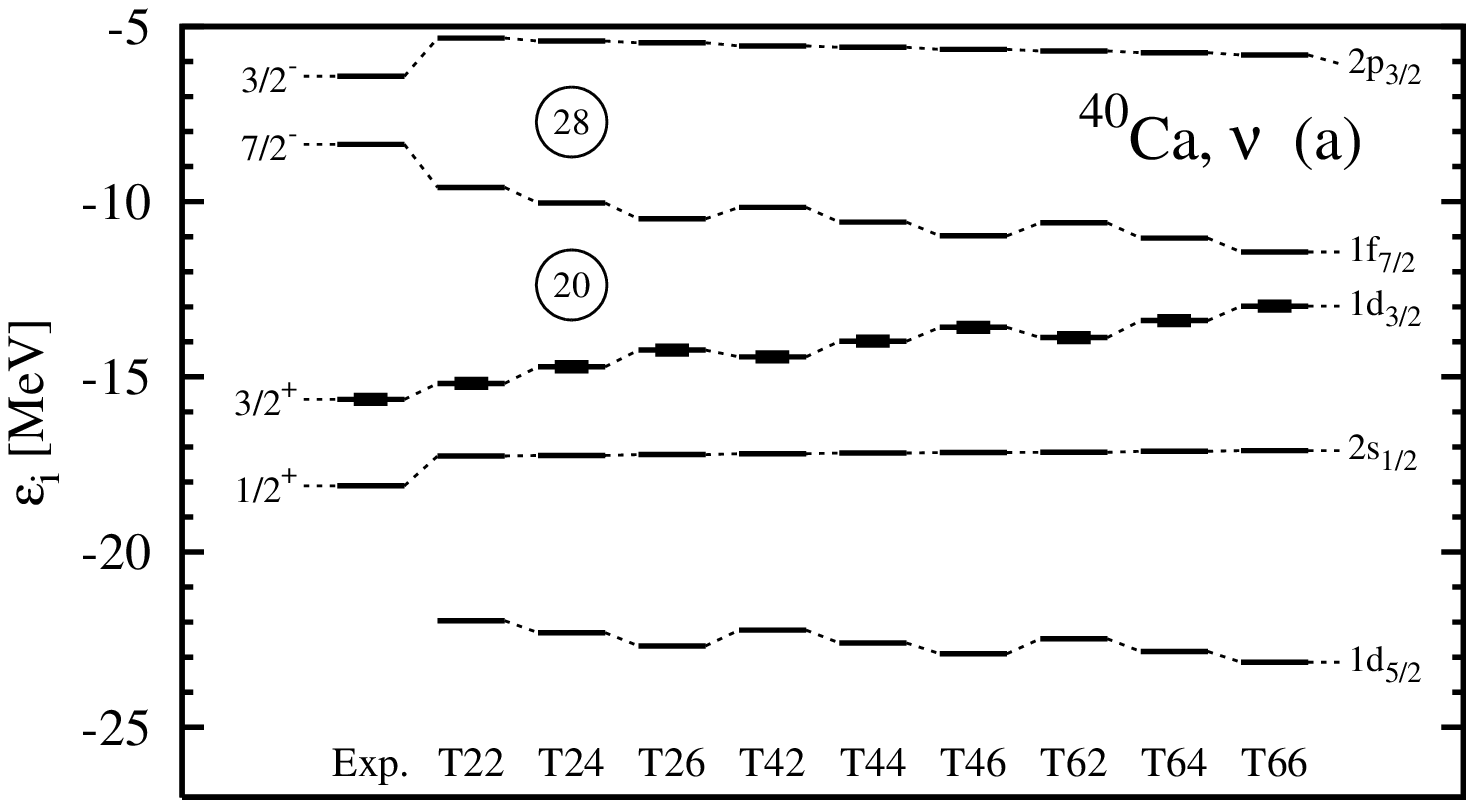}
  \\ \vspace{-12pt}
  \includegraphics[width=\columnwidth]{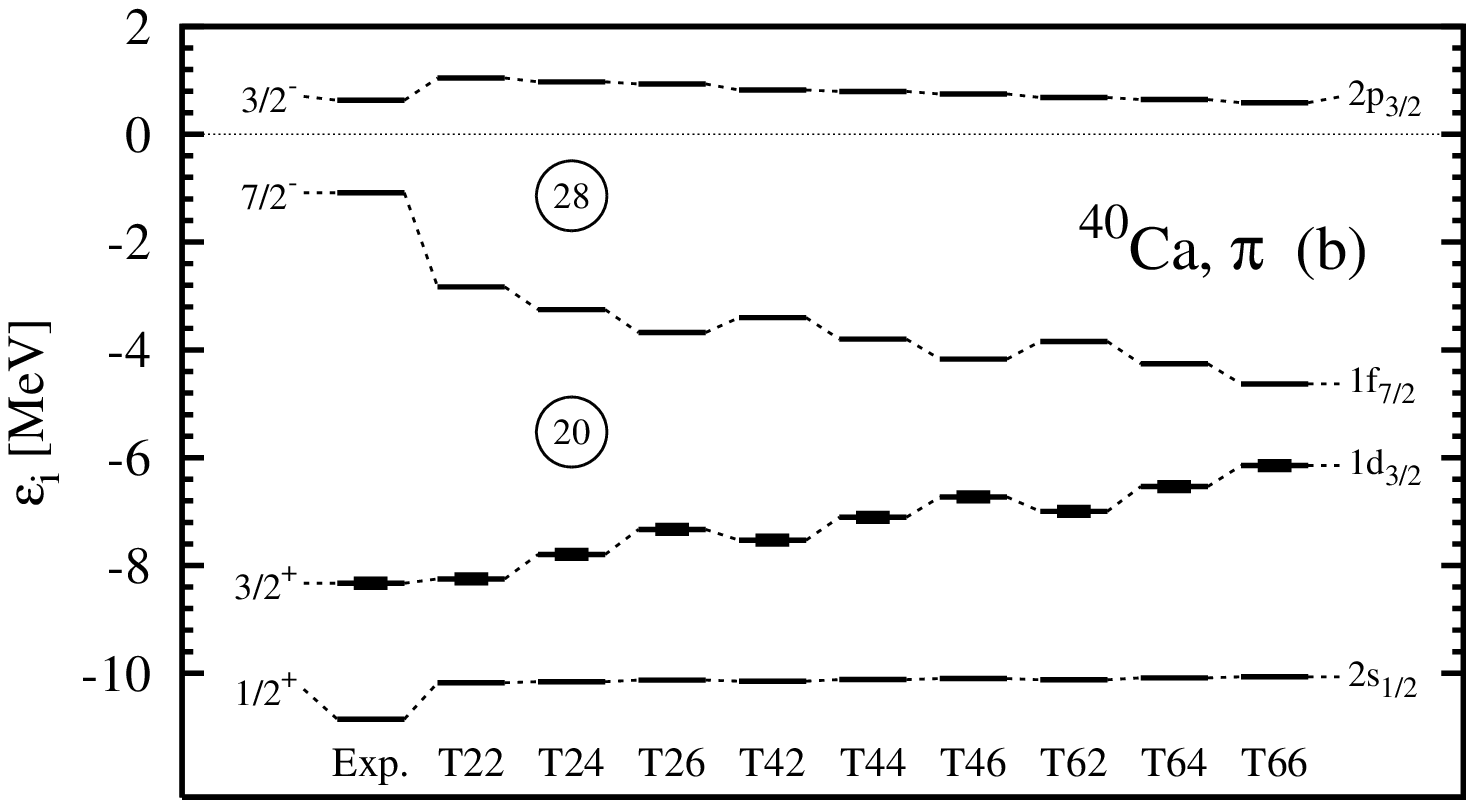}
  \caption{
    Same as Fig.~\ref{fig:spe-Sn132} for \nuc{40}{Ca}.
  }
  \label{fig:spe-Ca40}
\end{figure}

\begin{figure}[t!]
  \includegraphics[width=\columnwidth]{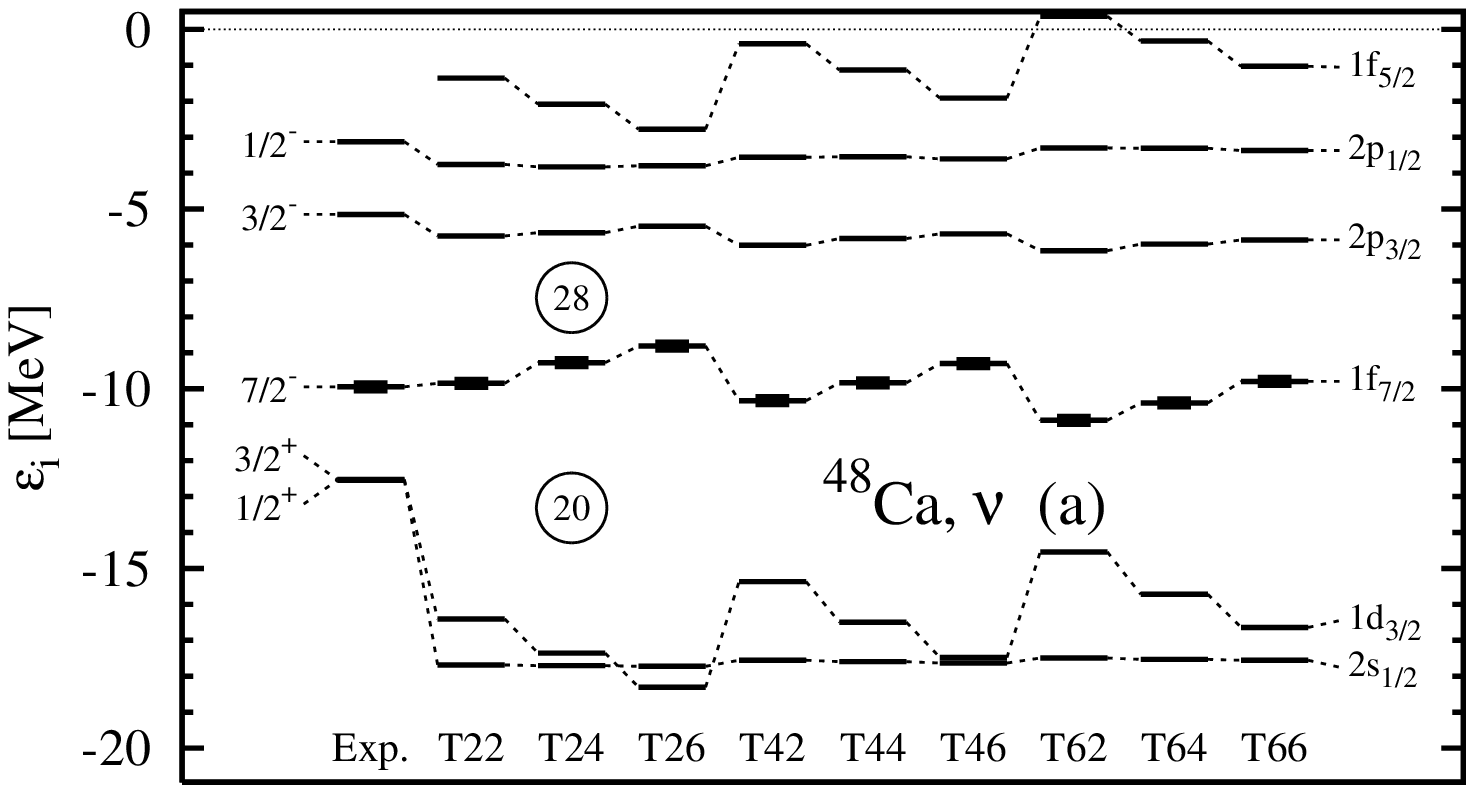}
  \\ \vspace{-12pt}
  \includegraphics[width=\columnwidth]{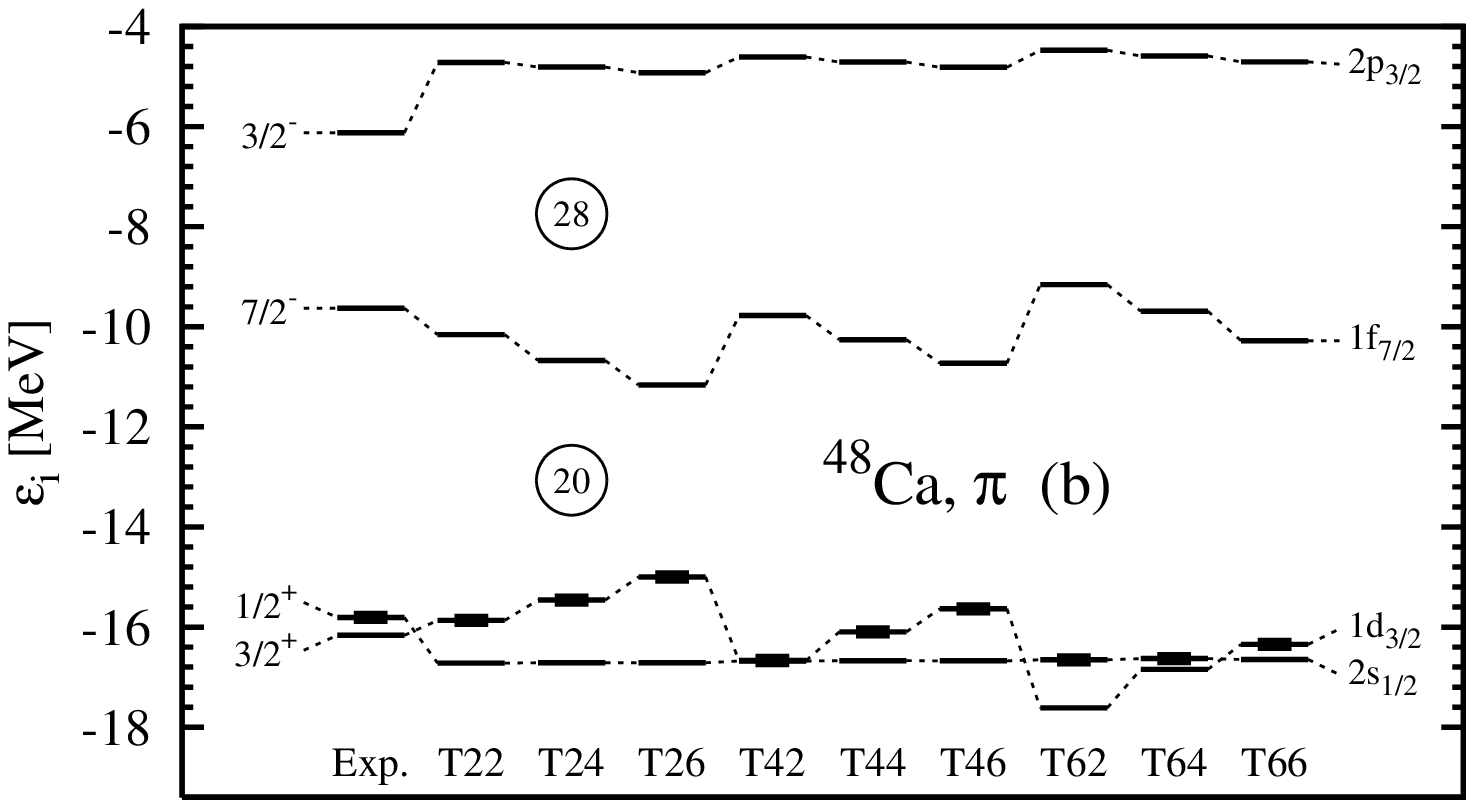}
  \caption{
    Same as Fig.~\ref{fig:spe-Sn132} for \nuc{48}{Ca}.
  }
  \label{fig:spe-Ca48}
\end{figure}

\begin{figure}[t!]
  \includegraphics[width=\columnwidth]{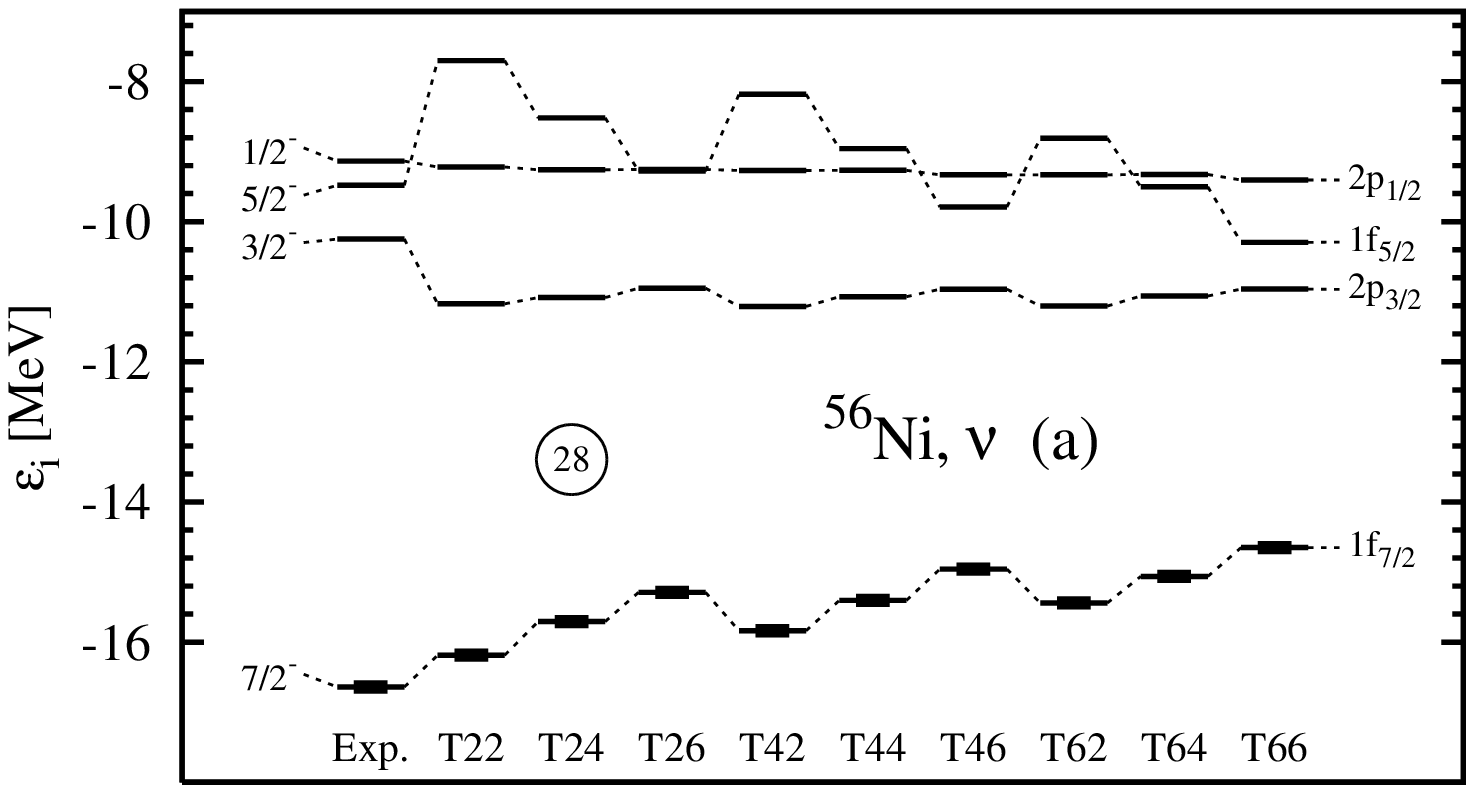}
  \\ \vspace{-12pt}
  \includegraphics[width=\columnwidth]{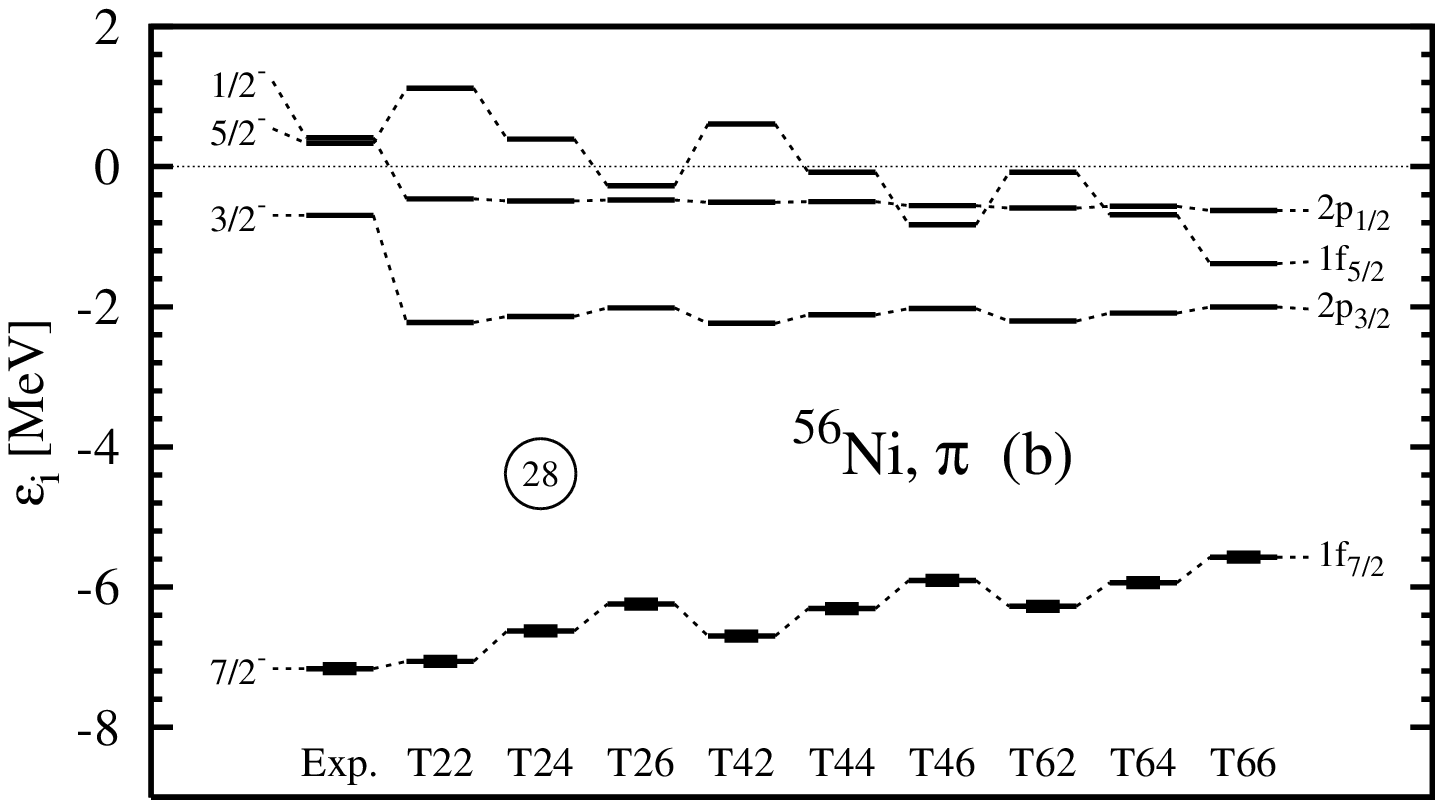}
  \caption{
    Same as Fig.~\ref{fig:spe-Sn132} for \nuc{56}{Ni}.
  }
  \label{fig:spe-Ni56}
\end{figure}

\begin{figure}[t!]
  \includegraphics[width=\columnwidth]{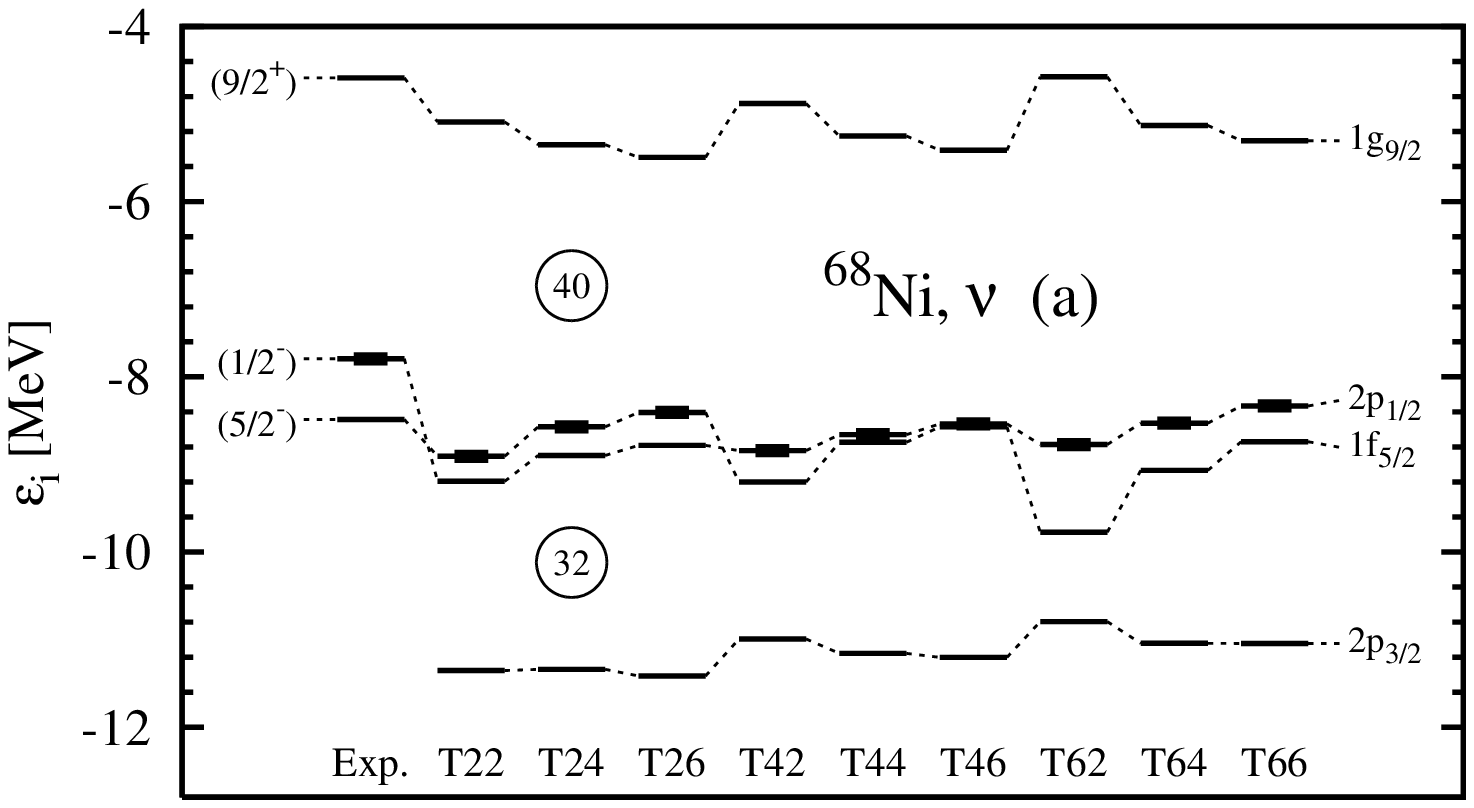}
  \\ \vspace{-12pt}
  \includegraphics[width=\columnwidth]{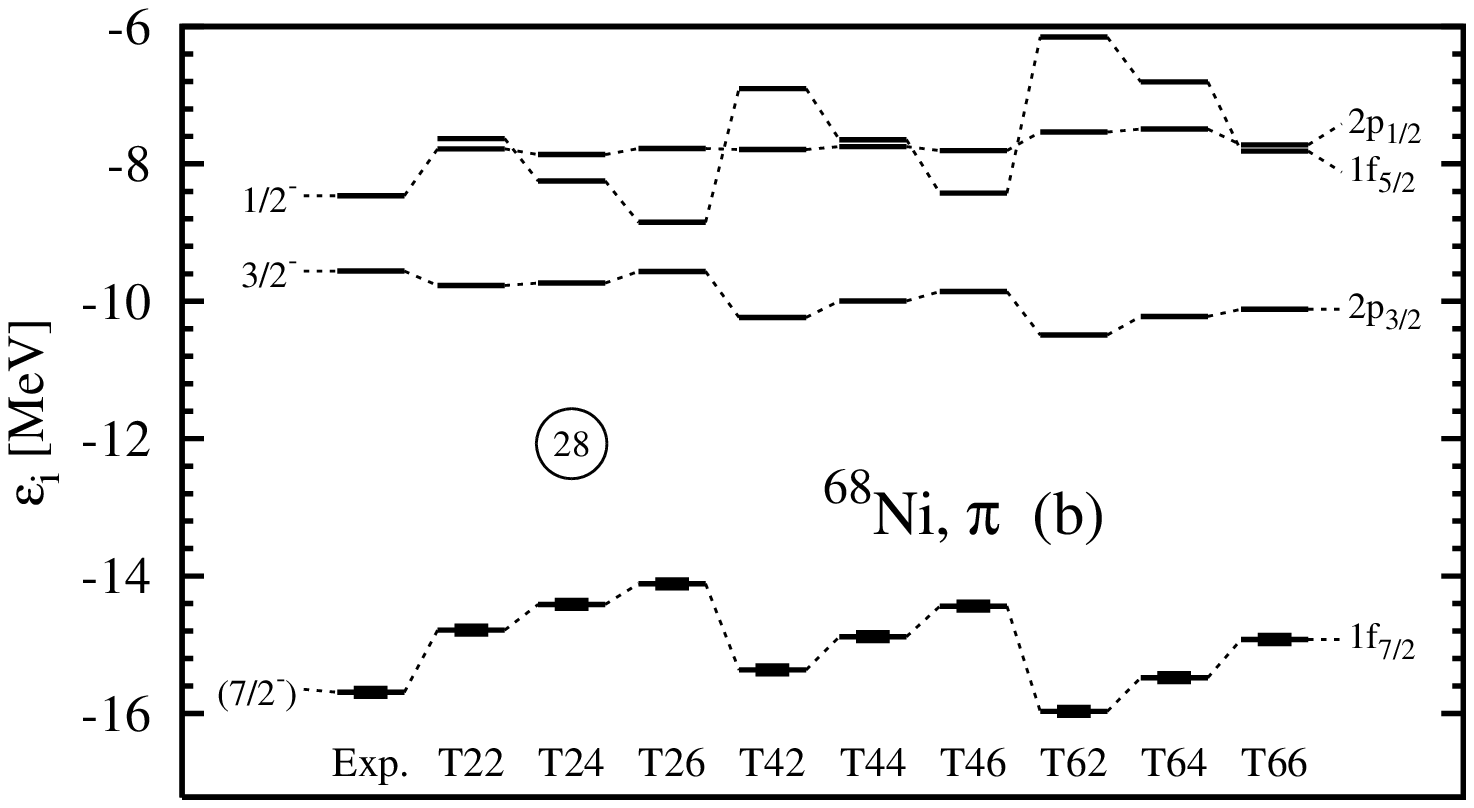}
  \caption{
    Same as Fig.~\ref{fig:spe-Sn132} for \nuc{68}{Ni}.
  }
  \label{fig:spe-Ni68}
\end{figure}

\begin{figure}[t!]
  \includegraphics[width=\columnwidth]{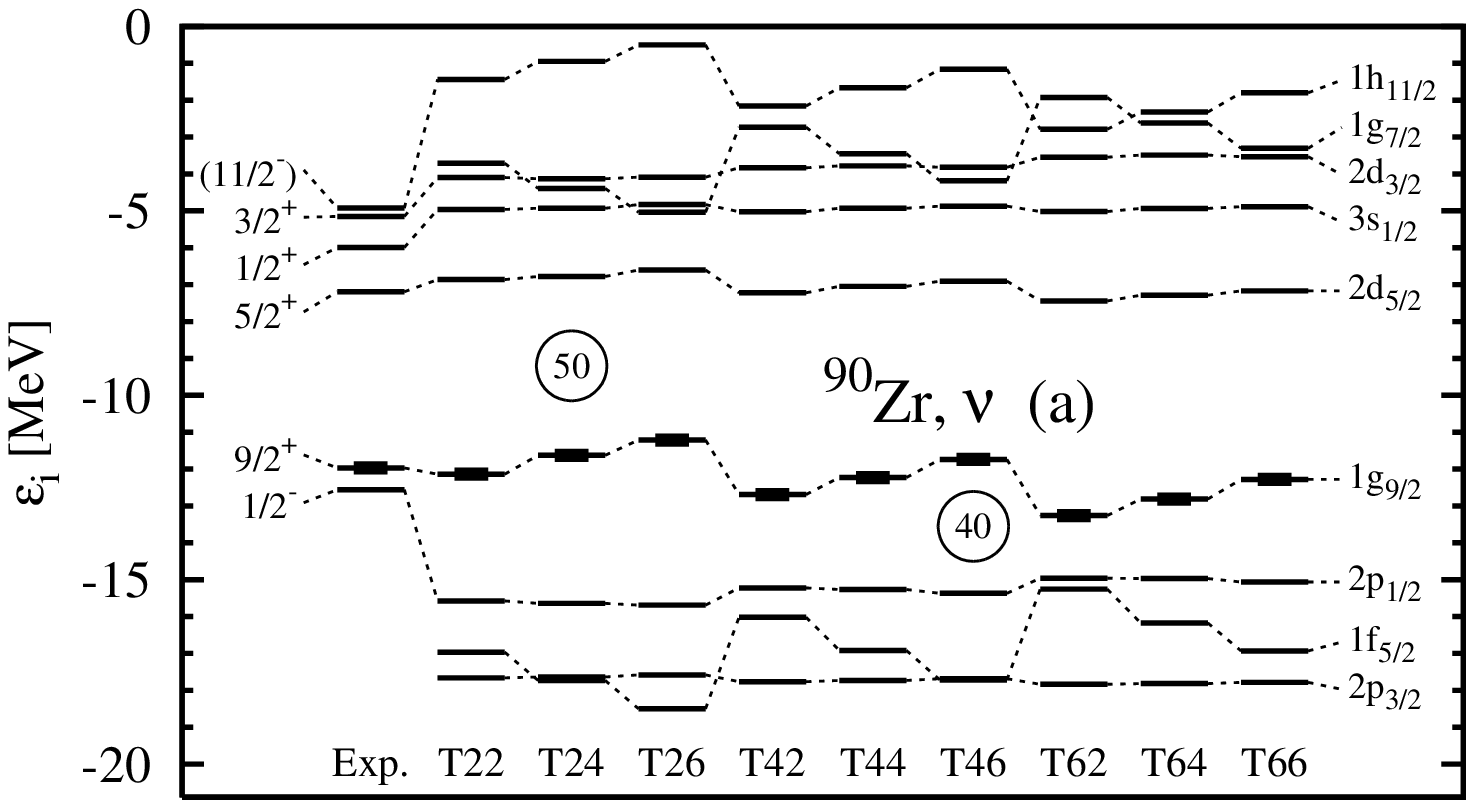}
  \\ \vspace{-12pt}
  \includegraphics[width=\columnwidth]{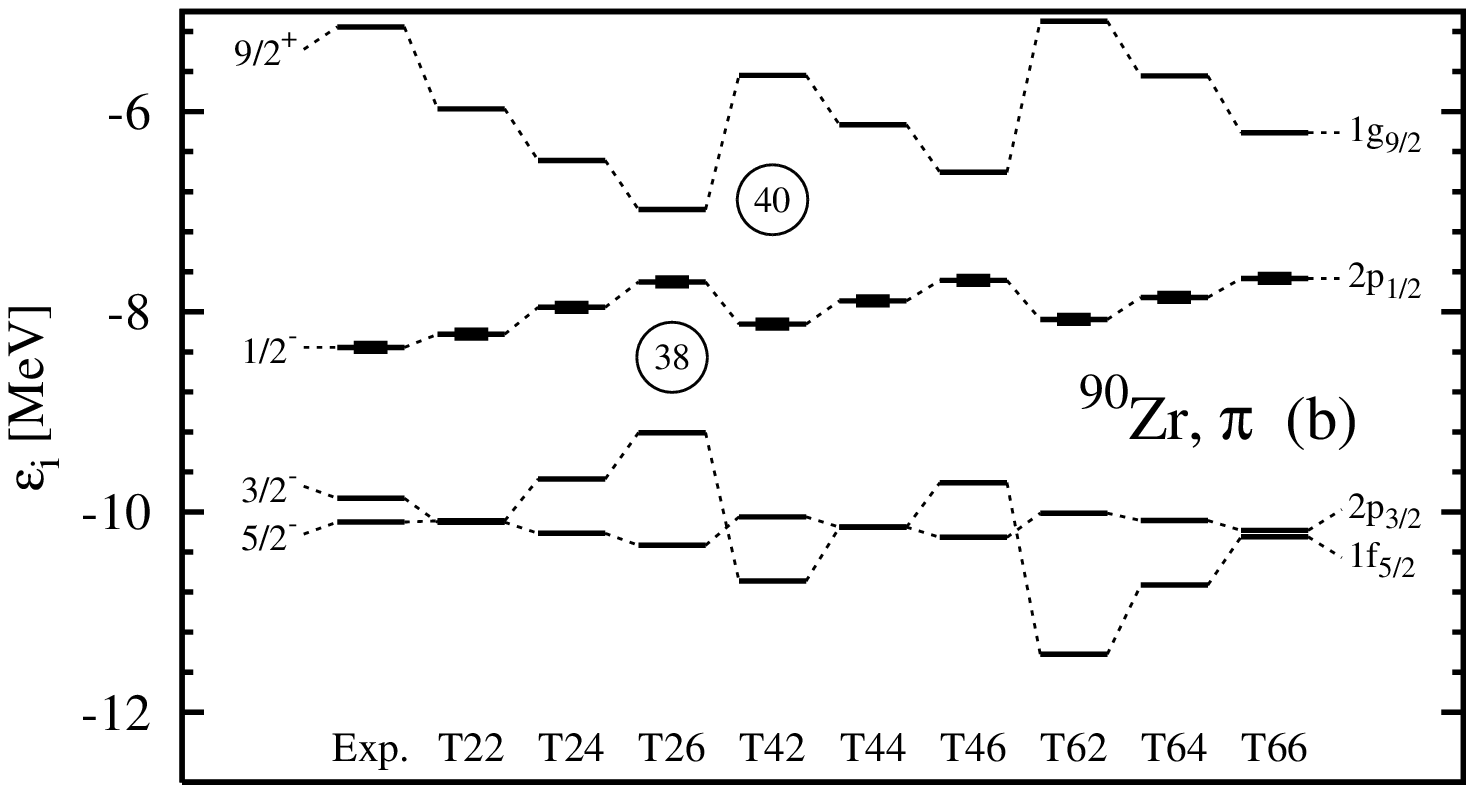}
  \caption{
    Same as Fig.~\ref{fig:spe-Sn132} for \nuc{90}{Zr}.
  }
  \label{fig:spe-Zr90}
\end{figure}

The single-particle spectra for the light doubly magic nuclei \nuc{40}{Ca}
(Fig. \ref{fig:spe-Ca40}), \nuc{48}{Ca} (Fig. \ref{fig:spe-Ca48}),
\nuc{56}{Ni} (Fig. \ref{fig:spe-Ni56}), \nuc{68}{Ni} (Fig. \ref{fig:spe-Ni68})
and \nuc{90}{Zr} (Fig. \ref{fig:spe-Zr90}),
all have in common that the relative impact of the
$\vec{J}^2$ terms on the ordering and relative distance of single-particle
levels is even stronger than for the heavy nuclei discussed above.
But not all of the strong dependence on the coupling constants of
the $\vec{J}^2$ terms that
we see in the figures is due to the actual contribution of the tensor terms to
the spin-orbit potential. This is most obvious for \nuc{40}{Ca}, where protons
and neutrons are spin-saturated so that the $\vec{J}^2$ terms do not contribute
to the spin-orbit potentials. Still, increasing their coupling constants
increases the spin-orbit splittings, which manifests the readjustment of the
spin-orbit force to a given set of $C^{J}_0$ and $C^{J}_1$ (see
Fig.~\ref{fig:w0}). The evolution of the spin-orbit splittings in \nuc{40}{Ca}
visible in Fig. \ref{fig:spe-Ca40} is the background which we have to keep in
mind when discussing the impact of the tensor terms on nuclei with
non-vanishing spin-orbit currents.
Note that the spin-orbit coupling constant $W_0$ is
correlated with isoscalar tensor coupling constant $C^{J}_0$, such that the
single-particle spectra obtained with T24 and T42 are very similar, as they are
for T26, T44 and T62.

For \nuc{48}{Ca}, Fig. \ref{fig:spe-Ca48}, the protons are still spin-saturated
with vanishing proton spin-orbit current $\vec{J}_\prm$, while for neutrons we
have a large $\vec{J}_\nrm$. Depending on the nature of the tensor terms in the
energy functional -- i.e.~like-particle or proton-neutron or a mixture of
both -- the spin-orbit current will either contribute to the spin-orbit
potential of the neutrons or that of the protons or both, see Eq.
(\ref{eq:wpot:tot}). For the parameterizations with dominating like-particle
$\vec{J}^2$ term, for example T24 and T26, the situation for the protons is the
same as for \nuc{40}{Ca}: there is no contribution from the tensor terms to the
proton spin-orbit splittings, but compared to T22 the proton $Z=20$ gap is
reduced through the readjustment of the spin-orbit force, leading to values that
are too small. For the same parameterizations, the large contribution from
$\vec{J}_\nrm$ to $W_\nrm$ opens up the $N=20$ gap to values that are
tentatively too large, as it reduces the neutron spin-orbit splittings and
thereby compensates, even overcompensates, the effect from the readjustment of
the spin-orbit force. At the same time the $N=28$ gap is reduced. The opposite
effect is seen for parameterizations with large proton-neutron tensor term, for
example T42 or T62. For those, the proton spin-orbit splitting is reduced,
opening up the $Z=20$ gap compared to T22, while the neutron spin-orbit
splittings are increased by the background effect from the readjusted spin-orbit
force.

For \nuc{56}{Ni}, Fig. \ref{fig:spe-Ni56}, we have large $\vec{J}_\nrm$ and
$\vec{J}_\prm$.
In this $N=Z$ nucleus, the like-particle or proton-neutron parts
of the tensor terms cannot be distinguished.  The spectra depend only on the
overall coupling constant of the isoscalar tensor term $C^{J}_0$, on the one
hand directly through the contribution of the tensor terms to the spin-orbit
potentials, and on the other hand through the background readjustment of $W_0$
that is correlated to $C^{J}_0$ as well. As already mentioned,
results for T24 and T42 are very similar, as they are for T26, T44 and T62.
All parameterizations
have in common that the proton and neutron gaps at 28 are too small. The
variation of the single-particle spectra among the parameterizations is smaller
than for \nuc{40}{Ca}, mainly because the tensor terms compensate the
background drift from the readjustment of $W_0$.

The slightly neutron-rich \nuc{68}{Ni} combines a spin-saturated sub-shell
closure $N=40$ that gives a vanishing neutron spin-orbit current with the magic
$Z=28$ that gives a strong proton spin-orbit current. The variation of the
single-particle spectra in dependence of the coupling constants
of the tensor terms is similar to those of \nuc{48}{Ca}, with the roles of
protons and neutrons exchanged.

The nucleus \nuc{90}{Zr} combines the spin-saturated proton sub-shell closure
$Z=40$ with the major neutron shell closure $N=50$. The high degeneracy of the
occupied $\nu$ $1g_{9/2}$ level leads to a very strong neutron spin-orbit
current, while the proton spin-orbit current is zero. Even in the absence
of a tensor term
contributing to their spin-orbit potential for parameterizations with pure
like-particle tensor terms, the proton single-particle spectra are dramatically
changed by the feedback effect from the readjusted spin-orbit force; see
the evolution from T22 to T26. The $\pi$~$1g_{9/2}$ comes down, and
closes the $Z=40$ sub-shell gap. For parameterizations with pure
proton-neutron tensor term, one has the opposite effect, this time because the
contribution from the tensor terms overcompensates the background effect from
the spin-orbit force. The effect of the tensor terms on the neutron
spin-orbit splittings is less dramatic, but still might be sizable.

We have to point out that the calculations displayed in
Fig.~\ref{fig:spe-Zr90} were performed without taking pairing into account,
as the HFB scheme breaks down in the weak pairing regime of doubly magic
nuclei. For some extreme (and unrealistic) parameterizations, however, the gaps
disappear which, in turn, would lead to strong pairing correlations if the
calculations were performed within the HFB scheme. This happens, for example,
for neutrons in \nuc{90}{Zr} when using T26 and T46. Interestingly, the pairing
correlations for neutrons break the spin saturation, which leads to a
substantial neutron spin-orbit current $\vec{J}_\nrm$.
As these parameterizations use values of the like-particle
coupling constant significantly larger than the neutron-proton one,
$\vec{J}_\nrm$ feeds back onto the neutron spin-orbit
potential only, Eq.~(\ref{eq:wpot:tot}). As the corresponding coupling
constant $\alpha$ is positive for T26 and T46, the contribution from the
tensor terms reduces the spin-orbit splittings, in particular those of the
$1g_{9/2}$ and $1f_{5/2}$. As a result, this counteracts the
reduction of the $N=40$ gap predicted by T26 and T46 in calculations without
pairing.


\subsubsection{Evolution along isotopic chains: $np$ coupling}
\label{sect:spectro-np}


\begin{figure}[t!]
  \includegraphics[width=\columnwidth]{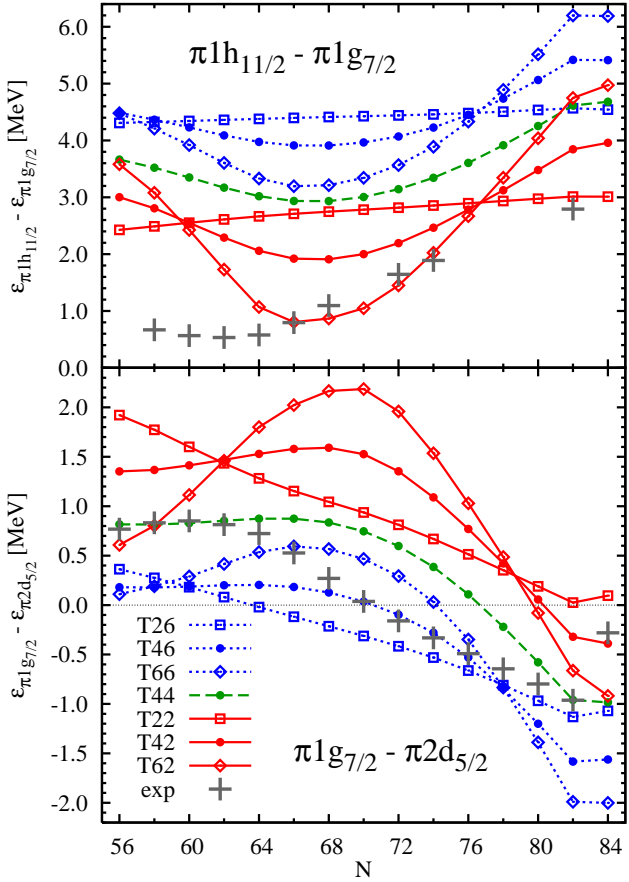}
  \caption{
    (Color online)
    Distance of the proton $1h_{11/2}$ and $1g_{7/2}$ levels (top) and of the
    proton $2d_{5/2}$ and $1g_{7/2}$ levels (bottom), for the chain of
    tin isotopes. The ``best'' parameterization cannot and should not be
    determined with a $\chi^2$ criterion, see text.
  }
  \label{fig:spli-Sn}
\end{figure}

In the preceding sections, we have analyzed characteristics of the
single-particle spectra for isolated doubly-magic nuclei. We found that larger
tensor terms do not lead to an overall improvement of the single-particle
spectra. However, we also argued that it might be essentially due to
deficiencies of the central (and possibly spin-orbit) interactions and that it
should not be used to discard the tensor terms as such. In any case, the results
gathered so far on single-particle spectra of doubly-magic nuclei do not permit to
narrow down a region of meaningful coupling constants of the tensor terms. The
analysis must be complemented by looking at other observables. A better suited
observable is provided by the evolution of spin-orbit splittings along an
isotopic or isotonic chain, which ideally reflects the nucleon-number-dependent
contribution from the $\vec{J}^2$ terms to the spin-orbit potentials.
Unfortunately, safe experimental data for the evolution of spin-orbit partners
are scarce; hence, one has to content oneself to the evolution of the energy
distance of levels with different $\ell$, assuming that the effect is primarily
caused by the evolution of the spin-orbit splittings of each level with its
respective partner. A popular playground for such studies is the chain of Sn
isotopes, where two such pairs of levels have gained attention; the
$\pi$~$2d_{5/2}$ and $\pi$~$1g_{7/2}$ on the one hand, and the $\pi$~$1g_{7/2}$
and $\pi$~$1h_{11/2}$ on the other hand. Figure~\ref{fig:spli-Sn} shows these
two sets of results for a selection of our parameterizations.

Experimentally, the $2d_{5/2}$ and $1g_{7/2}$ levels cross between $N=70$ and
72, such that the $2d_{5/2}$ provides the ground state of light odd-$A$ Sb
isotopes, and $1g_{7/2}$ that of the heavy ones, see for example
Ref.~\cite{She05aE}. The crossing as such is predicted by many mean-field
interactions and most of the parameterizations of the Skyrme interaction we use
here. It has also been studied in detail with the standard Gogny force (without
any tensor term) using elaborate blocking calculations of the odd-$A$
nuclei~\cite{Por05a}. The crossing, however, is never predicted at the right
neutron number, see Fig.~\ref{fig:spli-Sn}. As we have learned above, we should
not assume that the absolute distance of the two levels will be correctly
described by any of our parameterizations (as the centroids of the $\ell$ shells
will not have the proper distance and the spin-orbit splittings have a wrong
$\ell$ dependence within a given shell). Hence, the neutron number where the
crossing takes place cannot and should not be used as a quality criterion. What
does characterize the tensor terms is the bend of the curves in
Fig.~\ref{fig:spli-Sn}, as ideally it reflects how the spin-orbit splittings of
both levels change in the presence of the tensor terms. Similar caution has to
be exercised in the analysis of the unusual relative evolution of the proton
$1g_{7/2}$ and $1h_{11/2}$ levels that was brought to attention by Schieffer
\etal~\cite{Schi04aE}. Their spacing has been investigated in terms of the
tensor force before~\cite{Ots05a,Ots06a,Bro06a,Col07a}. Again, we pay attention
to the qualitative nature of the bend without focusing too much on the precise
value by which the splitting changes when going from $N\approx58$ to $N=82$.
Indeed, the matching of the lowest proton fragment with quantum number
$1h_{11/2}$ seen experimentally with the corresponding empirical single-particle
energy is unsafe because of the fractionization of the strength as discussed in
Ref.~\cite{Bro06a}.

For both pairs of levels, the evolution of their distance can be attributed to
the tensor coupling between the proton levels and neutrons filling the
$1h_{11/2}$ level below the $N=82$ gap. Unfortunately, this introduces an
additional source of uncertainty: as can be seen in Fig.~\ref{fig:spe-Sn132},
the ordering of the neutron levels in \nuc{132}{Sn} is not properly reproduced
by any of our parameterizations, with the $1h_{11/2}$ level being predicted
above the $2d_{3/2}$ level, while it is the other way round in experiment. This
means that in the calculations, the contribution from the $1h_{11/2}$ level to
the neutron spin-orbit current builds up at larger $N$ than what can be expected
in experiment. As a consequence, the prediction for the relative evolution of
the levels might be shifted by up to four mass units to the right compared to
experiment for both pairs of levels we examine here.

\begin{figure}[t!]
  \includegraphics[width=0.8\columnwidth]{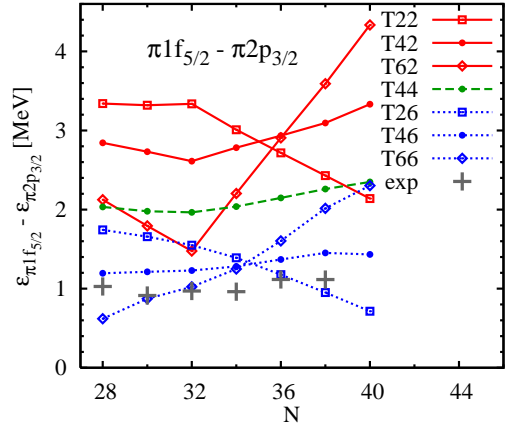}
  \caption{
    (Color online)
    Distance of the proton $1f_{5/2}$ and $2p_{3/2}$
    in the chain of Ni isotopes.
  }
  \label{fig:spli-Ni}
\end{figure}

In the end, the trend of both
splittings is best reproduced when using a positive value of the neutron-proton
$\vec{J}_\nrm \cdot \vec{J}_\prm$ coupling constant $\beta$ such that the
filling of the neutron $1h_{11/2}$ shell decreases the spin-orbit splittings of
the proton shells. The parameterizations from the T4$J$ and T6$J$ series
indeed do reproduce the bend of empirical data, with, however, a clear shift in
the neutron number where it occurs, as expected from the previous
discussion. A value of $\beta = 120$ MeV\,fm$^5$, which corresponds to the
series of T4$J$ parameterizations, matches its magnitude best (see for example
T44).

A similar analysis can be performed for the proton $1f_{5/2}$ and $2p_{3/2}$
levels in the chain of Ni isotopes, see Fig.~\ref{fig:spli-Ni}. This case is
interesting as no distinctive feature can be observed in the empirical spectra,
yet the standard parameterizations without tensor terms like T22 do not
reproduce them. In fact, to keep the $1f_{5/2}$ and $2p_{3/2}$ at a constant
distance, two competing effects have to cancel. First, the increasing
diffuseness of the neutron density with increasing neutron number diminishes the
proton spin-orbit splittings through its reduced gradient in the expression for
the proton spin-orbit potential when going from $N=32$ to $N=40$. Second, the
filling of the neutron $1f_{5/2}$ state reduces the neutron spin-orbit current
which in turn increases the proton spin-orbit splittings for interactions with
sizable proton-neutron tensor contribution to the proton spin-orbit potential
when going from $N=32$ to $N=40$. The former effect can be clearly seen for
parameterizations T2$J$ with vanishing proton-neutron tensor term, $\beta=0$.
Again, parameterizations of the T4$J$ series seem to be the most appropriate to
describe the evolution of these levels.

The evolution of single-particle levels is the tool of choice to determine the
sign and magnitude of the proton-neutron tensor coupling constant. The value
which we favor, as a result of our semi-qualitative analysis is $\beta=120$
MeV\,fm$^5$. This value is only slightly larger than the value of 94
to 96~MeV\,fm$^5$ advocated by Brown \etal\ in Ref.~\cite{Bro06a}, which was
adjusted to \emph{theoretical} level shifts in the chain of tin isotopes
obtained from a $G$-matrix interaction. We can consider this as a
reasonable agreement.

Let us defer the discussion of this value to the end of this section and study
in the next paragraph the like-particle tensor-term coupling constant $\alpha$.


\subsubsection{Evolution along isotopic chains: $nn$ coupling}
\label{sect:spectro-nn}


\begin{figure}[t!]
  \includegraphics[width=0.8\columnwidth]{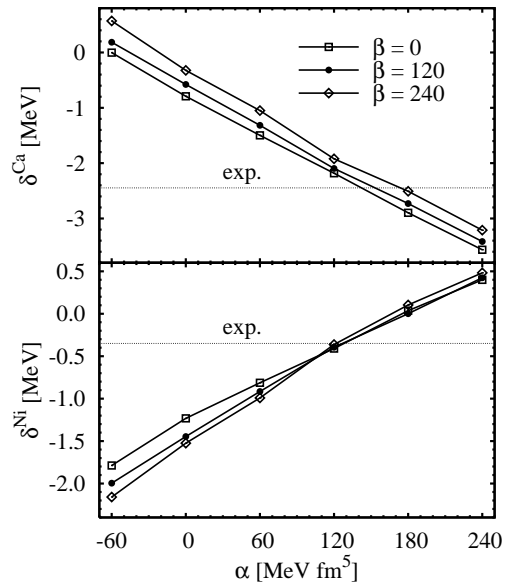}
  \caption{
    Shift of the distance between the neutron $1d_{3/2}$
    and $2s_{1/2}$ levels when going from  \nuc{40}{Ca} to \nuc{48}{Ca},
    Eq.~(\ref{eq:nnshift-Ca}) (top) and of the neutron $1f_{5/2}$ and
    $2p_{1/2}$ levels when going from \nuc{56}{Ni} and \nuc{68}{Ni},
    Eq.~(\ref{eq:nnshift-Ni}) (bottom).
  }
  \label{fig:spe-shift-Ca+Ni}
\end{figure}

In order to narrow down an empirical value for the neutron-neutron tensor
coupling constant, the ideal observable
would be the evolution of neutron single-particle levels along an isotopic
chain. Unfortunately, these are only
accessible at the respective shell closures. We shall therefore compare neutron
single-particle spectra of pairs
of doubly-magic nuclei belonging to the same isotopic chain. Again, the
necessity to extract pure single-particle
effects calls for precautions. We choose pairs of particle or hole levels which
are close enough in energy
that their absolute spacing is not much affected by particle-vibration coupling.
Of course, one also has to be
careful if both states appear at relatively high excitation energy in the
neighboring odd isotope because the
fractionization of their strength could again interfere with the analysis. In
the following, we choose pairs of
orbitals which are as safe as possible.

To remove the uncertainties from the deficiencies of the central and spin-orbit
parts of the effective interaction
that we have identified above, we will look at a double difference, where,
first, we construct the energy
difference between the neutron $1d_{3/2}$ and $2s_{1/2}$ levels separately for
\nuc{40}{Ca} and \nuc{48}{Ca}, and
then compare the value of this difference in both nuclei
\begin{equation}
\label{eq:nnshift-Ca}
\delta^\text{Ca}
= \left(   \varepsilon_{1d_{3/2}}^{\text{\nuc{48}{Ca}}}
         - \varepsilon_{2s_{1/2}}^{\text{\nuc{48}{Ca}}}
  \right)
  - \left(   \varepsilon_{1d_{3/2}}^{\text{\nuc{40}{Ca}}}
           - \varepsilon_{2s_{1/2}}^{\text{\nuc{40}{Ca}}}
    \right)
.
\end{equation}
Assuming that the problems from the central and spin-orbit forces
discussed in Sects.~\ref{sect:splittings} and~\ref{sect:doublymagic}
have the same effect in both nuclei, they will cancel out
in $\delta^\text{Ca}$.

The interesting feature of this pair of states is that they are separated by
more than 2~MeV in \nuc{40}{Ca}, while they are nearly degenerate in
\nuc{48}{Ca}, see Figs.~\ref{fig:spe-Ca40} and~\ref{fig:spe-Ca48}. Such a shift
can only be reproduced with a positive (140-180~MeV\,fm$^5$) value of $\alpha$,
which decreases the splitting of the neutron $1d$ shell when the neutron
$1f_{7/2}$ level is filled.

A similar analysis can be performed for the $1f_{5/2}$ and $2p_{1/2}$
neutron states in the Ni isotopes \nuc{56}{Ni} and \nuc{68}{Ni}
\begin{equation}
\label{eq:nnshift-Ni}
\delta^\text{Ni}
= \left(   \varepsilon_{1f_{5/2}}^{\text{\nuc{68}{Ni}}}
         - \varepsilon_{2p_{1/2}}^{\text{\nuc{68}{Ni}}}
  \right)
  - \left(   \varepsilon_{1f_{5/2}}^{\text{\nuc{56}{Ni}}}
           - \varepsilon_{2p_{1/2}}^{\text{\nuc{56}{Ni}}}
    \right)
.
\end{equation}
Going from \nuc{56}{Ni} to \nuc{68}{Ni}, the neutron $1f_{5/2}$ level comes
further down in energy than the $2p_{1/2}$ level for parameterizations without
tensor terms (T22), see Figs.~\ref{fig:spe-Ni56} and~\ref{fig:spe-Ni68}.
The reason for this trend is the geometrical growth of the nucleus, which on
the one hand lowers the centroid of the $1f$ levels in the widening potential
well, and on the other hand pushes the spin-orbit field to larger radii, which
has opposite effects on the splittings of $2p$ and $1f$ states.
The like-particle tensor terms can compensate this trend through a reduction
of the spin-orbit splitting of the $1f$ levels. The observed downward shift
by $0.3$~MeV can be recovered with a value of $\alpha$ around 120 MeV\,fm$^5$,
see Fig.~\ref{fig:spe-shift-Ca+Ni}.

It is also gratifying to see that the analysis of Ca and Ni isotopes suggests
nearly the same value for the like-particle tensor term coupling constant
$\alpha$.


\subsection{Binding energies}


\begin{figure*}[t!]
  \includegraphics[width=\textwidth]{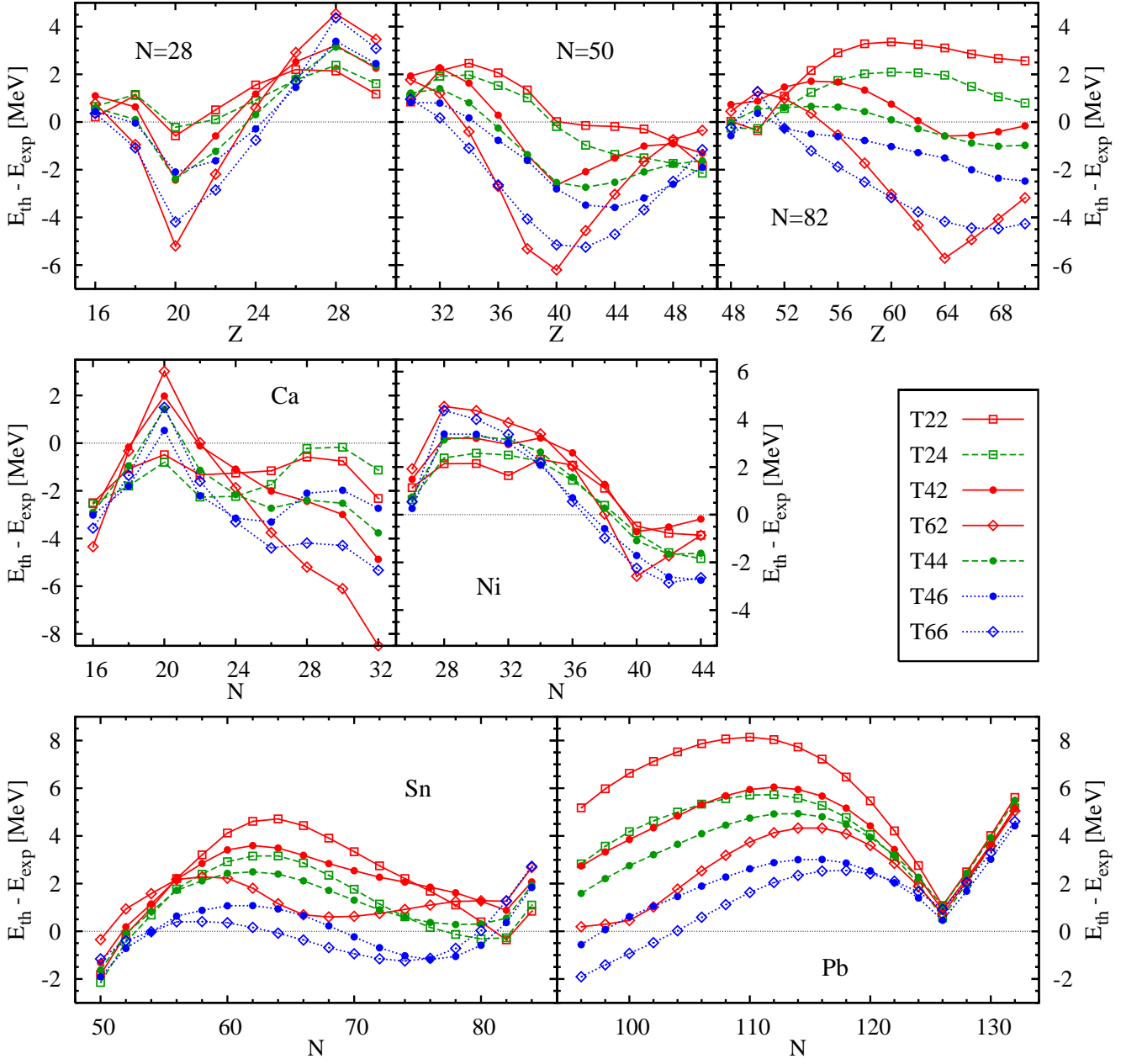}
  \caption{
    (Color online)
    Mass residuals $E_{\text{th}}-E_{\text{exp}}$ along selected
    isotopic and isotonic chains of semi-magic nuclei for the
    parameterizations as indicated. Positive values of
    $E_{\text{th}}-E_{\text{exp}}$ denote underbound nuclei,
    negative values overbound nuclei.
  }
  \label{fig:deltae}
\end{figure*}

Our ultimate goal, although far beyond the scope of the present paper, is the
construction of a universal nuclear energy density functional that
simultaneously describes  bulk properties like masses and radii, giant
resonances, and low-energy spectroscopy, such as quasiparticle configurations
and collective rotational and vibrational states. To crosscheck how our findings
on single-particle spectra and spin-orbit splittings translate into bulk
properties, we will now analyze the evolution of mass residuals and charge radii
along isotopic and isotonic chains. It has been repeatedly noted in the
literature that the mass residuals from mean-field calculations show
characteristic
arches~\cite{Dob84a,Fri86a,Cha98a,Pat99a,RMP,Lun03a,Dob04a,BBH06a}, where heavy
mid-shell nuclei are usually underbound compared to the doubly magic ones that
are located at the bottom of deep ravines. For light nuclei, the patterns are
often less obvious. Part of this effect can be explained and removed taking
large-amplitude correlations from collective shape degrees of freedom into
account through suitable beyond-mean-field methods. In turn, this means that the
mass residuals should leave room for the extra binding of mid-shell nuclei from
correlations. However, it turns out that for typical effective interactions the
amplitude of the arches is larger than what is brought by
correlations~\cite{BBH06a}. Furthermore, this effect seems not to be of the same
size for isotopic and isotonic chains, which altogether hints at deficiencies of
the current effective interactions.

Recently, Dobaczewski pointed out~\cite{Dobaczewskitalk} that the strongly
fluctuating contribution brought by the $\vec{J}^2$ terms to the total binding
energy could remove at least some of the ravines found in the mass residuals
around magic numbers. The hypothesis was motivated by calculations that evaluate
the tensor terms either perturbatively, or self-consistently, using in this
case an existing standard parameterization without tensor terms for the rest of
the energy functional. Our set of refitted parameterizations with varied
coupling constants of the tensor terms gives us a tool to check how much of the
argument persists to a full fit.


\subsubsection{Semi-magic series}
\label{sect:masses:chains}


Figure~\ref{fig:deltae} displays binding energy residuals along various
isotopic and isotonic chains of semi-magic nuclei for a selection of our
parameterizations: T22 is the reference with vanishing $\vec{J}^2$ terms at
sphericity; T24 has a substantial like-particle coupling constant $\alpha$ and
vanishing proton-neutron coupling constant $\beta$, which is similar to most of
the published  parameterizations which take the $\vec{J}^2$ terms from the
central Skyrme force into account; T42 and T62 are parameterizations with
substantial proton-neutron coupling constant $\beta$ and vanishing
like-particle coupling constant; T44 has a mixture of like-particle and
proton-neutron tensor terms that is close to what we found preferable for the
evolution of spin-orbit splittings above; and T46 is a parameterization that
gives the best root-mean-square residual of binding energies for spherical
nuclei, as we will see below. Finally, T66 is a parameterization with large
and equal proton-neutron and like-particle tensor-term coupling constants.

The tensor terms have opposite effects in light and heavy nuclei: The curves
obtained with T22, the parameterization without $\vec{J}^2$ term contribution
at sphericity, are relatively flat for the light isotopic and isotonic chains,
but show very pronounced arches with an amplitude of 5 or even more MeV for the
heavy Sn and Pb isotopic chains. By contrast, the most striking effect of the
$\vec{J}^2$ terms is that they induce large fluctuations of the mass residuals
in light nuclei, while they flatten the curves in the heavy ones.

\begin{figure}[t!]
  \includegraphics[width=0.8\columnwidth]{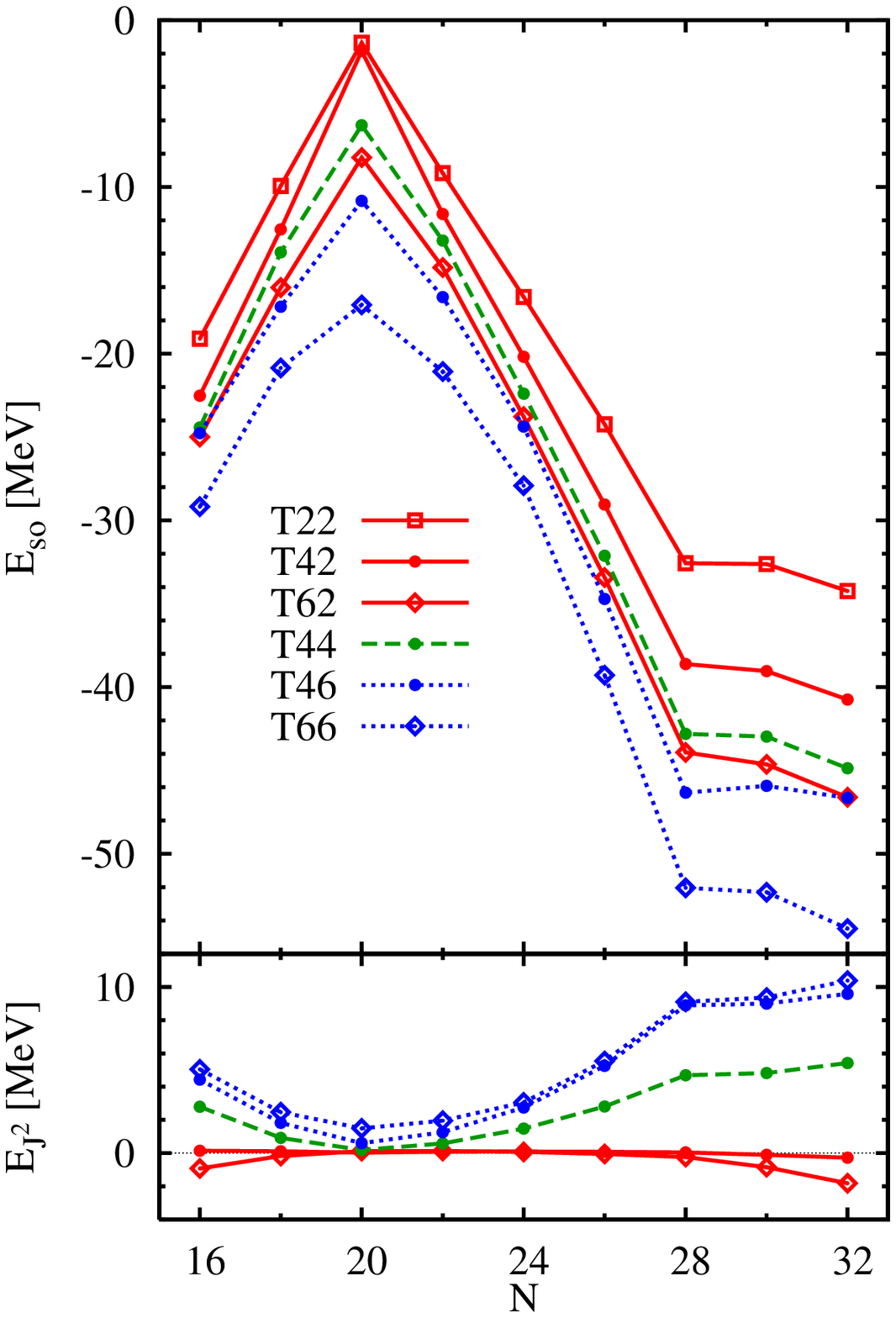}
  \caption{
    (Color online)
    Evolution of spin-orbit current ($J_t^2$) energy (bottom panel, zero by
    construction for T22) and spin-orbit energy (top panel) with neutron number
    $N$ in the chain of Ca isotopes ($Z=20$).
  }
  \label{fig:Ca-J2so}
\end{figure}

\begin{figure}[t!]
  \includegraphics[width=0.8\columnwidth]{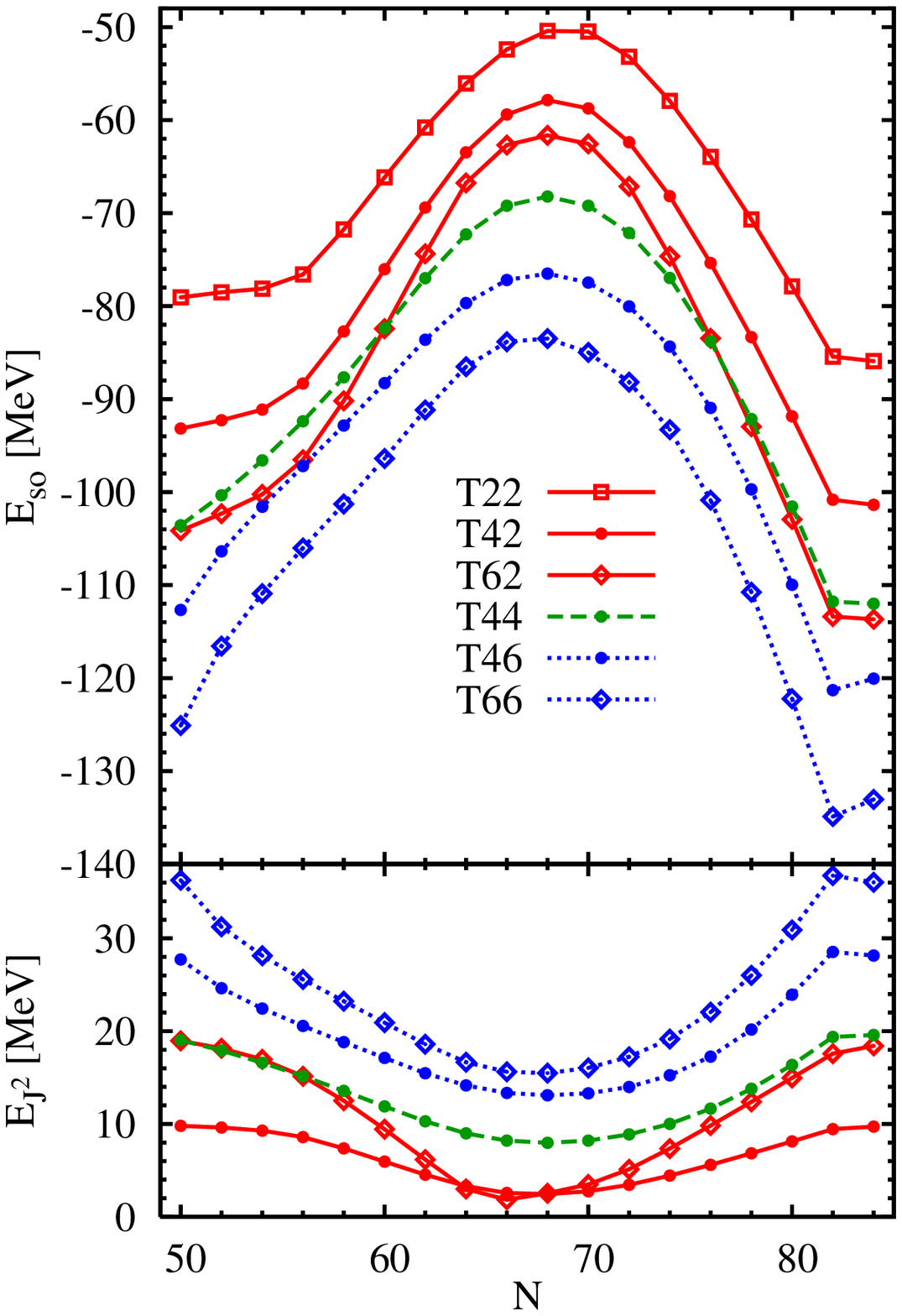}
  \caption{
    (Color online)
    Same as Fig.~\ref{fig:Ca-J2so} for tin isotopes ($Z=50$).
  }
  \label{fig:Sn-J2so}
\end{figure}

The strong variation between the parameter sets for light nuclei are of course
the direct consequence of the
strong variation of the spin-orbit current $\vec{J}$ that enters the spin-orbit
and tensor terms when going back
and forth between nuclei where the configuration of at least one nucleon species
is spin-saturated.
The variations seen are a result of the
modifications of tensor-term coupling constants and the associated readjustment
of the spin-orbit strength $W_0$.
For example, \nuc{48}{Ca} is overbound with respect to \nuc{40}{Ca} and
\nuc{56}{Ni} for parameterizations with a
proton-neutron coupling constant $\beta > 0$, while the like-particle coupling
constant $\alpha$ has a more
limited effect. Since only the neutron core is spin-unsaturated in this nucleus,
this must be attributed to the
increase in the readjusted spin-orbit strength $W_0$ (correlated with $C^J_0 =
\tfrac{1}{2}(\alpha + \beta)$)
which dominates when $\beta$ is increased and $\alpha$ kept at zero, and
counterbalances the effect of $\alpha$
when the latter varies. See the parameter sets T62 and T66 in
Figures~\ref{fig:deltae} and \ref{fig:Ca-J2so}.
The large overbinding of nuclei around \nuc{90}{Zr} ($Z=40$, $N=50$) for
parameterizations with large
proton-neutron tensor coupling constant has the same origin. For a given
parameterization and a given nucleus, the
energy gain from the spin-orbit term seems to be almost always larger than the
energy loss from the $\vec{J}^2$
one, see Fig.~\ref{fig:Ca-J2so} for Ca isotopes and Fig.~\ref{fig:Sn-J2so} for
Sn isotopes. Of course other terms
in the energy functional compensate for a part of the gain from the spin-orbit
term, but the overall trends of the
mass residuals suggest that the spin-orbit energy has a much larger contribution
to the differences between the
parameterizations visible in Fig.~\ref{fig:deltae} than the $\vec{J}^2$ terms.

We have to note that the spin-orbit current does not completely vanish for the
nominally proton and neutron spin-saturated \nuc{40}{Ca} for parameterizations
with large coupling constants of the $\vec{J}^2$ terms.
For those, the gap at 20
is strongly (and nonphysically) reduced, see Fig. \ref{fig:spe-Ca40}. The small
gap at 20 does not suppress pairing correlations anymore in our HFB approach.
The resulting scattering of particles from the $sd$ shell to the $fp$ shell
breaks the spin-saturation, such that there is a finite, in some cases quite
sizable, contribution from the spin-orbit term to the total binding energy.
Owing to the compensation between all contributions, the total energy gain
compared to a HF calculation without pairing is usually small and rests on the
order of 200 keV for the parameterizations shown in Fig. \ref{fig:deltae}.

It is also important to note that some of the light chains
in Fig.~\ref{fig:deltae} are sufficiently close to or even cross the $N=Z$
line that they are subject to the Wigner energy, which still lacks a
satisfying explanation, not to mention a description in the framework of
mean-field methods~\cite{Sat97a}. The Wigner energy is not taken into account
in our fits, while it turned out to be a crucial ingredient of any
HFB~\cite{Ton00a,Sam02a,Gor03b} or other mass formula.
In fact, as shown in Fig.~14 of Ref.~\cite{BBH06a}, the missing Wigner energy
clearly sticks out from the mass residuals for SLy4 (which is very similar to
T22) when they are plotted for isobaric chains. This local trend around $N=Z$
is, however, overlaced with a global trend with mass number, such that the
missing Wigner energy cannot be spotted anymore when looking at the mass
residuals for the isotopic chain of Ca isotopes, similar to what is seen for
T22
in Fig.~\ref{fig:deltae}. Within our fit protocol, the correlation between the
masses of \nuc{40}{Ca}, \nuc{48}{Ca} and \nuc{56}{Ni}, that is brought by the
spin-orbit force (see Sect.~\ref{sect:J2-so-rearrangement}) does not
tolerate a correction for the Wigner energy for standard central and spin-orbit
Skyrme forces, as this will lead to an unacceptable underbinding of
\nuc{48}{Ca}. This, however, might change when the $\vec{J}^2$ terms are added.
Indeed, Fig.~\ref{fig:deltae} suggests that adding a phenomenological Wigner
term around \nuc{40}{Ca} and \nuc{56}{Ni} to a parameter set like T44,
which is consistent with the evolution of single-particle levels, would flatten
the curves for the mass residuals in the Ca, Ni and $N=28$ chains. The mass
residuals for the chain of oxygen isotopes that are not shown here would be
improved in a similar manner. However, extreme caution should be exercised
before jumping to premature conclusions, as the spin-orbit splittings and level
distances in light nuclei are far from realistic for all our parameterizations;
as a consequence it is difficult to judge if the room we find for the Wigner
energy is fortuitous or indeed a feature of well-tuned $\vec{J}^2$ terms. Note
that the HFB mass formulas that do include a correction for the Wigner energy
side-by-side with the $\vec{J}^2$ terms from the central Skyrme force give
satisfying mass residuals for light nuclei~\cite{Ton00a,Sam02a,Gor03b}, but
have nuclear matter properties that are quite different from ours;
{\em cf.}~BSk1 and BSk6 with SLy4 in Table I of Ref.~\cite{Rei06a}. Our
constraints on the empirical nuclear matter properties (same as
those on SLy4) that are absent in these HFB mass fits might be the deeper
reason for this conflict.

Large tensor-term coupling constants straighten the arches in the mass
residuals in the heavy Sn and Pb isotopic chains, but the improvements are not
completely satisfactory. Large, combined proton-neutron and like-particle
coupling constants tend to transform the arch for the tin isotopic chain into a
an s-shaped curve, which is not very realistic from the standpoint of
expected corrections through collective effects. It can again be assumed that
the deficiencies of the single-particle spectra pointed out in
Fig.~\ref{fig:spe-Sn132} are responsible, where the $\nu$ $1h_{11/2}$ and $\pi$
$1g_{9/2}$ are placed too high above the rest of the single-particle spectra in
heavy Sn isotopes. For Pb isotopes, large values of the tensor terms tend to
overbind the neutron-deficient isotopes. It is noteworthy that the tensor terms
seem to not much affect the mass residuals of the heavy Pb isotopes above
$N=126$, which are on the flank of a very deep ravine that becomes visible
when going towards heavier elements, {\em cf.}~the SLy4 results
in Ref.~\cite{BBH06a}.

\begin{figure}[t!]
  \includegraphics[width=\columnwidth]{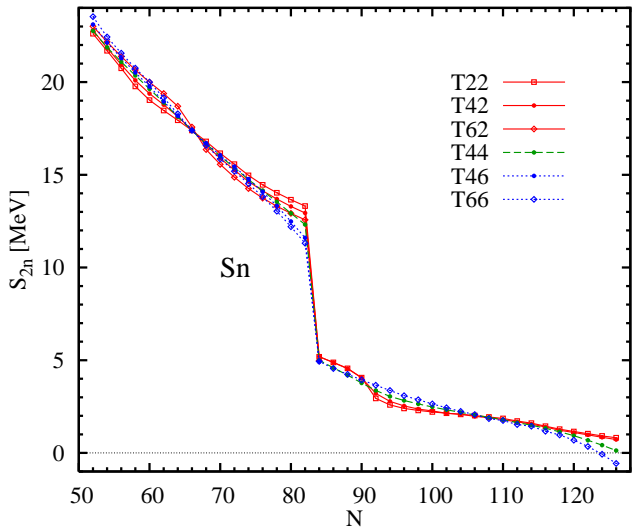}
  \caption{
    (Color online)
    Two-neutron separation energy along the chain of isotopes ($Z=50$).
  }
  \label{fig:Sn-S2n}
\end{figure}

It has been often noted that effective interactions that give a similar
satisfying description of masses close to the valley of stability give diverging
predictions when extrapolated to exotic nuclei. The standard example is the
two-neutron separation energy $S_{2n} (N,Z) = E(N,Z-2) -E(N,Z)$ for the chain of
Sn isotopes. Results obtained with a subset of our parameterizations are shown
in Fig. \ref{fig:Sn-S2n}. It is noteworthy that the differences for neutron-rich
nuclei beyond $N=82$ are not larger than those for the isotopes closer to
stability. Around the valley of stability, increasing the coupling constants of
tensor terms, in particular the like-particle ones, tilts the curve, pushing it
up for light isotopes and pulling it down it for heavy ones, which reflects of
course the position of the $\nu$ $1h_{11/2}$ level that is pushed into the
$N=82$ gap, see Fig. \ref{fig:spe-Sn132}. For the neutron-rich isotopes, small
differences appear around $N=90$, which reflects the change of level structure
above the $\nu$ $2f_{7/2}$ level and at the drip line, but they are much
smaller than the differences seen between parameterizations obtained with
different fit protocols, see Fig.~5 of Ref.~\cite{RMP}.


\subsubsection{Systematics}


In the preceding section we showed how the $\vec{J}^2$ terms in the energy
functional modify the trends of mass residuals along isotopic and isotonic
chains, in particular the amplitude of the arches between doubly-magic nuclei.
In this section, we want to examine how this translates into quality criteria
for the overall performance of the parameterizations for masses.

\begin{figure}[t!]
  \includegraphics[width=0.8\columnwidth]{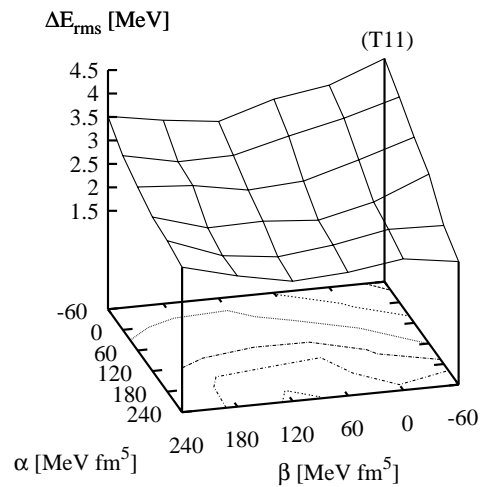}
  \caption{
    Root-mean-square deviation from experiment of the binding
    energies of a set of 134 spherical nuclei, for each of the forces
    T$IJ$, \emph{vs.} $\alpha$ and $\beta$ (The ``(T11)'' label indicates the
    position of this parameterization in the ($\alpha$, $\beta$)-plane).
    Contour
    lines at $\Delta E_\mathrm{rms} = 2.0, 2.25, 2.5, 3.0, 3.5, 4.0$~MeV.
    The minimal value is found for T46 ($\Delta E_\mathrm{rms} = 1.96$~MeV).
  }
  \label{fig:derms}
\end{figure}

Figure \ref{fig:derms} displays the root-mean-square deviation of the mass
residuals for all our 36 parameterizations, evaluated for a set of 134 nuclei
predicted to have spherical mean-field ground states when calculated with the
parameterizations SLy4~\cite{BBH06a}. One observes a clear minimum around T46,
\ie~$(\alpha, \beta) = (240, 120)$, with
$(E_\text{th}-E_\text{exp})_\text{r.m.s.}=1.96$ MeV, compared with $3.44$~MeV
for T22 ($\alpha=\beta=0$). We found even slightly better values with even more
repulsive isoscalar and isovector coupling constants, but the single-particle
spectra of these interactions turn out to be quite unrealistic,
{\em cf.}~Sect.~\ref{sect:splittings}. This already demonstrates that in the
presence of the $\vec{J}^2$ terms a good fit of masses does not necessarily
lead to satisfactory single-particle spectra.

\begin{figure}[t!]
  \includegraphics[width=0.8\columnwidth]{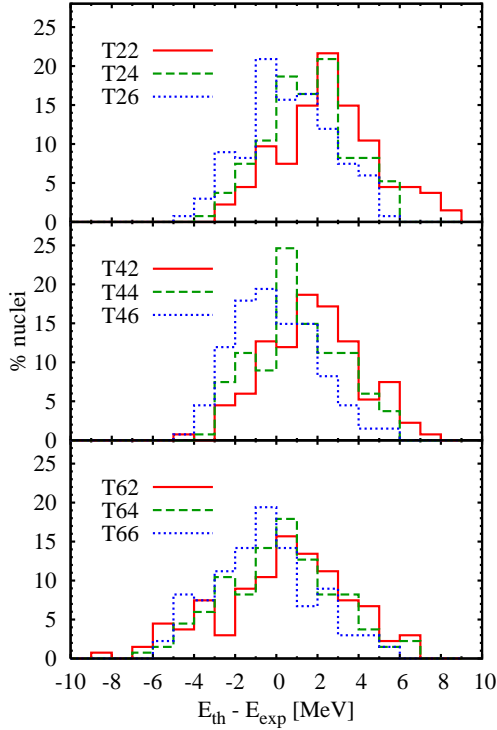}
  \caption{
    (Color online)
    Distribution of deviations from experiment of the binding
    energies of a set of 134 spherical nuclei (1~MeV bins) for a subset
    of parameterizations. Each panel corresponds to a given value of $\beta$
    (from top to bottom: $\beta = 0$, 120, 240~MeV\,fm$^5$).
  }
  \label{fig:dedist}
\end{figure}

Figure~\ref{fig:dedist} demonstrates how the distribution of the mass
residuals $E_\text{th}-E_\text{exp}$ affects the evolution of their r.m.s.
value for a subset of 9 parameterizations. For T22 ($\alpha = \beta = 0$), the
distribution is centered at positive mass residuals, with only very few nuclei
being overbound. Increasing $\beta$ to 120 MeV\,fm$^5$ (T42) or even
240 MeV\,fm$^5$ (T62) shifts the median of the distribution to smaller values,
which yields more and more overbound nuclei. For large values of $\beta$,
the distribution spreads out more, which diminishes the improvement from
centering the distribution closer to zero. For given $\beta$, increasing
$\alpha$ mainly shifts the median of the distribution without spreading out
its overall shape, which is preferable to optimize the r.m.s.~value.

These considerations, however, have to be taken with caution. As said above, we
aim at a model where certain correlations beyond the mean-field are treated
explicitly, which asks for a distribution of \emph{mean-field} mass residuals
with an asymmetric distribution towards positive mass residuals, and a width
that is similar to the difference between the maximum and minimum correlation
energies to be found.

\subsection{Radii}


The evolution of nuclear charge radii along isotopic chains reflects how
the mean field of the protons changes when neutrons are added in the system.
In the simplistic liquid-drop model, it just follows the geometrical growth
of the nucleus $\sim A^{1/3}$, but data show that there are many local
deviations from this global trend. On the one hand, radii are of course
subject to correlations beyond the mean
field~\cite{Rei79a,Gir82a,Bon91a,Hee93a,BBH06a}
On the other hand, they are also sensitive to the detailed shell structure,
which, in turn, might be influenced by tensor terms. We will concentrate
here on two anomalies of the evolution of charge radii, both of which are not
much influenced by collective correlations beyond the mean-field (at least in
calculations with the Skyrme interaction SLy4)~\cite{BBH06a}: that the
root-mean-square (r.m.s.) charge radius of \nuc{48}{Ca} is almost the same
as the one of the lighter \nuc{40}{Ca} or possibly slightly smaller,
and the kink in the isotopic shifts of
mean-square (m.s.) charge radii in the Pb isotopes, where Pb isotopes above
\nuc{208}{Pb} are larger than what could be expected from liquid-drop
systematics. In both cases it is plausible that shell effects are the
determining factor, although alternative explanations that involve pairing
effects have been put forward for the latter case as well~\cite{Taj93a,Fay00a}.

\begin{figure}[t!]
    \includegraphics[width=\columnwidth]{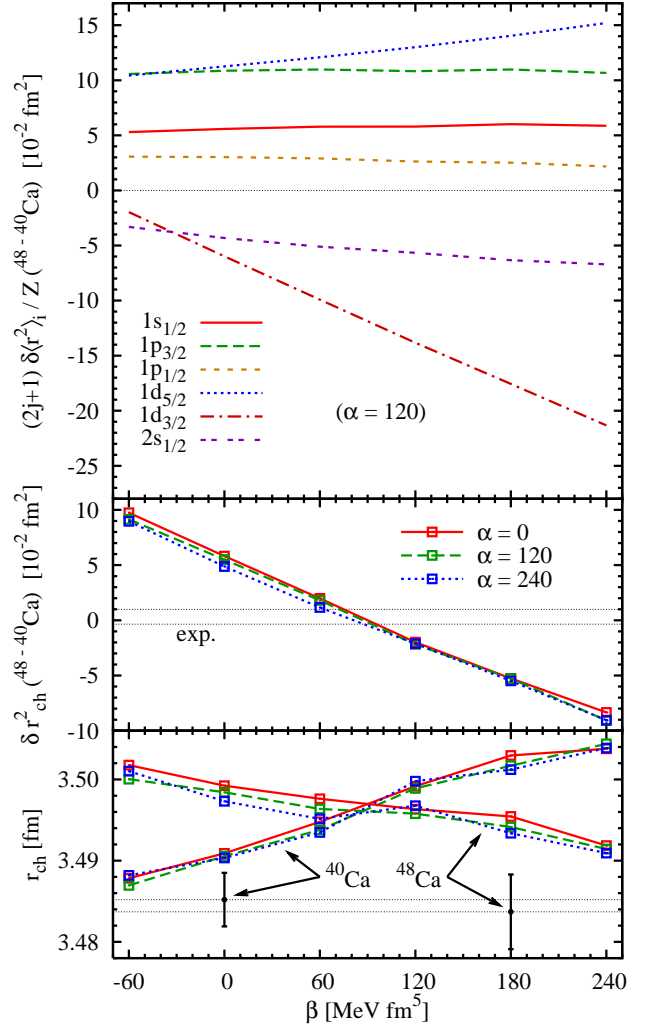}
    \caption{
    (Color online)
         Middle panel: Difference of mean-square charge radii between
         \nuc{40}{Ca} and \nuc{48}{Ca} as a function of the proton-neutron
         tensor term coupling constant $\beta$ for three values of $\alpha$.
         The experimental value (with error bar) is represented by the two
         horizontal black lines.
         Bottom panel: Root-mean-square charge radii of \nuc{40}{Ca} and
         \nuc{48}{Ca}.
         Top panel: Contribution of the single-particle proton states to the
         difference of the charge radii (mean square radius of the point
         proton distribution, see Eq.~(\ref{eq:rch})).
        }
    \label{fig:rad1}
\end{figure}

Charge radii have been calculated with the approximation used in
Ref.~\cite{Cha97a}\footnote{There is a typographical error in Eq.~(4.2) in
Ref.~\cite{Cha97a}, that was copied to Eq.~(110) in Ref.~\cite{RMP}: the
$\hbar/mc$ factor should be squared, as is trivially found by dimensional
analysis and confirmed by Ref.~\cite{Ber72a}.} and derived from
Ref.~\cite{Ber72a}
\begin{equation}
    r^2_\mathrm{ch}=\langle r^2\rangle_\prm + r^2_\prm
        + \frac{N}{Z}r^2_\nrm
        +\frac{1}{Z} \left(\frac{\hbar}{mc}\right)^2
            \sum_i v^2_i\mu_{q_i}
        \langle\boldsymbol\sigma\cdot\boldsymbol\ell\rangle_i\,,
\label{eq:rch}
\end{equation}
where the mean-square (m.s.) radius of the point-proton distribution
$\langle r^2\rangle_\prm$ is corrected by three terms: the first two estimate
the effects of the intrinsic charge distribution of the free proton and neutron
(with m.s.~radii $r^2_\prm$ and $r^2_\nrm$) and the third adds a correction from
the magnetic moments of the nucleons. Since we will consider the shift of charge
radii for different isotopes of the same series, the actual value of $r^2_\prm$
cancels out. For the second correction term, which is independent from the
interaction, we take $r^2_\nrm=-0.117$~fm$^2$~\cite{RMP}. Finally, the magnetic
correction can only depend weakly on the details of the interaction through the
occupation factors $v^2_i$ when non-magic nuclei are considered. The same
expressions had been used during the fit of our parameterizations.

We begin with the Ca isotopes. Most parameterizations of Skyrme's interaction
are not able to reproduce that the charge radius of \nuc{48}{Ca} has about the
same size as that of \nuc{40}{Ca}, see Fig.~11 in Ref.~\cite{RMP}. The middle
panel of Fig.~\ref{fig:rad1} shows the difference of the the m.s.~radii of
\nuc{48}{Ca} and \nuc{40}{Ca} in dependence of the tensor term coupling
constants $\alpha$ and $\beta$. First, this difference is almost independent
of $\alpha$, the strength of the like-particle tensor terms. Second, it is
strongly correlated with $\beta$, the strength of the proton-neutron tensor
term, with large positive values of $\beta$ bringing the difference of radii
into the domain of experimentally acceptable values \cite{Otten} or even below,
with a best match obtained for $\beta = 80$ MeV\,fm$^5$.
This effect can be explained by looking at the proton single-particle spectra
of \nuc{40}{Ca} (Fig.~\ref{fig:spe-Ca40}) and \nuc{48}{Ca}
(Fig.~\ref{fig:spe-Ca48}). Indeed, one observes that a positive neutron-proton
tensor coupling constant decreases the strength of the proton spin-orbit field
in \nuc{48}{Ca}, which in turn lowers the $\pi$ $1d_{3/2}$ level in
\nuc{48}{Ca} (compare the parameterizations T$IJ$ in Fig.~\ref{fig:spe-Ca48}
with increasing $I$ for given $J$). As a consequence, the  m.s.~radius of this
state decreases as it sinks deeper into the potential well of \nuc{48}{Ca}.
At the same time, this level is pushed up in \nuc{40}{Ca}, which slightly
increases the contribution of this state to the charge m.s.~radius of this
nucleus. This effect is demonstrated in the top panel of Fig.~\ref{fig:rad1},
which displays the degeneracy-weighted and normalized change of the
m.s.~radii of proton hole states between \nuc{40}{Ca} and \nuc{48}{Ca} as a
function of the proton-neutron tensor term coupling constant $\beta$ for forces
with a like-particle tensor term coupling constant $\alpha = 120$~MeV\,fm$^5$.
Indeed, the decreasing contribution from the $\pi 1d_{3/2}$ state to the m.s.
radius significantly decreases the isotopic shift between both Ca isotopes.
It has to be noted that the m.s. value of the charge radii of \nuc{40}{Ca}
and \nuc{48}{Ca} are almost independent of alpha and that their absolute
values are not reproduced for any of our parameterizations.

The latter study demonstrates the correlation between the isotopic shift of
m.s.\ charge radius between \nuc{40}{Ca} and \nuc{48}{Ca} and the absolute
single-particle energy of the proton $1d_{3/2}$ state. This level can be moved
around within the single-particle spectrum with the $\vec{J}^2$ terms. However,
the agreement of the calculated single-particle energy of the proton $1d_{3/2}$
state in both nuclei with experiment is not necessarily improved for the
parameterizations that reproduce the isotopic shift of the m.s.\ charge radius.
Furthermore, a good reproduction of the isotopic shift does not guarantee that
the absolute values of the charge radii are well reproduced, see the bottom
panel in Fig.~\ref{fig:rad1}. In fact, they are predicted too large for all of
our parameterizations, which again points to deficiencies of the central field.
Altogether, this suggests that in spite of its sensitivity to the coupling
constants of the $\vec{J}^2$ terms, the isotopic shift of m.s.\ charge radius
between \nuc{40}{Ca} and \nuc{48}{Ca} should not be used to constrain them
before one has gained sufficient control over the central interaction.

A few further words of caution are in place. The charge radii of
all light nuclei are significantly increased by dynamical quadrupole
correlations, see Fig.~23 of Ref.~\cite{BBH06a}. Correlations beyond the
static self-consistent mean field are also at the origin of the arch
of the ms charge radii between \nuc{40}{Ca} and \nuc{48}{Ca} that is
neither reproduced by any pure mean-field model, see again Fig.~11 in
Ref.~\cite{RMP}, nor by the beyond-mean-field calculations with SLy4
of Ref.~\cite{BBH06a}, while the shell model allows for a satisfactory
description \cite{Cau01a}.

\begin{figure}[t!]
    \includegraphics[width=0.8\columnwidth]{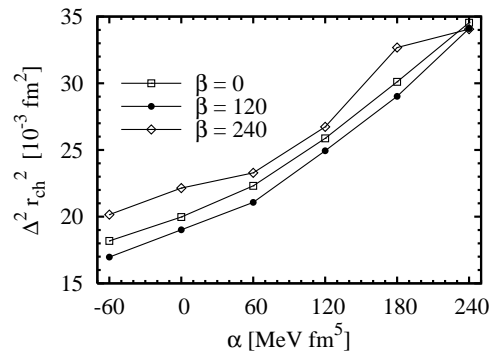}
    \caption{
    Change of slope in the m.s.\ charge radii $\Delta^2 r_\text{ch}^2$
    around \nuc{208}{Pb}, Eq.~(\ref{eq:pb:kink}),
    in fm$^2$ as a function of $\alpha$ for three
    values of $\beta$. The experimental value is
    about one and a half times as large as the largest theoretical value
    shown here, see text.
    }
\label{fig:rad2}
\end{figure}

Many explanations have been put forward to explain the kink in the isotopic
shifts of Pb radii. As it qualitatively appears in relativistic mean-field
models, but not in non-relativistic ones using the standard spin-orbit
interaction (\ref{eq:Skyrme:LS}), it has been used as a motivation to
generalize the isospin mix of the standard spin-orbit energy density
functional, Eq.~(\ref{eq:ef:ls}), to simulate the isospin dependence of
the relativistic Hartree models~\cite{Sha95a,Rei95a}. The resulting
parameterizations are not completely satisfactory, as the price for the
improvement of the radii is a further deterioration of spin-orbit
splittings~\cite{Ben99b}, while the relativistic mean field gives a
satisfactory description of both. Some standard Skyrme interactions that
take the tensor terms from the central Skyrme force into
account also give a kink, but it is by far too small to reproduce
the experimental values~\cite{Cha98a}.

Plotting the m.s.~radii along the chain of Pb isotopes as a function of $N$,
the slopes are nearly linear when looking separately at the isotopes below
and above \nuc{208}{Pb}. We will concentrate on the change in the slope
at \nuc{208}{Pb} that is brought by the tensor terms, which can be quantified
through the second finite difference of the m.s.~radii at \nuc{208}{Pb}
\begin{eqnarray}
\label{eq:pb:kink}
\lefteqn{\Delta^2 \langle r^2_{\text{ch}} \rangle (\text{\nuc{208}{Pb}})
}  \\
& = & \tfrac{1}{2} \, \big[        r^2_{\text{ch}} (\text{\nuc{206}{Pb}})
                            - 2 \, r^2_{\text{ch}} (\text{\nuc{208}{Pb}})
                            +      r^2_{\text{ch}} (\text{\nuc{210}{Pb}})
                      \big]\,.\nn
\end{eqnarray}
There are two conflicting values to be found in the literature, either
$46.4\pm 1.4$~fm$^2$ \cite{Otten} and the significantly larger
$59\pm 3$~fm$^2$ \cite{Ang04a}.
Figure~\ref{fig:rad2} shows the change of slope around \nuc{208}{Pb} as
defined through Eq.~(\ref{eq:pb:kink}) as a function of the like-particle
tensor coupling constant $\alpha$ and for three different values of $\beta$.
It is striking to see that this quantity is almost independent of the
neutron-proton tensor coupling constant $\beta$, so the change is mainly
induced by the tensor interaction between particles of the same kind.
It has been noted before that the kink in the isotopic shift
of the charge radii in Pb isotopes is correlated to the single-particle
spectrum of neutrons above $N=126$, in particular the position of the
$1i_{11/2}$ level. (This has to be contrasted with the Ca isotopic chain
discussed above, where the difference of charge radii between \nuc{40}{Ca}
and \nuc{48}{Ca} appears to be particularly sensitive to the single-particle
spectrum of the protons.)
The closer the $1i_{11/2}$ level is to the $2g_{9/2}$ level that is filled
above $N=126$, the more the $1i_{11/2}$ becomes occupied through
pairing correlations. Through the shape of its radial wave function,
the partial filling of the nodeless $1i_{11/2}$ increases the neutron
radius faster than filling only the $2g_{9/2}$, and in particular faster
than for the isotopes below $N=126$. As the protons follow the density
distribution of the neutrons, the charge radius grows rapidly beyond
$N=126$.
This offers an explanation why the kink increases with the like-particle
tensor term coupling constant $\alpha$: for large values of the weight
$\alpha$ of the neutron spin-orbit current in the neutron spin-orbit
potential, Eq.~(\ref{eq:wpot:tot}), the spin-orbit splitting of the
$\nu$ $1i$ levels is  reduced such that the $1i_{11/2}$ approaches
the $2g_{9/2}$ level in \nuc{208}{Pb}, see Fig.~\ref{fig:spe-Pb208}.

While the kink is clearly sensitive to the tensor terms, they cannot
be responsible for the entire effect, as even for extreme parameterizations
that give unrealistic single-particle spectra the calculated kink hardly
reaches about three quarters of its experimental value.


\section{Summary and conclusions}
\label{sect:conclusions}


We have reported a systematic study of the effects of the $\vec{J}^2$ (tensor)
terms in the Skyrme energy functional for spherical nuclei. The aim of the
present study was not to obtain a unique best fit of the Skyrme energy
functional with tensor terms, but to analyze the impact of the tensor terms on a
large variety of observables in calculations at a pure mean-field level and to
identify, if possible, observables that are particularly, even uniquely,
sensitive to the $\vec{J}^2$ terms. To reach our goal, we have built a set of 36
parameterizations that cover the two-dimensional parameter space of the coupling
constants of the $\vec{J}_t^2$ terms that does not give obviously unphysical
predictions for a wide variety of observables we have looked at. The fits were
performed using a protocol very similar to that of the SLy parameterizations
\cite{Cha97a,Cha98a}. The 36 actual sets of parameters can be found in the
\emph{Physical Review} archive
\cite{EPAPS}.

We use a formalism that explicitly relates the tensor terms in the energy
functional to underlying effective density-dependent central, spin-orbit and
tensor forces (or vertices) in the particle-hole channel. As has been known for
long, a zero-range tensor force gives no qualitatively new terms for spherical
mean-field states when combined with a central Skyrme force, but solely modifies
the coupling constants of the $\vec{J}^2$ terms that are already present. The
contribution  from the central Skyrme force to the coupling constants of the
$\vec{J}^2$ terms depends on the same parameters $t_1$, $x_1$, $t_2$ and $x_2$
that determine the effective mass and contribute to the surface terms. As the
latter terms are much more important for the description of bulk properties than
the $\vec{J}^2$ terms, the coupling constants of the $\vec{J}^2$ terms are
confined to a very small region of the parameter space. From this point of view,
adding a tensor force is necessary to explore it fully.

There is, however, the alternative interpretation of the Skyrme energy
functional from the density matrix  expansion, which in the absence of ab-initio
realizations so far is used as a motivation to set up energy functionals with
independent, and phenomenologically fitted, coupling constants of all terms not
constrained by symmetries. In particular, this can be used to set unwanted or
underconstrained terms to zero, as it is done for many existing
parameterizations of the (central) Skyrme interaction. For the ground states of
spherical nuclei, as discussed here, the frameworks cannot be distinguished. For
deformed nuclei and, in particular, polarized nuclear matter, this choice will
make a difference.

As a result of our study, we have obtained a long list of potential deficiencies
of the Skyrme energy functional, most of which can be expected to be related to
the properties of the central and spin-orbit interactions used. In fact, these
deficiencies become more obvious the moment one adds a tensor force, as it
appears that the presence of a tensor force unbalances a delicate compromise
within various terms of the Skyrme interaction that permits to get the global
trend of gross features of the shell structure right.

Our conclusions, however, have to be taken with a grain of salt. On the one
hand, some might depend on the fit protocol; and on the other hand, we have to
stress that (within the framework of our study -- and all others available so
far using mean-field methods) the comparison between calculated and empirical
single-particle energies is not straightforward and without the risk of being
misled. However, without even looking at single-particle spectra, we find that
\begin{enumerate}
\item
The presence of the tensor terms leads to a strong rearrangement of the other
coupling constants, most notably that of the spin-orbit force. In fact, we find
that the variation of the spin-orbit strength $W_0$ provoked by the presence of
tensor terms has a larger impact on the global systematics of single-particle
spectra than the tensor terms themselves. The rearrangement of the parameters of
the central and spin-orbit parts of the effective interaction suggests that
perturbative studies of the tensor terms, in which they are added to an existing
parameterization without readjustment, allow only very limited conclusions.
\item
In the Skyrme energy functional, the combined coupling constants of the
spin-orbit and tensor terms are nearly exclusively fixed by the mass differences
between \nuc{40}{Ca}, \nuc{48}{Ca} and \nuc{56}{Ni}. This correlation appears to
be (at least partly) spurious, the rapidly varying spin-orbit and tensor terms
being misused to simulate missing physics in the standard Skyrme functional.
\item
The cost function $\chi^2$ used in our fit protocol prefers parameterizations
with $\beta = 0$, i.e.\ pure like-particle tensor terms $\sim (\vec{J}^2_\nrm +
\vec{J}^2_\prm)$, without giving a clear preference for a value of the
corresponding coupling constant $\alpha$. By contrast, the mass residuals of 134
spherical even-even nuclei are minimized for interactions with large $\alpha$
and $\beta$. However, and as we will discuss in \cite{II}, the deformation
properties of many nuclei obtained with the latter parameterizations are
unrealistic, which disfavors this region of the parameter space.
\item
The difference of the charge radii of \nuc{40}{Ca} and \nuc{48}{Ca} turns out
to be particularly sensitive to the absolute single-particle energy of the
proton $1d_{3/2}$ level, which can be moved around by the $\vec{J}^2$ terms.
As the parameterizations that give the best agreement for the absolute
placement of this level do not necessarily give the best overall
single-particle spectra for these two nuclei, this quantity should not be used
to constrain the $\vec{J}^2$ terms.
\end{enumerate}
Concerning the global properties of the spin-orbit current
$\vec{J}$ and its contribution to the spin-orbit potential,
we have shown that
\begin{enumerate}
\item
The spin-orbit current $\vec{J}$ in non-spin-saturated doubly-magic nuclei as
\nuc{56}{Ni}, \nuc{100}{Sn}, \nuc{132}{Sn} or \nuc{208}{Pb} is dominated by the
nodeless intruder orbitals. Through the contribution of the tensor terms to the
spin-orbit field, the feedback effect on their own spin-orbit splitting is
maximized.
\item
In light nuclei, $\vec{J}$ and consequently the contribution of the $\vec{J}^2$
terms to the binding energy and the spin-orbit potential, vary rapidly between
near-zero and very large values when adding just a few nucleons to a given
nucleus. In heavy spherical nuclei, the variation becomes much slower and
smoother as on the one hand one does not encounter spin-saturated
configurations anymore, and on the other hand there are more and more
high-$\ell$ states with large degeneracy that require more nucleons to be
filled.
\item
The contribution from the zero-range spin-orbit force to the spin-orbit
potential is peaked at the nuclear surface, as it is proportional to the
gradient of the density. By contrast, the contribution from the zero-range
tensor terms is peaked further inside of the nucleus, modifying the width of
the spin-orbit potential with varying nucleon numbers. As shown in
Ref.~\cite{Bro06a}, experimental data tend to dislike such a modification.
\item
Large negative coupling constants of the tensor terms will lead to
instabilities, where a nucleus gains energy separating the levels from
many spin-orbit partners on both sides of the Fermi energy.
This process leads to unphysical single-particle spectra and rules out
a large part of the parameter space.
In particular cases, one might even obtain a (probably spurious)
coexistence of two spherical configurations with different shell
structure in the same nucleus, which are separated by a barrier.
\end{enumerate}
The main motivation to add $\vec{J}^2$ terms is of course to improve
the single-particle spectra. All observations and conclusions concerning
those have to be taken with care, as in this study we compare the eigenvalues
of a spherical single-particle Hamiltonian with the separation energy
to low-lying states in the odd-$A$ neighbors of doubly and semi-magic
nuclei (as was done in all existing earlier studies). When looking at
the single-particle spectra in doubly-magic nuclei (or semi-magic nuclei
combined with a strong subshell closure of the other species) we find
that
\begin{enumerate}
\item
The relative error of the spin-orbit splittings depends strongly on the
principal quantum number of the orbitals within a given shell, such that for
parameterizations without the tensor terms the splittings of the intruder state
(without nodes in the radial wave function) is tentatively too small, while it
becomes too large with increasing number of nodes. Adding the tensor terms
further increases the discrepancy. This problem can only be resolved by an
improved control over the shape of the spin-orbit potential. Indeed, the size of
the spin-orbit splittings is related to the overlap of the radial wave function
of a given single-particle state with the spin-orbit potential. The tensor terms
modify the width of the spin-orbit potential, but to cure this deficiency calls
for a large negative like-particle tensor coupling constant $\alpha$, which is
not consistent with the evolution of spin-orbit splittings along chains of
semi-magic nuclei, and will lead to instabilities.
\item
We also find that, in a given nucleus, the predicted spin-orbit splittings
of neutron levels are larger than those of the protons when both are
compared to experiment, which hints at an unresolved isospin trend
in the spin-orbit interaction.
\item
For spin-saturated doubly-magic nuclei as \nuc{16}{O} and \nuc{40}{Ca}, the
spin-orbit splittings of the spin-saturated species of nucleons depends strongly
on the coupling constants of the $\vec{J}^2$ terms, although they do not
contribute to the spin-orbit field. This is a consequence of the strong
correlation between the spin-orbit and tensor term coupling constants, which try
to compensate each other in spin-unsaturated nuclei. For parameterizations with
strong tensor-term coupling constants, the resulting spin-orbit force leads to
unrealistic single-particle spectra of spin-saturated configurations.
\item
The centroid of the spin-orbit partners that give the intruder state
is tentatively too high compared to the major shell below.
\end{enumerate}
The main effect of the tensor terms, that most of the recent studies
concentrate on, is the evolution of spin-orbit splittings with $N$
and $Z$. Unfortunately, there are no data for the splittings themselves,
such that one relies on data for the evolution of the distance of
two levels with different $\ell$. The comparison is compromised by
the global deficiencies of the single-particle spectra listed above.
Still, a careful comparison of calculations and experiment suggests that
\begin{enumerate}
\item
The evolution of the proton $1h_{11/2}$, $1g_{7/2}$ and $2d_{5/2}$ levels in
the chain of Sn isotopes and that of the proton $1f_{5/2}$ and $2p_{3/2}$
levels in Ni isotopes call for a positive proton-neutron tensor coupling
constant $\beta$
with a value around 120 MeV fm$^{5}$, consistent with the findings of
Refs.~\cite{Bro06a,Col07a,Bri07a}.
\item
The evolution of the neutron $1d_{3/2}$ and $2s_{1/2}$ levels between
\nuc{40}{Ca} and \nuc{48}{Ca} calls for a like-particle tensor coupling
constant $\alpha$ with a similar value around 120 MeV fm$^{5}$. This it at
variance to the findings of Refs.~\cite{Bro06a,Col07a,Bri07a},
but in qualitative agreement with the parameterization skxta
of Brown~\etal~\cite{Bro06a}
for which the tensor terms were derived from a realistic interaction but
disregarded thereafter because of its poor description of spin-orbit
splittings.
\item
Combined this leads to a dominantly isoscalar tensor term with
a coupling constant $C^{J}_0$ around 120 MeV fm$^{5}$, while the
isovector coupling constant will have a small, near-zero, value.
\end{enumerate}
Our study is obviously only a stepping stone towards improved parameterizations
of the Skyrme energy density functional. There are a number of necessary
further studies and future theoretical developments
\begin{enumerate}
\item
The deformation properties of selected parameterizations T$IJ$ from this
study will be discussed in a forthcoming paper \cite{II}.
\item
The influence of the terms depending on time-odd densities and currents
in the complete energy functional (\ref{eq:EF:full}) on nuclear matter
and finite nuclei (rotational bands etc) is under investigation as well.
The existing stability criteria of polarized matter have to be generalized
as the tensor force introduces new unique terms, for example in the
Landau parameters \cite{Hae82a}.
\item
It is well known that the strength of the spin-orbit force has to scale
with the effective mass of an interaction, which in turn determines
the average density of single-particle levels. All parameterizations
discussed here have a similar effective mass close to $m^*_0/m = 0.7$
that was already used for the SLy parameterizations. This value is
somewhat smaller than the one obtained from \emph{ab-initio}
calculations. We have checked that increasing the effective
isoscalar mass to the more realistic $m^*_0/m = 0.8$ (which within
our fit protocol requires to use two density dependent terms \cite{Les06a})
does not significantly affect any of our conclusions.
\item
It is evident that improvements of the central and spin-orbit parts of the
energy density functional are necessary, which will require a generalization of
its functional form. Other motivations were found recently to
perform such a generalization~\cite{Les06a}.
\item
The only quantity that we found sufficiently sensitive to the tensor terms is
the evolution of the distance between single-particle levels in isotopic or
isotonic chains of semi-magic nuclei. The distance between the levels that can
be used for such studies is so large, that it might be compromised by their
coupling to collective excitations. Reliable calculations including pairing,
polarization as well as particle-vibration coupling effects~\cite{Ber80a,Lit06a}
along isotopic and isotonic chains are needed to test the quality, reliability
and limits of the simplistic identification of the eigenvalues of the spherical
mean-field Hamiltonian in an even-even nucleus with the separation energy to or
from low-lying states in the adjacent odd-$A$ nuclei.
\end{enumerate}


\section*{Acknowledgments}


We thank P.~Bonche, H.~Flocard, P.-H.~Heenen and B.~A.~Brown for stimulating and
encouraging discussions. Work by M.~B.\ and K.~B.\ was performed within the
framework of the Espace de Structure Nucl{\'e}aire Th{\'e}orique (ESNT).
T.\ L. acknowledges the hospitality of the SPhN and ESNT on many occasions
during the realization of this work. This work was supported by the
U.S.\ National Science Foundation under Grant No.\ PHY-0456903.


\begin{appendix}


\section{Coupling constants of the Skyrme energy functional}
\label{app:sect:cpl}


The coupling constants of the central Skyrme energy density
functional in terms of the parameters of the
central Skyrme force are given by
\begin{eqnarray}
\label{eq:cpl:SF}
A_0^{\rho}
& = &   \tfrac{3}{8}  t_0
      + \tfrac{3}{48} t_3 \, \rho_0^\alpha (\vec{r})
      \nn \\
A_1^{\rho}
& = & - \tfrac{1}{4}  t_0 \big( \tfrac{1}{2} + x_0 \big)
      - \tfrac{1}{24} t_3 \big( \tfrac{1}{2} + x_3 \big)
      \, \rho_0^\alpha (\vec{r})
      \nn \\
A_0^{s}
& = & - \tfrac{1}{4} t_0 \big( \tfrac{1}{2} - x_0 \big)
      - \tfrac{1}{24} t_3 \big( \tfrac{1}{2} - x_3 \big)
      \, \rho_0^\alpha (\vec{r})
     \nn \\
A_1^{s}
& = & - \tfrac{1}{8} t_0
      - \tfrac{1}{48} t_3  \, \rho_0^\alpha (\vec{r})
      \nn \\
A_0^{\tau}
& = &   \tfrac{3}{16} \, t_1
      + \tfrac{1}{4} t_2 \; \big( \tfrac{5}{4} + x_2 \big)
      \nn \\
A_1^{\tau}
& = & - \tfrac{1}{8} t_1 \big(\tfrac{1}{2} + x_1 \big)
      + \tfrac{1}{8} t_2 \big(\tfrac{1}{2} + x_2 \big)
      \nn \\
A_0^{T}
& = & - \tfrac{1}{8} t_1 \big( \tfrac{1}{2} - x_1 \big) \,
      + \tfrac{1}{8} t_2 \big( \tfrac{1}{2} + x_2 \big)
      \nn \\
A_1^{T}
& = & - \tfrac{1}{16} t_1
      + \tfrac{1}{16} t_2
      \nn \\
A_0^{\Delta \rho}
& = & - \tfrac{9}{64} t_1
      + \tfrac{1}{16}  t_2 \big( \tfrac{5}{4} + x_2 \big)
      \nn \\
A_1^{\Delta \rho}
& = &   \tfrac{3}{32} t_1 \big( \tfrac{1}{2} + x_1 \big)
      + \tfrac{1}{32} t_2 \big( \tfrac{1}{2} + x_2 \big)
      \nn \\
A_0^{\Delta s}
& = &   \tfrac{3}{32} t_1 \big( \tfrac{1}{2} - x_1 \big)
      + \tfrac{1}{32} t_2 \big( \tfrac{1}{2} + x_2 \big)
      \nn \\
A_1^{\Delta s}
& = & \tfrac{3}{64} t_1 + \tfrac{1}{64} t_2\,.
\end{eqnarray}
The coupling constants of the spin-orbit energy density
functional in terms of the parameters of the
spin-orbit force are given by
\begin{eqnarray}
\label{eq:cpl:LSF}
A_0^{\nabla J}
& = & - \tfrac{3}{4} W_0\,,
       \nn \\
A_1^{\nabla J}
& = & - \tfrac{1}{4} W_0\,.
\end{eqnarray}
The coupling constants of the tensor energy density
functional in terms of the parameters of
Skyrme's tensor force are given by (Table I in~\cite{Per04a})
\begin{alignat}{4}
\label{eq:cpl:tensor}
B^{T}_0
& =   - \tfrac{1}{8} (t_e + 3 t_o)
      & \qquad
B^{T}_1
& =   \phantom{-} \tfrac{1}{8} (t_e - t_o)
      \\
B^{F}_0
& =   \phantom{-} \tfrac{3}{8} (t_e + 3 t_o)
      & \qquad
B^{F}_1
& =   - \tfrac{3}{8} (t_e - t_o)
      \\
B^{\Delta s}_0
& =   \phantom{-} \tfrac{3}{32} (t_e - t_o)
      & \qquad
B^{\Delta s}_1
& =   - \tfrac{1}{32} (3 t_e + t_o)
      \\
B^{\nabla s}_0
& =   \phantom{-} \tfrac{9}{32} (t_e - t_o)
      & \qquad
B^{\nabla s}_1
& =  - \tfrac{3}{32} (3t_e + t_o)
\,.
\end{alignat}
%


\section{Phase transitions}
\label{sect:app:instability}
\label{app:instab}


The densities $\rho$ and $\tau$ entering the energy functional
(\ref{eq:EF:sphere}) vary smoothly with nucleon numbers as they
follow the geometric growth of the nucleus.
As a result, a functional depending only on $\rho$ and $\tau$ usually
shows a unique minimum for given $N$, $Z$ and shape.
The situation is quite different when the tensor terms are taken into
account. Indeed, the amplitude of the spin-orbit current density
$\vec{J}$ (\ref{eq:j:radial}) depends on the number of spin-unsaturated
single-particle states in the nucleus; it varies from (almost) zero in
spin-saturated nuclei to large finite values as a consequence of shell
and finite-size effects, see Fig.~\ref{fig:Ni-Jn}.

\begin{figure}[t!]
  \includegraphics[width=\columnwidth]{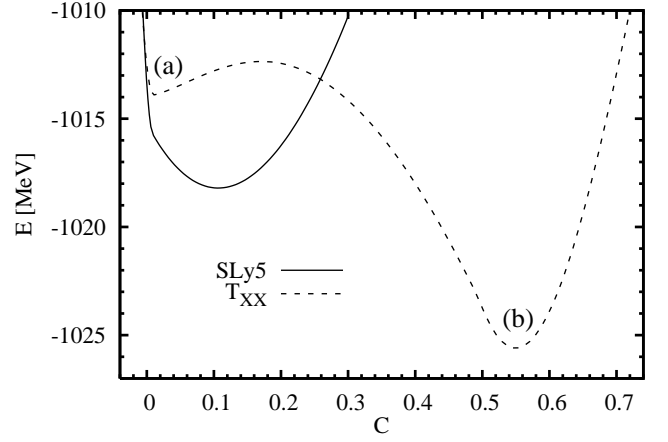}
  \caption{
    Total binding energy of \nuc{120}{Sn} as a function of
    $C = \int \! d^3r \; \vec{J}_\nrm \cdot \boldsymbol{\nabla} \rho_\nrm$
    in a constrained calculation. The dashed curve shows results obtained
    with the parameterization mentioned in the text, while the
    solid curve shows results obtained with SLy5.
  }
  \label{fig:cstr}
\end{figure}

This behavior poses the risk of an instability, which was already reported
in~\cite{Bei75a}: multiplying $\vec{J}$ with a
large coupling constant in the spin-orbit potential (\ref{eq:wpot:tot})
might, for certain combinations of the signs of the coupling constant and
the spin-orbit currents of protons and neutrons, increase the spin-orbit
splittings. In some nuclei, this will cause two levels originating from
different $\ell$ shells to approach the Fermi energy, one from above and
the other from below, or even to cross. In that situation, their
occupation numbers will change such that $\vec{J}$ increases further,
which feeds back onto the spin-orbit potential and ultimately leads to
a dramatic rearrangement of the single-particle spectrum.

We faced this problem when attempting to fit parameter sets with
large negative $C_0^{J}$ and $C_1^{J}$. During the fit, some nuclei
sometimes fell into the instability, depending on the values of the
other coupling constants. As this is a highly nonlinear threshold effect
that results in a very large energy gain from tiny modifications
of the coupling constants, the corresponding fits did not, and could
not, converge.

\begin{figure}[t!]
  \includegraphics[width=\columnwidth]{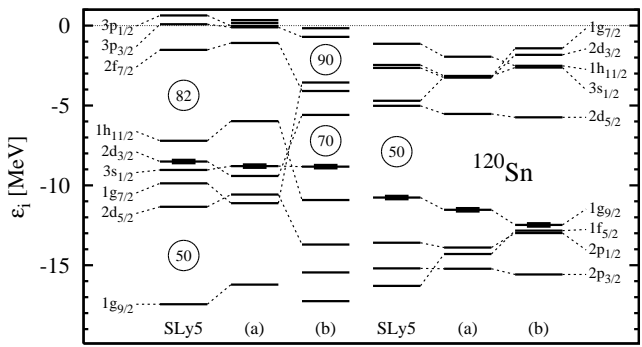}
  \caption{Single-particle spectra corresponding to the minimum found with
    SLy5 and (a) the secondary minimum found with T${XX}$, (b) the
    absolute minimum (see Fig.~\ref{fig:cstr}; left: neutron levels,
    right: proton levels).
  }
  \label{fig:cstr-spe}
\end{figure}

In special cases, one might even run into a situation with two coexisting
minima, where as a function of a suitable coordinate the configuration with
regular shell structure is separated from a configuration with unphysical large
spin-orbit splittings by a barrier. In such a case, a calculation of the ground
state might converge into one or the other minimum depending on the initial
conditions chosen for the iterative solution of the HFB equations. In a
calculation along an isotopic or isotonic chain, the coexistence will reveal
itself through a large scattering of the mass residuals, which will fall on two
distinct curves. We illustrate this phenomenon in Fig.~\ref{fig:cstr} for
\nuc{120}{Sn} using a parameter set denoted ``T$XX$'' with
$C_0^{J}=-157.57$~MeV\,fm$^5$ and $C_1^{J}=-114.88$~MeV\,fm$^5$, which is
located outside the parameter space shown in Fig.~\ref{fig:cjplane-tot}, to its
lower left. Among the various possible recipes for a constraint on the
spin-orbit current density, we chose to minimize the following quantity:
\begin{equation}
E[\rho]
- \mu \left[ \int \! d^3r \; \vec{J}_\nrm \cdot \mathbf{\nabla} \rho_\nrm - C
      \right]^2
\end{equation}
where $\mu$ is a Lagrange parameter and $C$ is a constant used to tune the
constraint. The energy curve exhibits two minima denoted $(a)$ and $(b)$. The
corresponding single-particle spectra are shown in Fig.~\ref{fig:cstr-spe}
along with those obtained for SLy5. The minimum $(a)$ corresponds to an almost
spin saturated neutron configuration where both spin partners are either
occupied or empty,\footnote{
Note that the spin-saturation is a consequence of the wrong ordering
of the $\nu$ $1h_{11/2}$, $2d_{3/2}$ and $1s_{1/2}$ levels compared to
empirical data that is found for all our parameterizations,
see Fig.~\ref{fig:Sn132-field}, and in fact virtually all mean-field
models~\cite{RMP}.
}
which is very similar to what is found using SLy5. In the minimum $(b)$,
which is deeper by more than 7~MeV,
the single-particle spectrum is completely reorganized in order to maximize the
spin-orbit current density and take advantage of its contribution in the
functional. In this situation the neutron spin doublets $2d$, $1g$ and $1h$
split on both sides of the Fermi surface and generate a large spin-orbit
current density.

This clearly shows that the parameter sets with large and negative coupling
constants of the $\vec{J}^2$ terms must be discarded since for many nuclei
they lead to ground states with unrealistic single-particle structure.

Note that this kind of instability does not appear for the spin-orbit term:
although  its contribution to the energy functional (\ref{eq:EF:sphere}) also
varies between small, sometimes near-zero, and very large values, see
Figs.~\ref{fig:Ca-J2so} and~\ref{fig:Sn-J2so}, it is only linear in $\vec{J}$.
As a consequence, its contribution to the spin-orbit potential
(\ref{eq:wpot:tot}) lacks the feedback mechanism outlined above as it does not
scale  with $\vec{J}$. Still, its contribution to the total energy is usually
much larger than that of the $\vec{J}^2$ terms, so it plays a decisive role for
the absolute energy gained when varying $\vec{J}$.

\end{appendix}


\bibliography{tensor}

\begin{thebibliography}{121}
\expandafter\ifx\csname natexlab\endcsname\relax\def\natexlab#1{#1}\fi
\expandafter\ifx\csname bibnamefont\endcsname\relax
  \def\bibnamefont#1{#1}\fi
\expandafter\ifx\csname bibfnamefont\endcsname\relax
  \def\bibfnamefont#1{#1}\fi
\expandafter\ifx\csname citenamefont\endcsname\relax
  \def\citenamefont#1{#1}\fi
\expandafter\ifx\csname url\endcsname\relax
  \def\url#1{\texttt{#1}}\fi
\expandafter\ifx\csname urlprefix\endcsname\relax\def\urlprefix{URL }\fi
\providecommand{\bibinfo}[2]{#2}
\providecommand{\eprint}[2][]{\url{#2}}

\bibitem[{\citenamefont{Goeppert~Mayer}(1948)}]{GM48a}
\bibinfo{author}{\bibfnamefont{M.}~\bibnamefont{Goeppert~Mayer}},
  \bibinfo{journal}{Phys. Rev.} \textbf{\bibinfo{volume}{74}},
  \bibinfo{pages}{235} (\bibinfo{year}{1948}).

\bibitem[{\citenamefont{Haxel et~al.}(1949)\citenamefont{Haxel, Jensen, and
  Suess}}]{Hax49a}
\bibinfo{author}{\bibfnamefont{O.}~\bibnamefont{Haxel}},
  \bibinfo{author}{\bibfnamefont{J.~H.~D.} \bibnamefont{Jensen}},
  \bibnamefont{and} \bibinfo{author}{\bibfnamefont{H.~E.} \bibnamefont{Suess}},
  \bibinfo{journal}{Phys. Rev.} \textbf{\bibinfo{volume}{75}},
  \bibinfo{pages}{1766} (\bibinfo{year}{1949}).

\bibitem[{\citenamefont{Feenberg and Hammack}(1949)}]{Fee49a}
\bibinfo{author}{\bibfnamefont{E.}~\bibnamefont{Feenberg}} \bibnamefont{and}
  \bibinfo{author}{\bibfnamefont{K.~C.} \bibnamefont{Hammack}},
  \bibinfo{journal}{Phys. Rev.} \textbf{\bibinfo{volume}{75}},
  \bibinfo{pages}{1877} (\bibinfo{year}{1949}).

\bibitem[{\citenamefont{Goeppert~Mayer}(1949)}]{GM49a}
\bibinfo{author}{\bibfnamefont{M.}~\bibnamefont{Goeppert~Mayer}},
  \bibinfo{journal}{Phys. Rev.} \textbf{\bibinfo{volume}{75}},
  \bibinfo{pages}{1969} (\bibinfo{year}{1949}).

\bibitem[{\citenamefont{Beiner et~al.}(1975)\citenamefont{Beiner, Flocard,
  {Nguyen Van Giai}, and Quentin}}]{Bei75a}
\bibinfo{author}{\bibfnamefont{M.}~\bibnamefont{Beiner}},
  \bibinfo{author}{\bibfnamefont{H.}~\bibnamefont{Flocard}},
  \bibinfo{author}{\bibnamefont{{Nguyen Van Giai}}}, \bibnamefont{and}
  \bibinfo{author}{\bibfnamefont{P.}~\bibnamefont{Quentin}},
  \bibinfo{journal}{Nucl. Phys.} \textbf{\bibinfo{volume}{A238}},
  \bibinfo{pages}{29} (\bibinfo{year}{1975}).

\bibitem[{\citenamefont{Dobaczewski et~al.}(1994)\citenamefont{Dobaczewski,
  Hamamoto, Nazarewicz, and Sheikh}}]{Dob94a}
\bibinfo{author}{\bibfnamefont{J.}~\bibnamefont{Dobaczewski}},
  \bibinfo{author}{\bibfnamefont{I.}~\bibnamefont{Hamamoto}},
  \bibinfo{author}{\bibfnamefont{W.}~\bibnamefont{Nazarewicz}},
  \bibnamefont{and} \bibinfo{author}{\bibfnamefont{J.~A.}
  \bibnamefont{Sheikh}}, \bibinfo{journal}{Phys. Rev. Lett.}
  \textbf{\bibinfo{volume}{72}}, \bibinfo{pages}{981} (\bibinfo{year}{1994}).

\bibitem[{\citenamefont{Lalazissis
  et~al.}(1998{\natexlab{a}})\citenamefont{Lalazissis, Vretenar, P\"oschl, and
  Ring}}]{Lal98a}
\bibinfo{author}{\bibfnamefont{G.~A.} \bibnamefont{Lalazissis}},
  \bibinfo{author}{\bibfnamefont{D.}~\bibnamefont{Vretenar}},
  \bibinfo{author}{\bibfnamefont{W.}~\bibnamefont{P\"oschl}}, \bibnamefont{and}
  \bibinfo{author}{\bibfnamefont{P.}~\bibnamefont{Ring}},
  \bibinfo{journal}{Phys. Lett.} \textbf{\bibinfo{volume}{B418}},
  \bibinfo{pages}{7} (\bibinfo{year}{1998}{\natexlab{a}}).

\bibitem[{\citenamefont{Lalazissis
  et~al.}(1998{\natexlab{b}})\citenamefont{Lalazissis, Vretenar, P\"oschl, and
  Ring}}]{Lal98b}
\bibinfo{author}{\bibfnamefont{G.~A.} \bibnamefont{Lalazissis}},
  \bibinfo{author}{\bibfnamefont{D.}~\bibnamefont{Vretenar}},
  \bibinfo{author}{\bibfnamefont{W.}~\bibnamefont{P\"oschl}}, \bibnamefont{and}
  \bibinfo{author}{\bibfnamefont{P.}~\bibnamefont{Ring}},
  \bibinfo{journal}{Nucl. Phys.} \textbf{\bibinfo{volume}{A632}},
  \bibinfo{pages}{363} (\bibinfo{year}{1998}{\natexlab{b}}).

\bibitem[{\citenamefont{Chen et~al.}(1995)\citenamefont{Chen, Dobaczewski,
  Kratz, Langanke, Pfeiffer, Thielemann, and Vogel}}]{Che95a}
\bibinfo{author}{\bibfnamefont{B.}~\bibnamefont{Chen}},
  \bibinfo{author}{\bibfnamefont{J.}~\bibnamefont{Dobaczewski}},
  \bibinfo{author}{\bibfnamefont{K.~L.} \bibnamefont{Kratz}},
  \bibinfo{author}{\bibfnamefont{K.}~\bibnamefont{Langanke}},
  \bibinfo{author}{\bibfnamefont{B.}~\bibnamefont{Pfeiffer}},
  \bibinfo{author}{\bibfnamefont{F.-K.} \bibnamefont{Thielemann}},
  \bibnamefont{and} \bibinfo{author}{\bibfnamefont{P.}~\bibnamefont{Vogel}},
  \bibinfo{journal}{Phys. Lett.} \textbf{\bibinfo{volume}{B355}},
  \bibinfo{pages}{37} (\bibinfo{year}{1995}).

\bibitem[{\citenamefont{Dobaczewski et~al.}(1995)\citenamefont{Dobaczewski,
  Nazarewicz, and Werner}}]{Dob95b}
\bibinfo{author}{\bibfnamefont{J.}~\bibnamefont{Dobaczewski}},
  \bibinfo{author}{\bibfnamefont{A.}~\bibnamefont{Nazarewicz}},
  \bibnamefont{and} \bibinfo{author}{\bibfnamefont{T.~R.}
  \bibnamefont{Werner}}, \bibinfo{journal}{Phys. Scr.}
  \textbf{\bibinfo{volume}{T56}}, \bibinfo{pages}{15} (\bibinfo{year}{1995}).

\bibitem[{\citenamefont{Pearson et~al.}(1996)\citenamefont{Pearson, Nayak, and
  Goriely}}]{Pea96a}
\bibinfo{author}{\bibfnamefont{J.~M.} \bibnamefont{Pearson}},
  \bibinfo{author}{\bibfnamefont{R.~C.} \bibnamefont{Nayak}}, \bibnamefont{and}
  \bibinfo{author}{\bibfnamefont{S.}~\bibnamefont{Goriely}},
  \bibinfo{journal}{Phys. Lett.} \textbf{\bibinfo{volume}{B387}},
  \bibinfo{pages}{455} (\bibinfo{year}{1996}).

\bibitem[{\citenamefont{Pfeiffer et~al.}(1997)\citenamefont{Pfeiffer, Kratz,
  and Thielemann}}]{Pfe97a}
\bibinfo{author}{\bibfnamefont{B.}~\bibnamefont{Pfeiffer}},
  \bibinfo{author}{\bibfnamefont{K.-L.} \bibnamefont{Kratz}}, \bibnamefont{and}
  \bibinfo{author}{\bibfnamefont{F.-K.} \bibnamefont{Thielemann}},
  \bibinfo{journal}{Z. Phys.} \textbf{\bibinfo{volume}{A357}},
  \bibinfo{pages}{235} (\bibinfo{year}{1997}).

\bibitem[{\citenamefont{Decharg\'e et~al.}(1999)\citenamefont{Decharg\'e,
  Berger, Dietrich, and Weiss}}]{Dec99a}
\bibinfo{author}{\bibfnamefont{J.}~\bibnamefont{Decharg\'e}},
  \bibinfo{author}{\bibfnamefont{J.~F.} \bibnamefont{Berger}},
  \bibinfo{author}{\bibfnamefont{K.}~\bibnamefont{Dietrich}}, \bibnamefont{and}
  \bibinfo{author}{\bibfnamefont{M.~S.} \bibnamefont{Weiss}},
  \bibinfo{journal}{Phys. Lett.} \textbf{\bibinfo{volume}{B451}},
  \bibinfo{pages}{275} (\bibinfo{year}{1999}).

\bibitem[{\citenamefont{Bender et~al.}(1999)\citenamefont{Bender, Rutz,
  Reinhard, Maruhn, and Greiner}}]{Ben99b}
\bibinfo{author}{\bibfnamefont{M.}~\bibnamefont{Bender}},
  \bibinfo{author}{\bibfnamefont{K.}~\bibnamefont{Rutz}},
  \bibinfo{author}{\bibfnamefont{P.-G.} \bibnamefont{Reinhard}},
  \bibinfo{author}{\bibfnamefont{J.~A.} \bibnamefont{Maruhn}},
  \bibnamefont{and} \bibinfo{author}{\bibfnamefont{W.}~\bibnamefont{Greiner}},
  \bibinfo{journal}{Phys. Rev. C} \textbf{\bibinfo{volume}{60}},
  \bibinfo{pages}{034304} (\bibinfo{year}{1999}).

\bibitem[{\citenamefont{Langanke et~al.}(2003)\citenamefont{Langanke, Terasaki,
  Nowacki, Dean, and Nazarewicz}}]{Lan03a}
\bibinfo{author}{\bibfnamefont{K.}~\bibnamefont{Langanke}},
  \bibinfo{author}{\bibfnamefont{J.}~\bibnamefont{Terasaki}},
  \bibinfo{author}{\bibfnamefont{F.}~\bibnamefont{Nowacki}},
  \bibinfo{author}{\bibfnamefont{D.~J.} \bibnamefont{Dean}}, \bibnamefont{and}
  \bibinfo{author}{\bibfnamefont{W.}~\bibnamefont{Nazarewicz}},
  \bibinfo{journal}{Phys. Rev. C} \textbf{\bibinfo{volume}{67}},
  \bibinfo{pages}{044314} (\bibinfo{year}{2003}).

\bibitem[{\citenamefont{Wiringa et~al.}(1995)\citenamefont{Wiringa, Stoks, and
  Schiavilla}}]{Wir95a}
\bibinfo{author}{\bibfnamefont{R.~B.} \bibnamefont{Wiringa}},
  \bibinfo{author}{\bibfnamefont{V.~G.~J.} \bibnamefont{Stoks}},
  \bibnamefont{and}
  \bibinfo{author}{\bibfnamefont{R.}~\bibnamefont{Schiavilla}},
  \bibinfo{journal}{Phys. Rev. C} \textbf{\bibinfo{volume}{51}},
  \bibinfo{pages}{38} (\bibinfo{year}{1995}).

\bibitem[{\citenamefont{Machleidt}(2001)}]{Mac01a}
\bibinfo{author}{\bibfnamefont{R.}~\bibnamefont{Machleidt}},
  \bibinfo{journal}{Phys. Rev. C} \textbf{\bibinfo{volume}{63}},
  \bibinfo{pages}{024001} (\bibinfo{year}{2001}).

\bibitem[{\citenamefont{Pieper and Wiringa}(2001)}]{Pie01a}
\bibinfo{author}{\bibfnamefont{S.~C.} \bibnamefont{Pieper}} \bibnamefont{and}
  \bibinfo{author}{\bibfnamefont{R.~B.} \bibnamefont{Wiringa}},
  \bibinfo{journal}{Ann. Rev. Nucl. Part. Sci.} \textbf{\bibinfo{volume}{51}},
  \bibinfo{pages}{53} (\bibinfo{year}{2001}).

\bibitem[{\citenamefont{Navr{\'a}til and Ormand}(2003)}]{Nav03a}
\bibinfo{author}{\bibfnamefont{P.}~\bibnamefont{Navr{\'a}til}}
  \bibnamefont{and} \bibinfo{author}{\bibfnamefont{W.~E.}
  \bibnamefont{Ormand}}, \bibinfo{journal}{Phys. Rev. C}
  \textbf{\bibinfo{volume}{68}}, \bibinfo{pages}{034305}
  (\bibinfo{year}{2003}).

\bibitem[{\citenamefont{Eisenberg and Greiner}(1972)}]{EGIII}
\bibinfo{author}{\bibfnamefont{J.~M.} \bibnamefont{Eisenberg}}
  \bibnamefont{and} \bibinfo{author}{\bibfnamefont{W.}~\bibnamefont{Greiner}},
  \emph{\bibinfo{title}{Nuclear Theory. III. Microscopic theory of the
  nucleus}} (\bibinfo{publisher}{Second printing, North Holland Physics Publ.,
  Elsevier Science Publishers, Amsterdam},
  \bibinfo{year}{1976}).

\bibitem[{\citenamefont{Nilsson and Ragnarsson}(1995)}]{Nil95a}
\bibinfo{author}{\bibfnamefont{S.~G.} \bibnamefont{Nilsson}} \bibnamefont{and}
  \bibinfo{author}{\bibfnamefont{I.}~\bibnamefont{Ragnarsson}},
  \emph{\bibinfo{title}{Shapes and Shells in Nuclear Structure}}
  (\bibinfo{publisher}{Cambridge University Press},
  \bibinfo{address}{Cambridge, England}, \bibinfo{year}{1995}).

\bibitem[{\citenamefont{Neff and Feldmeier}(2003)}]{Nef03a}
\bibinfo{author}{\bibfnamefont{T.}~\bibnamefont{Neff}} \bibnamefont{and}
  \bibinfo{author}{\bibfnamefont{H.}~\bibnamefont{Feldmeier}},
  \bibinfo{journal}{Nucl. Phys.} \textbf{\bibinfo{volume}{A713}},
  \bibinfo{pages}{311} (\bibinfo{year}{2003}).

\bibitem[{\citenamefont{Roth et~al.}(2004)\citenamefont{Roth, Neff, Hergert,
  and Feldmeier}}]{Rot04a}
\bibinfo{author}{\bibfnamefont{R.}~\bibnamefont{Roth}},
  \bibinfo{author}{\bibfnamefont{T.}~\bibnamefont{Neff}},
  \bibinfo{author}{\bibfnamefont{H.}~\bibnamefont{Hergert}}, \bibnamefont{and}
  \bibinfo{author}{\bibfnamefont{H.}~\bibnamefont{Feldmeier}},
  \bibinfo{journal}{Nucl. Phys.} \textbf{\bibinfo{volume}{A745}},
  \bibinfo{pages}{3} (\bibinfo{year}{2004}).

\bibitem[{\citenamefont{Bethe}(1968)}]{Bet68a}
\bibinfo{author}{\bibfnamefont{H.~A.} \bibnamefont{Bethe}},
  \bibinfo{journal}{Phys. Rev.} \textbf{\bibinfo{volume}{167}},
  \bibinfo{pages}{879} (\bibinfo{year}{1968}).

\bibitem[{\citenamefont{Negele}(1970)}]{Neg70a}
\bibinfo{author}{\bibfnamefont{J.~W.} \bibnamefont{Negele}},
  \bibinfo{journal}{Phys. Rev. C} \textbf{\bibinfo{volume}{1}},
  \bibinfo{pages}{1260} (\bibinfo{year}{1970}).

\bibitem[{\citenamefont{Scheerbaum}(1976)}]{Sche76a}
\bibinfo{author}{\bibfnamefont{R.~R.} \bibnamefont{Scheerbaum}},
  \bibinfo{journal}{Phys. Lett.} \textbf{\bibinfo{volume}{B63}},
  \bibinfo{pages}{381} (\bibinfo{year}{1976}).

\bibitem[{\citenamefont{Goodman and Borysowicz}(1978)}]{Goo78a}
\bibinfo{author}{\bibfnamefont{A.~L.} \bibnamefont{Goodman}} \bibnamefont{and}
  \bibinfo{author}{\bibfnamefont{J.}~\bibnamefont{Borysowicz}},
  \bibinfo{journal}{Nucl. Phys.} \textbf{\bibinfo{volume}{A295}},
  \bibinfo{pages}{333} (\bibinfo{year}{1978}).

\bibitem[{\citenamefont{Zheng and Zamick}(1991)}]{Zhe91a}
\bibinfo{author}{\bibfnamefont{D.~C.} \bibnamefont{Zheng}} \bibnamefont{and}
  \bibinfo{author}{\bibfnamefont{L.}~\bibnamefont{Zamick}},
  \bibinfo{journal}{Ann. Phys. (NY)} \textbf{\bibinfo{volume}{206}},
  \bibinfo{pages}{106} (\bibinfo{year}{1991}).

\bibitem[{\citenamefont{Bender et~al.}(2003{\natexlab{a}})\citenamefont{Bender,
  Heenen, and Reinhard}}]{RMP}
\bibinfo{author}{\bibfnamefont{M.}~\bibnamefont{Bender}},
  \bibinfo{author}{\bibfnamefont{P.-H.} \bibnamefont{Heenen}},
  \bibnamefont{and} \bibinfo{author}{\bibfnamefont{P.-G.}
  \bibnamefont{Reinhard}}, \bibinfo{journal}{Rev. Mod. Phys.}
  \textbf{\bibinfo{volume}{75}}, \bibinfo{pages}{121}
  (\bibinfo{year}{2003}{\natexlab{a}}).

\bibitem[{\citenamefont{Skyrme}(1956)}]{Sky56a}
\bibinfo{author}{\bibfnamefont{T.~H.~R.} \bibnamefont{Skyrme}},
  \bibinfo{journal}{Phil. Mag.} \textbf{\bibinfo{volume}{1}},
  \bibinfo{pages}{1043} (\bibinfo{year}{1956}).

\bibitem[{\citenamefont{Skyrme}(1958{\natexlab{a}})}]{Sky59a}
\bibinfo{author}{\bibfnamefont{T.~H.~R.} \bibnamefont{Skyrme}},
  \bibinfo{journal}{Nucl. Phys.} \textbf{\bibinfo{volume}{9}},
  \bibinfo{pages}{615} (\bibinfo{year}{1958}{\natexlab{a}}).

\bibitem[{\citenamefont{Bell and Skyrme}(1956)}]{Bel56a}
\bibinfo{author}{\bibfnamefont{J.~S.} \bibnamefont{Bell}} \bibnamefont{and}
  \bibinfo{author}{\bibfnamefont{T.~H.~R.} \bibnamefont{Skyrme}},
  \bibinfo{journal}{Phil. Mag.} \textbf{\bibinfo{volume}{1}},
  \bibinfo{pages}{1055} (\bibinfo{year}{1956}).

\bibitem[{\citenamefont{Skyrme}(1958{\natexlab{b}})}]{Sky59b}
\bibinfo{author}{\bibfnamefont{T.~H.~R.} \bibnamefont{Skyrme}},
  \bibinfo{journal}{Nucl. Phys.} \textbf{\bibinfo{volume}{9}},
  \bibinfo{pages}{635} (\bibinfo{year}{1958}{\natexlab{b}}).

\bibitem[{\citenamefont{Gogny}(1975)}]{Gog75a}
\bibinfo{author}{\bibfnamefont{D.}~\bibnamefont{Gogny}},
  \bibinfo{journal}{Nucl. Phys.} \textbf{\bibinfo{volume}{A237}},
  \bibinfo{pages}{399} (\bibinfo{year}{1975}).

\bibitem[{\citenamefont{Decharg{\'e} and Gogny}(1980)}]{Dec80a}
\bibinfo{author}{\bibfnamefont{J.}~\bibnamefont{Decharg{\'e}}}
  \bibnamefont{and} \bibinfo{author}{\bibfnamefont{D.}~\bibnamefont{Gogny}},
  \bibinfo{journal}{Phys. Rev. C} \textbf{\bibinfo{volume}{21}},
  \bibinfo{pages}{1568} (\bibinfo{year}{1980}).

\bibitem[{\citenamefont{Vautherin and Brink}(1972)}]{Vau72a}
\bibinfo{author}{\bibfnamefont{D.}~\bibnamefont{Vautherin}} \bibnamefont{and}
  \bibinfo{author}{\bibfnamefont{D.~M.} \bibnamefont{Brink}},
  \bibinfo{journal}{Phys. Rev. C} \textbf{\bibinfo{volume}{5}},
  \bibinfo{pages}{626} (\bibinfo{year}{1972}).

\bibitem[{\citenamefont{Stancu et~al.}(1977)\citenamefont{Stancu, Brink, and
  Flocard}}]{Sta77a}
\bibinfo{author}{\bibfnamefont{F.}~\bibnamefont{Stancu}},
  \bibinfo{author}{\bibfnamefont{D.~M.} \bibnamefont{Brink}}, \bibnamefont{and}
  \bibinfo{author}{\bibfnamefont{H.}~\bibnamefont{Flocard}},
  \bibinfo{journal}{Phys. Lett.} \textbf{\bibinfo{volume}{B68}},
  \bibinfo{pages}{108} (\bibinfo{year}{1977}).

\bibitem[{\citenamefont{Tondeur}(1983)}]{Ton83a}
\bibinfo{author}{\bibfnamefont{F.}~\bibnamefont{Tondeur}},
  \bibinfo{journal}{Phys. Lett.} \textbf{\bibinfo{volume}{B123}},
  \bibinfo{pages}{139} (\bibinfo{year}{1983}).

\bibitem[{\citenamefont{Liu et~al.}(1991)\citenamefont{Liu, Luo, Ma, Shen, and
  Moszkowski}}]{Liu91a}
\bibinfo{author}{\bibfnamefont{K.-F.} \bibnamefont{Liu}},
  \bibinfo{author}{\bibfnamefont{H.}~\bibnamefont{Luo}},
  \bibinfo{author}{\bibfnamefont{Z.}~\bibnamefont{Ma}},
  \bibinfo{author}{\bibfnamefont{Q.}~\bibnamefont{Shen}}, \bibnamefont{and}
  \bibinfo{author}{\bibfnamefont{S.~A.} \bibnamefont{Moszkowski}},
  \bibinfo{journal}{Nucl. Phys.} \textbf{\bibinfo{volume}{A534}},
  \bibinfo{pages}{1} (\bibinfo{year}{1991}).

\bibitem[{\citenamefont{Onishi and Negele}(1978)}]{Oni78a}
\bibinfo{author}{\bibfnamefont{N.}~\bibnamefont{Onishi}} \bibnamefont{and}
  \bibinfo{author}{\bibfnamefont{J.~W.} \bibnamefont{Negele}},
  \bibinfo{journal}{Nucl. Phys.} \textbf{\bibinfo{volume}{A301}},
  \bibinfo{pages}{336} (\bibinfo{year}{1978}).

\bibitem[{\citenamefont{Bender et~al.}(2007)\citenamefont{Bender, Bennaceur,
  Duguet, Heenen, Lesinski, and Meyer}}]{II}
\bibinfo{author}{\bibfnamefont{M.}~\bibnamefont{Bender}},
  \bibinfo{author}{\bibfnamefont{K.}~\bibnamefont{Bennaceur}},
  \bibinfo{author}{\bibfnamefont{T.}~\bibnamefont{Duguet}},
  \bibinfo{author}{\bibfnamefont{P.-H.} \bibnamefont{Heenen}},
  \bibinfo{author}{\bibfnamefont{T.}~\bibnamefont{Lesinski}}, \bibnamefont{and}
  \bibinfo{author}{\bibfnamefont{J.}~\bibnamefont{Meyer}}
  (\bibinfo{year}{2007}), \bibinfo{note}{companion paper, in preparation}.

\bibitem[{\citenamefont{Long et~al.}(2006{\natexlab{a}})\citenamefont{Long,
  {Nguyen Van Giai}, and Meng}}]{Lon06a}
\bibinfo{author}{\bibfnamefont{W.-H.} \bibnamefont{Long}},
  \bibinfo{author}{\bibnamefont{{Nguyen Van Giai}}}, \bibnamefont{and}
  \bibinfo{author}{\bibfnamefont{J.}~\bibnamefont{Meng}},
  \bibinfo{journal}{Phys. Lett.} \textbf{\bibinfo{volume}{B640}},
  \bibinfo{pages}{150} (\bibinfo{year}{2006}{\natexlab{a}}).

\bibitem[{\citenamefont{Fayache et~al.}(1997)\citenamefont{Fayache, Zamick, and
  Castel}}]{Fay97a}
\bibinfo{author}{\bibfnamefont{M.~S.} \bibnamefont{Fayache}},
  \bibinfo{author}{\bibfnamefont{L.}~\bibnamefont{Zamick}}, \bibnamefont{and}
  \bibinfo{author}{\bibfnamefont{B.}~\bibnamefont{Castel}},
  \bibinfo{journal}{Phys. Rep.} \textbf{\bibinfo{volume}{290}},
  \bibinfo{pages}{201} (\bibinfo{year}{1997}).

\bibitem[{\citenamefont{Otsuka et~al.}(2005)\citenamefont{Otsuka, Suzuki,
  Fujimoto, Grawe, and Akaishi}}]{Ots05a}
\bibinfo{author}{\bibfnamefont{T.}~\bibnamefont{Otsuka}},
  \bibinfo{author}{\bibfnamefont{T.}~\bibnamefont{Suzuki}},
  \bibinfo{author}{\bibfnamefont{R.}~\bibnamefont{Fujimoto}},
  \bibinfo{author}{\bibfnamefont{H.}~\bibnamefont{Grawe}}, \bibnamefont{and}
  \bibinfo{author}{\bibfnamefont{Y.}~\bibnamefont{Akaishi}},
  \bibinfo{journal}{Phys. Rev. Lett.} \textbf{\bibinfo{volume}{95}},
  \bibinfo{pages}{232502} (\bibinfo{year}{2005}).

\bibitem[{\citenamefont{Schiffer et~al.}(2004)\citenamefont{Schiffer, Freeman,
  Caggiano, Deibel, Heinz, Jiang, Lewis, Parikh, Parker, Rehm
  et~al.}}]{Schi04aE}
\bibinfo{author}{\bibfnamefont{J.~P.} \bibnamefont{Schiffer}},
  \bibinfo{author}{\bibfnamefont{S.~J.} \bibnamefont{Freeman}},
  \bibinfo{author}{\bibfnamefont{J.~A.} \bibnamefont{Caggiano}},
  \bibinfo{author}{\bibfnamefont{C.}~\bibnamefont{Deibel}},
  \bibinfo{author}{\bibfnamefont{A.}~\bibnamefont{Heinz}},
  \bibinfo{author}{\bibfnamefont{C.-L.} \bibnamefont{Jiang}},
  \bibinfo{author}{\bibfnamefont{R.}~\bibnamefont{Lewis}},
  \bibinfo{author}{\bibfnamefont{A.}~\bibnamefont{Parikh}},
  \bibinfo{author}{\bibfnamefont{P.~D.} \bibnamefont{Parker}},
  \bibinfo{author}{\bibfnamefont{K.~E.} \bibnamefont{Rehm}}
  \bibnamefont{et~al.}, \bibinfo{journal}{Phys. Rev. Lett.}
  \textbf{\bibinfo{volume}{92}}, \bibinfo{eid}{162501} (\bibinfo{year}{2004}).

\bibitem[{\citenamefont{Otsuka et~al.}(2006)\citenamefont{Otsuka, Matsuo, and
  Abe}}]{Ots06a}
\bibinfo{author}{\bibfnamefont{T.}~\bibnamefont{Otsuka}},
  \bibinfo{author}{\bibfnamefont{T.}~\bibnamefont{Matsuo}}, \bibnamefont{and}
  \bibinfo{author}{\bibfnamefont{D.}~\bibnamefont{Abe}},
  \bibinfo{journal}{Phys. Rev. Lett.} \textbf{\bibinfo{volume}{97}},
  \bibinfo{pages}{162501} (\bibinfo{year}{2006}).

\bibitem[{\citenamefont{Dobaczewski}(2006)}]{Dobaczewskitalk}
\bibinfo{author}{\bibfnamefont{J.}~\bibnamefont{Dobaczewski}}, in
  \emph{\bibinfo{booktitle}{Proceedings of the Third ANL/MSU/JINA/INT RIA
  Workshop}}, edited by
  \bibinfo{editor}{\bibfnamefont{T.}~\bibnamefont{Duguet}},
  \bibinfo{editor}{\bibfnamefont{H.}~\bibnamefont{Esbensen}},
  \bibinfo{editor}{\bibfnamefont{K.~M.} \bibnamefont{Nollett}},
  \bibnamefont{and} \bibinfo{editor}{\bibfnamefont{C.~D.}
  \bibnamefont{Roberts}} (\bibinfo{publisher}{World Scientific},
  \bibinfo{year}{2006}), vol.~\bibinfo{volume}{15} of
  \emph{\bibinfo{series}{Proceedings from the Institute for Nuclear Theory}},
  \bibinfo{note}{preprint nucl-th/0604043}.

\bibitem[{\citenamefont{Brown et~al.}(2006)\citenamefont{Brown, Duguet, Otsuka,
  Abe, and Suzuki}}]{Bro06a}
\bibinfo{author}{\bibfnamefont{B.~A.} \bibnamefont{Brown}},
  \bibinfo{author}{\bibfnamefont{T.}~\bibnamefont{Duguet}},
  \bibinfo{author}{\bibfnamefont{T.}~\bibnamefont{Otsuka}},
  \bibinfo{author}{\bibfnamefont{D.}~\bibnamefont{Abe}}, \bibnamefont{and}
  \bibinfo{author}{\bibfnamefont{T.}~\bibnamefont{Suzuki}},
  \bibinfo{journal}{Phys. Rev. C} \textbf{\bibinfo{volume}{74}},
  \bibinfo{pages}{061303(R)} (\bibinfo{year}{2006}).

\bibitem[{\citenamefont{Col{\`o} et~al.}(2007)\citenamefont{Col{\`o}, Sagawa,
  Fracasso, and Bortignon}}]{Col07a}
\bibinfo{author}{\bibfnamefont{G.}~\bibnamefont{Col{\`o}}},
  \bibinfo{author}{\bibfnamefont{H.}~\bibnamefont{Sagawa}},
  \bibinfo{author}{\bibfnamefont{S.}~\bibnamefont{Fracasso}}, \bibnamefont{and}
  \bibinfo{author}{\bibfnamefont{P.~F.} \bibnamefont{Bortignon}},
  \bibinfo{journal}{Phys. Lett.} \textbf{\bibinfo{volume}{B646}},
  \bibinfo{pages}{227} (\bibinfo{year}{2007}).

\bibitem[{\citenamefont{Brink and Stancu}(2007)}]{Bri07a}
\bibinfo{author}{\bibfnamefont{D.~M.} \bibnamefont{Brink}} \bibnamefont{and}
  \bibinfo{author}{\bibfnamefont{F.}~\bibnamefont{Stancu}},
  \bibinfo{journal}{Phys. Rev. C} \textbf{\bibinfo{volume}{75}},
  \bibinfo{pages}{064311}
  (\bibinfo{year}{2007}).

\bibitem[{\citenamefont{Chabanat et~al.}(1997)\citenamefont{Chabanat, Bonche,
  Haensel, Meyer, and Schaeffer}}]{Cha97a}
\bibinfo{author}{\bibfnamefont{E.}~\bibnamefont{Chabanat}},
  \bibinfo{author}{\bibfnamefont{P.}~\bibnamefont{Bonche}},
  \bibinfo{author}{\bibfnamefont{P.}~\bibnamefont{Haensel}},
  \bibinfo{author}{\bibfnamefont{J.}~\bibnamefont{Meyer}}, \bibnamefont{and}
  \bibinfo{author}{\bibfnamefont{R.}~\bibnamefont{Schaeffer}},
  \bibinfo{journal}{Nucl. Phys.} \textbf{\bibinfo{volume}{A627}},
  \bibinfo{pages}{710} (\bibinfo{year}{1997}).

\bibitem[{\citenamefont{Chabanat et~al.}(1998)\citenamefont{Chabanat, Bonche,
  Haensel, Meyer, and Schaeffer}}]{Cha98a}
\bibinfo{author}{\bibfnamefont{E.}~\bibnamefont{Chabanat}},
  \bibinfo{author}{\bibfnamefont{P.}~\bibnamefont{Bonche}},
  \bibinfo{author}{\bibfnamefont{P.}~\bibnamefont{Haensel}},
  \bibinfo{author}{\bibfnamefont{J.}~\bibnamefont{Meyer}}, \bibnamefont{and}
  \bibinfo{author}{\bibfnamefont{R.}~\bibnamefont{Schaeffer}},
  \bibinfo{journal}{Nucl. Phys.} \textbf{\bibinfo{volume}{A635}},
  \bibinfo{pages}{231} (\bibinfo{year}{1998}), \bibinfo{note}{erratum Nucl.
  Phys. {\bf A643}, 441 (1998)}.

\bibitem[{\citenamefont{Long et~al.}(2006{\natexlab{b}})\citenamefont{Long,
  Sagawa, Meng, and {Nguyen Van Giai}}}]{Lon06b}
\bibinfo{author}{\bibfnamefont{W.~H.} \bibnamefont{Long}},
  \bibinfo{author}{\bibfnamefont{H.}~\bibnamefont{Sagawa}},
  \bibinfo{author}{\bibfnamefont{J.}~\bibnamefont{Meng}}, \bibnamefont{and}
  \bibinfo{author}{\bibnamefont{{Nguyen Van Giai}}}
  (\bibinfo{year}{2006}{\natexlab{b}}), \bibinfo{note}{preprint
  nucl-th/0609076}.

\bibitem[{\citenamefont{Bender et~al.}(2006{\natexlab{a}})\citenamefont{Bender,
  Bertsch, and Heenen}}]{BBH06a}
\bibinfo{author}{\bibfnamefont{M.}~\bibnamefont{Bender}},
  \bibinfo{author}{\bibfnamefont{G.~F.} \bibnamefont{Bertsch}},
  \bibnamefont{and} \bibinfo{author}{\bibfnamefont{P.-H.}
  \bibnamefont{Heenen}}, \bibinfo{journal}{Phys. Rev. C}
  \textbf{\bibinfo{volume}{73}}, \bibinfo{eid}{034322}
  (\bibinfo{year}{2006}{\natexlab{a}}).

\bibitem[{\citenamefont{Bender et~al.}(2003{\natexlab{b}})\citenamefont{Bender,
  Bonche, Duguet, and Heenen}}]{Ben03c}
\bibinfo{author}{\bibfnamefont{M.}~\bibnamefont{Bender}},
  \bibinfo{author}{\bibfnamefont{P.}~\bibnamefont{Bonche}},
  \bibinfo{author}{\bibfnamefont{T.}~\bibnamefont{Duguet}}, \bibnamefont{and}
  \bibinfo{author}{\bibfnamefont{P.~H.} \bibnamefont{Heenen}},
  \bibinfo{journal}{Nucl. Phys.} \textbf{\bibinfo{volume}{A723}},
  \bibinfo{pages}{354} (\bibinfo{year}{2003}{\natexlab{b}}).

\bibitem[{\citenamefont{Bender et~al.}(2006{\natexlab{b}})\citenamefont{Bender,
  Bonche, and Heenen}}]{Ben06a}
\bibinfo{author}{\bibfnamefont{M.}~\bibnamefont{Bender}},
  \bibinfo{author}{\bibfnamefont{P.}~\bibnamefont{Bonche}}, \bibnamefont{and}
  \bibinfo{author}{\bibfnamefont{P.~H.} \bibnamefont{Heenen}},
  \bibinfo{journal}{Phys. Rev. C} \textbf{\bibinfo{volume}{74}},
  \bibinfo{pages}{024312} (\bibinfo{year}{2006}{\natexlab{b}}).

\bibitem[{\citenamefont{Chatillon et~al.}(2006)\citenamefont{Chatillon,
  Theisen, Greenlees, Auger, Bastin, Bouchez, Bouriquet, Casandjian, Cee,
  Cl{\'e}ment et~al.}}]{Cha06a}
\bibinfo{author}{\bibfnamefont{A.}~\bibnamefont{Chatillon}},
  \bibinfo{author}{\bibfnamefont{C.}~\bibnamefont{Theisen}},
  \bibinfo{author}{\bibfnamefont{P.~T.} \bibnamefont{Greenlees}},
  \bibinfo{author}{\bibfnamefont{G.}~\bibnamefont{Auger}},
  \bibinfo{author}{\bibfnamefont{J.~E.} \bibnamefont{Bastin}},
  \bibinfo{author}{\bibfnamefont{E.}~\bibnamefont{Bouchez}},
  \bibinfo{author}{\bibfnamefont{B.}~\bibnamefont{Bouriquet}},
  \bibinfo{author}{\bibfnamefont{J.~M.} \bibnamefont{Casandjian}},
  \bibinfo{author}{\bibfnamefont{R.}~\bibnamefont{Cee}},
  \bibinfo{author}{\bibfnamefont{E.}~\bibnamefont{Cl{\'e}ment}}
  \bibnamefont{et~al.}, \bibinfo{journal}{Eur. Phys. J.}
  \textbf{\bibinfo{volume}{A30}}, \bibinfo{pages}{397} (\bibinfo{year}{2006}).

\bibitem[{\citenamefont{Slater}(1951)}]{Sla51}
\bibinfo{author}{\bibfnamefont{J.~C.} \bibnamefont{Slater}},
  \bibinfo{journal}{Phys. Rev.} \textbf{\bibinfo{volume}{81}},
  \bibinfo{pages}{385} (\bibinfo{year}{1951}).

\bibitem[{\citenamefont{Perli{\'n}ska et~al.}(2004)\citenamefont{Perli{\'n}ska,
  Rohozi{\'n}ski, Dobaczewski, and Nazarewicz}}]{Per04a}
\bibinfo{author}{\bibfnamefont{E.}~\bibnamefont{Perli{\'n}ska}},
  \bibinfo{author}{\bibfnamefont{S.~G.} \bibnamefont{Rohozi{\'n}ski}},
  \bibinfo{author}{\bibfnamefont{J.}~\bibnamefont{Dobaczewski}},
  \bibnamefont{and}
  \bibinfo{author}{\bibfnamefont{W.}~\bibnamefont{Nazarewicz}},
  \bibinfo{journal}{Phys. Rev. C} \textbf{\bibinfo{volume}{69}},
  \bibinfo{eid}{014316} (\bibinfo{year}{2004}).

\bibitem[{\citenamefont{Dobaczewski et~al.}(2000)\citenamefont{Dobaczewski,
  Dudek, Rohozi{\'n}ski, and Werner}}]{Dob00a}
\bibinfo{author}{\bibfnamefont{J.}~\bibnamefont{Dobaczewski}},
  \bibinfo{author}{\bibfnamefont{J.}~\bibnamefont{Dudek}},
  \bibinfo{author}{\bibfnamefont{S.~G.} \bibnamefont{Rohozi{\'n}ski}},
  \bibnamefont{and} \bibinfo{author}{\bibfnamefont{T.~R.}
  \bibnamefont{Werner}}, \bibinfo{journal}{Phys. Rev. C}
  \textbf{\bibinfo{volume}{62}}, \bibinfo{pages}{014310}
  (\bibinfo{year}{2000}).

\bibitem[{\citenamefont{Dobaczewski and Dudek}(1996)}]{Dob96a}
\bibinfo{author}{\bibfnamefont{J.}~\bibnamefont{Dobaczewski}} \bibnamefont{and}
  \bibinfo{author}{\bibfnamefont{J.}~\bibnamefont{Dudek}},
  \bibinfo{journal}{Acta Phys. Pol.} \textbf{\bibinfo{volume}{B27}},
  \bibinfo{pages}{45} (\bibinfo{year}{1996}).

\bibitem[{\citenamefont{Engel et~al.}(1975)\citenamefont{Engel, Brink, Goeke,
  Krieger, and Vautherin}}]{Eng75a}
\bibinfo{author}{\bibfnamefont{Y.~M.} \bibnamefont{Engel}},
  \bibinfo{author}{\bibfnamefont{D.~M.} \bibnamefont{Brink}},
  \bibinfo{author}{\bibfnamefont{K.}~\bibnamefont{Goeke}},
  \bibinfo{author}{\bibfnamefont{S.~J.} \bibnamefont{Krieger}},
  \bibnamefont{and}
  \bibinfo{author}{\bibfnamefont{D.}~\bibnamefont{Vautherin}},
  \bibinfo{journal}{Nucl. Phys.} \textbf{\bibinfo{volume}{A249}},
  \bibinfo{pages}{215} (\bibinfo{year}{1975}).

\bibitem[{\citenamefont{Dobaczewski and Dudek}(1995)}]{Dob95a}
\bibinfo{author}{\bibfnamefont{J.}~\bibnamefont{Dobaczewski}} \bibnamefont{and}
  \bibinfo{author}{\bibfnamefont{J.}~\bibnamefont{Dudek}},
  \bibinfo{journal}{Phys. Rev. C} \textbf{\bibinfo{volume}{52}},
  \bibinfo{pages}{1827} (\bibinfo{year}{1995}), \bibinfo{note}{erratum Phys.
  Rev. C {\bf 55}, 3177 (1997)}.

\bibitem[{\citenamefont{Flocard}(1975)}]{Flothesis}
\bibinfo{author}{\bibfnamefont{H.}~\bibnamefont{Flocard}}, Ph.D. thesis,
  \bibinfo{school}{Orsay, S{\'e}rie A, No. 1543, Universit{\'e} Paris Sud}
  (\bibinfo{year}{1975}).

\bibitem[{\citenamefont{Dobaczewski et~al.}(1984)\citenamefont{Dobaczewski,
  Flocard, and Treiner}}]{Dob84a}
\bibinfo{author}{\bibfnamefont{J.}~\bibnamefont{Dobaczewski}},
  \bibinfo{author}{\bibfnamefont{H.}~\bibnamefont{Flocard}}, \bibnamefont{and}
  \bibinfo{author}{\bibfnamefont{J.}~\bibnamefont{Treiner}},
  \bibinfo{journal}{Nucl. Phys.} \textbf{\bibinfo{volume}{A422}},
  \bibinfo{pages}{103} (\bibinfo{year}{1984}).

\bibitem[{\citenamefont{Negele and Vautherin}(1972)}]{Neg72a}
\bibinfo{author}{\bibfnamefont{J.~W.} \bibnamefont{Negele}} \bibnamefont{and}
  \bibinfo{author}{\bibfnamefont{D.}~\bibnamefont{Vautherin}},
  \bibinfo{journal}{Phys. Rev. C} \textbf{\bibinfo{volume}{5}},
  \bibinfo{pages}{1472} (\bibinfo{year}{1972}).

\bibitem[{\citenamefont{Negele and Vautherin}(1975)}]{Neg75a}
\bibinfo{author}{\bibfnamefont{J.~W.} \bibnamefont{Negele}} \bibnamefont{and}
  \bibinfo{author}{\bibfnamefont{D.}~\bibnamefont{Vautherin}},
  \bibinfo{journal}{Phys. Rev. C} \textbf{\bibinfo{volume}{11}},
  \bibinfo{pages}{1031} (\bibinfo{year}{1975}).

\bibitem[{\citenamefont{Campi and Bouyssy}(1978)}]{Cam78a}
\bibinfo{author}{\bibfnamefont{X.}~\bibnamefont{Campi}} \bibnamefont{and}
  \bibinfo{author}{\bibfnamefont{A.}~\bibnamefont{Bouyssy}},
  \bibinfo{journal}{Phys. Lett.} \textbf{\bibinfo{volume}{73B}},
  \bibinfo{pages}{263} (\bibinfo{year}{1978}).

\bibitem[{\citenamefont{Krivine et~al.}(1980)\citenamefont{Krivine, Treiner,
  and Bohigas}}]{Kri80a}
\bibinfo{author}{\bibfnamefont{H.}~\bibnamefont{Krivine}},
  \bibinfo{author}{\bibfnamefont{J.}~\bibnamefont{Treiner}}, \bibnamefont{and}
  \bibinfo{author}{\bibfnamefont{O.}~\bibnamefont{Bohigas}},
  \bibinfo{journal}{Nucl. Phys.} \textbf{\bibinfo{volume}{A336}},
  \bibinfo{pages}{155} (\bibinfo{year}{1980}).

\bibitem[{\citenamefont{Bartel et~al.}(1982)\citenamefont{Bartel, Quentin,
  Brack, Guet, and Hakansson}}]{Bar82a}
\bibinfo{author}{\bibfnamefont{J.}~\bibnamefont{Bartel}},
  \bibinfo{author}{\bibfnamefont{P.}~\bibnamefont{Quentin}},
  \bibinfo{author}{\bibfnamefont{M.}~\bibnamefont{Brack}},
  \bibinfo{author}{\bibfnamefont{C.}~\bibnamefont{Guet}}, \bibnamefont{and}
  \bibinfo{author}{\bibfnamefont{H.~B.} \bibnamefont{Hakansson}},
  \bibinfo{journal}{Nucl. Phys.} \textbf{\bibinfo{volume}{A386}},
  \bibinfo{pages}{79} (\bibinfo{year}{1982}).

\bibitem[{\citenamefont{Tondeur et~al.}(1984)\citenamefont{Tondeur, Brack,
  Farine, and Pearson}}]{Ton84a}
\bibinfo{author}{\bibfnamefont{F.}~\bibnamefont{Tondeur}},
  \bibinfo{author}{\bibfnamefont{M.}~\bibnamefont{Brack}},
  \bibinfo{author}{\bibfnamefont{M.}~\bibnamefont{Farine}}, \bibnamefont{and}
  \bibinfo{author}{\bibfnamefont{J.~M.} \bibnamefont{Pearson}},
  \bibinfo{journal}{Nucl. Phys.} \textbf{\bibinfo{volume}{A420}},
  \bibinfo{pages}{297} (\bibinfo{year}{1984}).

\bibitem[{\citenamefont{Friedrich and Reinhard}(1986)}]{Fri86a}
\bibinfo{author}{\bibfnamefont{J.}~\bibnamefont{Friedrich}} \bibnamefont{and}
  \bibinfo{author}{\bibfnamefont{P.~G.} \bibnamefont{Reinhard}},
  \bibinfo{journal}{Phys. Rev. C} \textbf{\bibinfo{volume}{33}},
  \bibinfo{pages}{335} (\bibinfo{year}{1986}).

\bibitem[{\citenamefont{Reinhard et~al.}(1999)\citenamefont{Reinhard, Dean,
  Nazarewicz, Dobaczewski, Maruhn, and Strayer}}]{Rei99a}
\bibinfo{author}{\bibfnamefont{P.~G.} \bibnamefont{Reinhard}},
  \bibinfo{author}{\bibfnamefont{D.~J.} \bibnamefont{Dean}},
  \bibinfo{author}{\bibfnamefont{W.}~\bibnamefont{Nazarewicz}},
  \bibinfo{author}{\bibfnamefont{J.}~\bibnamefont{Dobaczewski}},
  \bibinfo{author}{\bibfnamefont{J.~A.} \bibnamefont{Maruhn}},
  \bibnamefont{and} \bibinfo{author}{\bibfnamefont{M.~R.}
  \bibnamefont{Strayer}}, \bibinfo{journal}{Phys. Rev. C}
  \textbf{\bibinfo{volume}{60}}, \bibinfo{pages}{014316}
  (\bibinfo{year}{1999}).

\bibitem[{\citenamefont{Bonche et~al.}(1987)\citenamefont{Bonche, Flocard, and
  Heenen}}]{Bon87a}
\bibinfo{author}{\bibfnamefont{P.}~\bibnamefont{Bonche}},
  \bibinfo{author}{\bibfnamefont{H.}~\bibnamefont{Flocard}}, \bibnamefont{and}
  \bibinfo{author}{\bibfnamefont{P.~H.} \bibnamefont{Heenen}},
  \bibinfo{journal}{Nucl. Phys.} \textbf{\bibinfo{volume}{A467}},
  \bibinfo{pages}{115} (\bibinfo{year}{1987}).

\bibitem[{\citenamefont{Engel et~al.}(1999)\citenamefont{Engel, Bender,
  Dobaczewski, Nazarewicz, and Surman}}]{Eng99a}
\bibinfo{author}{\bibfnamefont{J.}~\bibnamefont{Engel}},
  \bibinfo{author}{\bibfnamefont{M.}~\bibnamefont{Bender}},
  \bibinfo{author}{\bibfnamefont{J.}~\bibnamefont{Dobaczewski}},
  \bibinfo{author}{\bibfnamefont{W.}~\bibnamefont{Nazarewicz}},
  \bibnamefont{and} \bibinfo{author}{\bibfnamefont{R.}~\bibnamefont{Surman}},
  \bibinfo{journal}{Phys. Rev. C} \textbf{\bibinfo{volume}{60}},
  \bibinfo{pages}{014302} (\bibinfo{year}{1999}).

\bibitem[{\citenamefont{Bender et~al.}(2002)\citenamefont{Bender, Dobaczewski,
  Engel, and Nazarewicz}}]{Ben02a}
\bibinfo{author}{\bibfnamefont{M.}~\bibnamefont{Bender}},
  \bibinfo{author}{\bibfnamefont{J.}~\bibnamefont{Dobaczewski}},
  \bibinfo{author}{\bibfnamefont{J.}~\bibnamefont{Engel}}, \bibnamefont{and}
  \bibinfo{author}{\bibfnamefont{W.}~\bibnamefont{Nazarewicz}},
  \bibinfo{journal}{Phys. Rev. C} \textbf{\bibinfo{volume}{65}},
  \bibinfo{pages}{054322} (\bibinfo{year}{2002}).

\bibitem[{\citenamefont{Terasaki et~al.}(2005)\citenamefont{Terasaki, Engel,
  Bender, Dobaczewski, Nazarewicz, and Stoitsov}}]{Ter05a}
\bibinfo{author}{\bibfnamefont{J.}~\bibnamefont{Terasaki}},
  \bibinfo{author}{\bibfnamefont{J.}~\bibnamefont{Engel}},
  \bibinfo{author}{\bibfnamefont{M.}~\bibnamefont{Bender}},
  \bibinfo{author}{\bibfnamefont{J.}~\bibnamefont{Dobaczewski}},
  \bibinfo{author}{\bibfnamefont{W.}~\bibnamefont{Nazarewicz}},
  \bibnamefont{and} \bibinfo{author}{\bibfnamefont{M.~V.}
  \bibnamefont{Stoitsov}}, \bibinfo{journal}{Phys. Rev. C}
  \textbf{\bibinfo{volume}{71}}, \bibinfo{pages}{034310}
  (\bibinfo{year}{2005}).

\bibitem[{\citenamefont{Sharma et~al.}(1995)\citenamefont{Sharma, Lalazissis,
  K\"onig, and Ring}}]{Sha95a}
\bibinfo{author}{\bibfnamefont{M.~M.} \bibnamefont{Sharma}},
  \bibinfo{author}{\bibfnamefont{G.}~\bibnamefont{Lalazissis}},
  \bibinfo{author}{\bibfnamefont{J.}~\bibnamefont{K\"onig}}, \bibnamefont{and}
  \bibinfo{author}{\bibfnamefont{P.}~\bibnamefont{Ring}},
  \bibinfo{journal}{Phys. Rev. Lett.} \textbf{\bibinfo{volume}{74}},
  \bibinfo{pages}{3744} (\bibinfo{year}{1995}).

\bibitem[{\citenamefont{Reinhard and Flocard}(1995)}]{Rei95a}
\bibinfo{author}{\bibfnamefont{P.~G.} \bibnamefont{Reinhard}} \bibnamefont{and}
  \bibinfo{author}{\bibfnamefont{H.}~\bibnamefont{Flocard}},
  \bibinfo{journal}{Nucl. Phys.} \textbf{\bibinfo{volume}{A584}},
  \bibinfo{pages}{467} (\bibinfo{year}{1995}).

\bibitem[{\citenamefont{Akmal et~al.}(1998)\citenamefont{Akmal, Pandharipande,
  and Ravenhall}}]{akmal98a}
\bibinfo{author}{\bibfnamefont{A.}~\bibnamefont{Akmal}},
  \bibinfo{author}{\bibfnamefont{V.~R.} \bibnamefont{Pandharipande}},
  \bibnamefont{and} \bibinfo{author}{\bibfnamefont{D.~G.}
  \bibnamefont{Ravenhall}}, \bibinfo{journal}{Phys. Rev.}
  \textbf{\bibinfo{volume}{C58}}, \bibinfo{pages}{1804} (\bibinfo{year}{1998}).

\bibitem[{\citenamefont{Goriely et~al.}(2005)\citenamefont{Goriely, Samyn,
  Pearson, and Onsi}}]{Gor05a}
\bibinfo{author}{\bibfnamefont{S.}~\bibnamefont{Goriely}},
  \bibinfo{author}{\bibfnamefont{M.}~\bibnamefont{Samyn}},
  \bibinfo{author}{\bibfnamefont{J.~M.} \bibnamefont{Pearson}},
  \bibnamefont{and} \bibinfo{author}{\bibfnamefont{M.}~\bibnamefont{Onsi}},
  \bibinfo{journal}{Nucl. Phys.} \textbf{\bibinfo{volume}{A750}},
  \bibinfo{pages}{425} (\bibinfo{year}{2005}).

\bibitem[{\citenamefont{Brown}()}]{BrownPrivComm}
\bibinfo{author}{\bibfnamefont{B.~A.} \bibnamefont{Brown}},
  \bibinfo{note}{private communication}.

\bibitem[{\citenamefont{Liu and Brown}(1976)}]{Liu76a}
\bibinfo{author}{\bibfnamefont{K.~F.} \bibnamefont{Liu}} \bibnamefont{and}
  \bibinfo{author}{\bibfnamefont{G.~E.} \bibnamefont{Brown}},
  \bibinfo{journal}{Nucl. Phys.} \textbf{\bibinfo{volume}{A265}},
  \bibinfo{pages}{385} (\bibinfo{year}{1976}).

\bibitem[{\citenamefont{Lesinski et~al.}(2006)\citenamefont{Lesinski,
  Bennaceur, Duguet, and Meyer}}]{Les06a}
\bibinfo{author}{\bibfnamefont{T.}~\bibnamefont{Lesinski}},
  \bibinfo{author}{\bibfnamefont{K.}~\bibnamefont{Bennaceur}},
  \bibinfo{author}{\bibfnamefont{T.}~\bibnamefont{Duguet}}, \bibnamefont{and}
  \bibinfo{author}{\bibfnamefont{J.}~\bibnamefont{Meyer}},
  \bibinfo{journal}{Phys. Rev. C} \textbf{\bibinfo{volume}{74}},
  \bibinfo{eid}{044315} (\bibinfo{year}{2006}).

\bibitem[{EPA()}]{EPAPS}
\bibinfo{note}{See EPAPS Document No. ???? for the coupling constants of the 36
  T$IJ$ parametrizations}.

\bibitem[{\citenamefont{Duguet et~al.}(2006)\citenamefont{Duguet, Bennaceur,
  and Bonche}}]{Dug07}
\bibinfo{author}{\bibfnamefont{T.}~\bibnamefont{Duguet}},
  \bibinfo{author}{\bibfnamefont{K.}~\bibnamefont{Bennaceur}},
  \bibnamefont{and} \bibinfo{author}{\bibfnamefont{P.}~\bibnamefont{Bonche}}
  (\bibinfo{year}{2006}), \bibinfo{note}{invited talk at the {\em Workshop on
  New developments in Nuclear Self-Consistent Mean-Field Theories}, Yukawa
  Institute for Theoretical Physics, Kyoto, Japan, May 30 - June 1, 2005,
  nucl-th/0508054}.

\bibitem[{\citenamefont{Bennaceur and Dobaczewski}(2005)}]{Ben05a}
\bibinfo{author}{\bibfnamefont{K.}~\bibnamefont{Bennaceur}} \bibnamefont{and}
  \bibinfo{author}{\bibfnamefont{J.}~\bibnamefont{Dobaczewski}},
  \bibinfo{journal}{Comp. Phys. Comm.} \textbf{\bibinfo{volume}{168}},
  \bibinfo{pages}{96} (\bibinfo{year}{2005}).

\bibitem[{\citenamefont{Rutz et~al.}(1998)\citenamefont{Rutz, Bender, Reinhard,
  Maruhn, and Greiner}}]{Rut98a}
\bibinfo{author}{\bibfnamefont{K.}~\bibnamefont{Rutz}},
  \bibinfo{author}{\bibfnamefont{M.}~\bibnamefont{Bender}},
  \bibinfo{author}{\bibfnamefont{P.~G.} \bibnamefont{Reinhard}},
  \bibinfo{author}{\bibfnamefont{J.~A.} \bibnamefont{Maruhn}},
  \bibnamefont{and} \bibinfo{author}{\bibfnamefont{W.}~\bibnamefont{Greiner}},
  \bibinfo{journal}{Nucl. Phys.} \textbf{\bibinfo{volume}{A634}},
  \bibinfo{pages}{67} (\bibinfo{year}{1998}).

\bibitem[{\citenamefont{Bernard and {Nguyen Van Giai}}(1980)}]{Ber80a}
\bibinfo{author}{\bibfnamefont{V.}~\bibnamefont{Bernard}} \bibnamefont{and}
  \bibinfo{author}{\bibnamefont{{Nguyen Van Giai}}}, \bibinfo{journal}{Nucl.
  Phys.} \textbf{\bibinfo{volume}{A348}}, \bibinfo{pages}{75}
  (\bibinfo{year}{1980}).

\bibitem[{\citenamefont{Litvinova and Ring}(2006)}]{Lit06a}
\bibinfo{author}{\bibfnamefont{E.}~\bibnamefont{Litvinova}} \bibnamefont{and}
  \bibinfo{author}{\bibfnamefont{P.}~\bibnamefont{Ring}},
  \bibinfo{journal}{Phys. Rev. C} \textbf{\bibinfo{volume}{73}},
  \bibinfo{eid}{044328} (\bibinfo{year}{2006}).

\bibitem[{\citenamefont{Bertsch}(1972)}]{Ber72aB}
\bibinfo{author}{\bibfnamefont{G.~F.} \bibnamefont{Bertsch}},
  \emph{\bibinfo{title}{The Practitioner's Shell model}}
  (\bibinfo{publisher}{North Holland, Amsterdam}, \bibinfo{year}{1972}).

\bibitem[{\citenamefont{Ring and Schuck}(1980)}]{Rin80aB}
\bibinfo{author}{\bibfnamefont{P.}~\bibnamefont{Ring}} \bibnamefont{and}
  \bibinfo{author}{\bibfnamefont{P.}~\bibnamefont{Schuck}},
  \emph{\bibinfo{title}{The Nuclear Many Body Problem}}
  (\bibinfo{publisher}{Springer, Berlin}, \bibinfo{year}{1980}).

\bibitem[{\citenamefont{Caurier et~al.}(2005)\citenamefont{Caurier,
  Martinez-Pinedo, Nowacki, Poves, and Zuker}}]{Cau05a}
\bibinfo{author}{\bibfnamefont{E.}~\bibnamefont{Caurier}},
  \bibinfo{author}{\bibfnamefont{G.}~\bibnamefont{Martinez-Pinedo}},
  \bibinfo{author}{\bibfnamefont{F.}~\bibnamefont{Nowacki}},
  \bibinfo{author}{\bibfnamefont{A.}~\bibnamefont{Poves}}, \bibnamefont{and}
  \bibinfo{author}{\bibfnamefont{A.~P.} \bibnamefont{Zuker}},
  \bibinfo{journal}{Rev. Mod. Phys.} \textbf{\bibinfo{volume}{77}},
  \bibinfo{pages}{427} (\bibinfo{year}{2005}).

\bibitem[{\citenamefont{Duflo and Zuker}(1999)}]{Duf99a}
\bibinfo{author}{\bibfnamefont{J.}~\bibnamefont{Duflo}} \bibnamefont{and}
  \bibinfo{author}{\bibfnamefont{A.~P.} \bibnamefont{Zuker}},
  \bibinfo{journal}{Phys. Rev. C} \textbf{\bibinfo{volume}{59}},
  \bibinfo{pages}{R2347} (\bibinfo{year}{1999}).

\bibitem[{\citenamefont{Grawe et~al.}(2005)\citenamefont{Grawe, Blazhev,
  G{\'o}rska, Mukha, Plettner, Roeckl, Nowacki, Grzywacz, and
  Sawicka}}]{Gra05a}
\bibinfo{author}{\bibfnamefont{H.}~\bibnamefont{Grawe}},
  \bibinfo{author}{\bibfnamefont{A.}~\bibnamefont{Blazhev}},
  \bibinfo{author}{\bibfnamefont{M.}~\bibnamefont{G{\'o}rska}},
  \bibinfo{author}{\bibfnamefont{I.}~\bibnamefont{Mukha}},
  \bibinfo{author}{\bibfnamefont{C.}~\bibnamefont{Plettner}},
  \bibinfo{author}{\bibfnamefont{E.}~\bibnamefont{Roeckl}},
  \bibinfo{author}{\bibfnamefont{F.}~\bibnamefont{Nowacki}},
  \bibinfo{author}{\bibfnamefont{R.}~\bibnamefont{Grzywacz}}, \bibnamefont{and}
  \bibinfo{author}{\bibfnamefont{M.}~\bibnamefont{Sawicka}},
  \bibinfo{journal}{Eur. Phys. J.} \textbf{\bibinfo{volume}{A25}},
  \bibinfo{pages}{357} (\bibinfo{year}{2005}).

\bibitem[{\citenamefont{Grawe et~al.}(2006)\citenamefont{Grawe, Blazhev,
  G{\'o}rska, Grzywacz, Mach, and Mukha}}]{Gra06a}
\bibinfo{author}{\bibfnamefont{H.}~\bibnamefont{Grawe}},
  \bibinfo{author}{\bibfnamefont{A.}~\bibnamefont{Blazhev}},
  \bibinfo{author}{\bibfnamefont{M.}~\bibnamefont{G{\'o}rska}},
  \bibinfo{author}{\bibfnamefont{R.}~\bibnamefont{Grzywacz}},
  \bibinfo{author}{\bibfnamefont{H.}~\bibnamefont{Mach}}, \bibnamefont{and}
  \bibinfo{author}{\bibfnamefont{I.}~\bibnamefont{Mukha}},
  \bibinfo{journal}{Eur. Phys. J.} \textbf{\bibinfo{volume}{A27}},
  \bibinfo{pages}{257} (\bibinfo{year}{2006}).

\bibitem[{\citenamefont{Brown}(1998)}]{Bro98a}
\bibinfo{author}{\bibfnamefont{B.~A.} \bibnamefont{Brown}},
  \bibinfo{journal}{Phys. Rev. C} \textbf{\bibinfo{volume}{58}},
  \bibinfo{pages}{220} (\bibinfo{year}{1998}).

\bibitem[{\citenamefont{L\'opez-Quelle
  et~al.}(2000)\citenamefont{L\'opez-Quelle, {Nguyen Van Giai}, Marcos, and
  Savushkin}}]{Lop00a}
\bibinfo{author}{\bibfnamefont{M.}~\bibnamefont{L\'opez-Quelle}},
  \bibinfo{author}{\bibnamefont{{Nguyen Van Giai}}},
  \bibinfo{author}{\bibfnamefont{S.}~\bibnamefont{Marcos}}, \bibnamefont{and}
  \bibinfo{author}{\bibfnamefont{L.~N.} \bibnamefont{Savushkin}},
  \bibinfo{journal}{Phys. Rev. C} \textbf{\bibinfo{volume}{61}},
  \bibinfo{pages}{064321} (\bibinfo{year}{2000}).

\bibitem[{\citenamefont{Skalski}(2001)}]{Ska01a}
\bibinfo{author}{\bibfnamefont{J.}~\bibnamefont{Skalski}},
  \bibinfo{journal}{Phys. Rev. C} \textbf{\bibinfo{volume}{63}},
  \bibinfo{pages}{024312} (\bibinfo{year}{2001}).

\bibitem[{\citenamefont{Hauschild et~al.}(2001)\citenamefont{Hauschild,
  Rejmund, Grawe, Caurier, Nowacki, Becker, Le~Coz, Korten, D{\"o}ring,
  G{\'o}rska et~al.}}]{Hau01aE}
\bibinfo{author}{\bibfnamefont{K.}~\bibnamefont{Hauschild}},
  \bibinfo{author}{\bibfnamefont{M.}~\bibnamefont{Rejmund}},
  \bibinfo{author}{\bibfnamefont{H.}~\bibnamefont{Grawe}},
  \bibinfo{author}{\bibfnamefont{E.}~\bibnamefont{Caurier}},
  \bibinfo{author}{\bibfnamefont{F.}~\bibnamefont{Nowacki}},
  \bibinfo{author}{\bibfnamefont{F.}~\bibnamefont{Becker}},
  \bibinfo{author}{\bibfnamefont{Y.}~\bibnamefont{Le~Coz}},
  \bibinfo{author}{\bibfnamefont{W.}~\bibnamefont{Korten}},
  \bibinfo{author}{\bibfnamefont{J.}~\bibnamefont{D{\"o}ring}},
  \bibinfo{author}{\bibfnamefont{M.}~\bibnamefont{G{\'o}rska}}
  \bibnamefont{et~al.}, \bibinfo{journal}{Phys. Rev. Lett.}
  \textbf{\bibinfo{volume}{87}}, \bibinfo{pages}{072501}
  (\bibinfo{year}{2001}).

\bibitem[{\citenamefont{Shergur et~al.}(2005)\citenamefont{Shergur, Dean,
  Seweryniak, Walters, Wohr, Boutachkov, Davids, Dillmann, Juodagalvis,
  Mukherjee et~al.}}]{She05aE}
\bibinfo{author}{\bibfnamefont{J.}~\bibnamefont{Shergur}},
  \bibinfo{author}{\bibfnamefont{D.~J.} \bibnamefont{Dean}},
  \bibinfo{author}{\bibfnamefont{D.}~\bibnamefont{Seweryniak}},
  \bibinfo{author}{\bibfnamefont{W.~B.} \bibnamefont{Walters}},
  \bibinfo{author}{\bibfnamefont{A.}~\bibnamefont{Wohr}},
  \bibinfo{author}{\bibfnamefont{P.}~\bibnamefont{Boutachkov}},
  \bibinfo{author}{\bibfnamefont{C.~N.} \bibnamefont{Davids}},
  \bibinfo{author}{\bibfnamefont{I.}~\bibnamefont{Dillmann}},
  \bibinfo{author}{\bibfnamefont{A.}~\bibnamefont{Juodagalvis}},
  \bibinfo{author}{\bibfnamefont{G.}~\bibnamefont{Mukherjee}}
  \bibnamefont{et~al.}, \bibinfo{journal}{Phys. Rev. C}
  \textbf{\bibinfo{volume}{71}}, \bibinfo{pages}{064323}
  (\bibinfo{year}{2005}).

\bibitem[{\citenamefont{Porquet et~al.}(2005)\citenamefont{Porquet, P{\'e}ru,
  and Girod}}]{Por05a}
\bibinfo{author}{\bibfnamefont{M.~G.} \bibnamefont{Porquet}},
  \bibinfo{author}{\bibfnamefont{S.}~\bibnamefont{P{\'e}ru}}, \bibnamefont{and}
  \bibinfo{author}{\bibfnamefont{M.}~\bibnamefont{Girod}},
  \bibinfo{journal}{Eur. Phys. J.} \textbf{\bibinfo{volume}{A25}},
  \bibinfo{pages}{319} (\bibinfo{year}{2005}).

\bibitem[{\citenamefont{Patyk et~al.}(1999)\citenamefont{Patyk, Baran, Berger,
  Decharg\'e, Dobaczewski, Ring, and Sobiczewski}}]{Pat99a}
\bibinfo{author}{\bibfnamefont{Z.}~\bibnamefont{Patyk}},
  \bibinfo{author}{\bibfnamefont{A.}~\bibnamefont{Baran}},
  \bibinfo{author}{\bibfnamefont{J.~F.} \bibnamefont{Berger}},
  \bibinfo{author}{\bibfnamefont{J.}~\bibnamefont{Decharg\'e}},
  \bibinfo{author}{\bibfnamefont{J.}~\bibnamefont{Dobaczewski}},
  \bibinfo{author}{\bibfnamefont{P.}~\bibnamefont{Ring}}, \bibnamefont{and}
  \bibinfo{author}{\bibfnamefont{A.}~\bibnamefont{Sobiczewski}},
  \bibinfo{journal}{Phys. Rev. C} \textbf{\bibinfo{volume}{59}},
  \bibinfo{pages}{704} (\bibinfo{year}{1999}).

\bibitem[{\citenamefont{Lunney et~al.}(2003)\citenamefont{Lunney, Pearson, and
  Thibault}}]{Lun03a}
\bibinfo{author}{\bibfnamefont{D.}~\bibnamefont{Lunney}},
  \bibinfo{author}{\bibfnamefont{J.~M.} \bibnamefont{Pearson}},
  \bibnamefont{and} \bibinfo{author}{\bibfnamefont{C.}~\bibnamefont{Thibault}},
  \bibinfo{journal}{Rev. Mod. Phys.} \textbf{\bibinfo{volume}{75}},
  \bibinfo{pages}{1021} (\bibinfo{year}{2003}).

\bibitem[{\citenamefont{Dobaczewski et~al.}(2004)\citenamefont{Dobaczewski,
  Stoitsov, and Nazarewicz}}]{Dob04a}
\bibinfo{author}{\bibfnamefont{J.}~\bibnamefont{Dobaczewski}},
  \bibinfo{author}{\bibfnamefont{M.~V.} \bibnamefont{Stoitsov}},
  \bibnamefont{and}
  \bibinfo{author}{\bibfnamefont{W.}~\bibnamefont{Nazarewicz}}, in
  \emph{\bibinfo{booktitle}{Proc. Int. Conf. on Nuclear Physics, Large and
  Small, Cocoyoc, Mexico, April 19-22, 2004}}, edited by
  \bibinfo{editor}{\bibfnamefont{R.}~\bibnamefont{Bijker}},
  \bibinfo{editor}{\bibfnamefont{R.~F.} \bibnamefont{Casten}},
  \bibnamefont{and} \bibinfo{editor}{\bibfnamefont{A.}~\bibnamefont{Frank}}
  (American Institute of Physics, Melville, NY, \bibinfo{year}{2004}),
  pp. \bibinfo{pages}{51--56},
  \bibinfo{note}{nucl-th/0404077}.

\bibitem[{\citenamefont{Satu{\l}a et~al.}(1997)\citenamefont{Satu{\l}a, Dean,
  Gary, Mizutori, and Nazarewicz}}]{Sat97a}
\bibinfo{author}{\bibfnamefont{W.}~\bibnamefont{Satu{\l}a}},
  \bibinfo{author}{\bibfnamefont{D.~J.} \bibnamefont{Dean}},
  \bibinfo{author}{\bibfnamefont{J.}~\bibnamefont{Gary}},
  \bibinfo{author}{\bibfnamefont{S.}~\bibnamefont{Mizutori}}, \bibnamefont{and}
  \bibinfo{author}{\bibfnamefont{W.}~\bibnamefont{Nazarewicz}},
  \bibinfo{journal}{Phys. Lett.} \textbf{\bibinfo{volume}{B407}},
  \bibinfo{pages}{103} (\bibinfo{year}{1997}).

\bibitem[{\citenamefont{Tondeur et~al.}(2000)\citenamefont{Tondeur, Goriely,
  Pearson, and Onsi}}]{Ton00a}
\bibinfo{author}{\bibfnamefont{F.}~\bibnamefont{Tondeur}},
  \bibinfo{author}{\bibfnamefont{S.}~\bibnamefont{Goriely}},
  \bibinfo{author}{\bibfnamefont{J.~M.} \bibnamefont{Pearson}},
  \bibnamefont{and} \bibinfo{author}{\bibfnamefont{M.}~\bibnamefont{Onsi}},
  \bibinfo{journal}{Phys. Rev. C} \textbf{\bibinfo{volume}{62}},
  \bibinfo{pages}{024308} (\bibinfo{year}{2000}).

\bibitem[{\citenamefont{Samyn et~al.}(2002)\citenamefont{Samyn, Goriely,
  Heenen, Pearson, and Tondeur}}]{Sam02a}
\bibinfo{author}{\bibfnamefont{M.}~\bibnamefont{Samyn}},
  \bibinfo{author}{\bibfnamefont{S.}~\bibnamefont{Goriely}},
  \bibinfo{author}{\bibfnamefont{P.~H.} \bibnamefont{Heenen}},
  \bibinfo{author}{\bibfnamefont{J.~M.} \bibnamefont{Pearson}},
  \bibnamefont{and} \bibinfo{author}{\bibfnamefont{F.}~\bibnamefont{Tondeur}},
  \bibinfo{journal}{Nucl. Phys.} \textbf{\bibinfo{volume}{A700}},
  \bibinfo{pages}{142} (\bibinfo{year}{2002}).

\bibitem[{\citenamefont{Goriely et~al.}(2003)\citenamefont{Goriely, Samyn,
  Bender, and Pearson}}]{Gor03b}
\bibinfo{author}{\bibfnamefont{S.}~\bibnamefont{Goriely}},
  \bibinfo{author}{\bibfnamefont{M.}~\bibnamefont{Samyn}},
  \bibinfo{author}{\bibfnamefont{M.}~\bibnamefont{Bender}}, \bibnamefont{and}
  \bibinfo{author}{\bibfnamefont{J.~M.} \bibnamefont{Pearson}},
  \bibinfo{journal}{Phys. Rev. C} \textbf{\bibinfo{volume}{68}},
  \bibinfo{pages}{054325} (\bibinfo{year}{2003}).

\bibitem[{\citenamefont{Reinhard et~al.}(2006)\citenamefont{Reinhard, Bender,
  Nazarewicz, and Vertse}}]{Rei06a}
\bibinfo{author}{\bibfnamefont{P.-G.} \bibnamefont{Reinhard}},
  \bibinfo{author}{\bibfnamefont{M.}~\bibnamefont{Bender}},
  \bibinfo{author}{\bibfnamefont{W.}~\bibnamefont{Nazarewicz}},
  \bibnamefont{and} \bibinfo{author}{\bibfnamefont{T.}~\bibnamefont{Vertse}},
  \bibinfo{journal}{Phys. Rev. C} \textbf{\bibinfo{volume}{73}},
  \bibinfo{eid}{014309} (\bibinfo{year}{2006}).

\bibitem[{\citenamefont{Reinhard and Drechsel}(1979)}]{Rei79a}
\bibinfo{author}{\bibfnamefont{P.-G.} \bibnamefont{Reinhard}} \bibnamefont{and}
  \bibinfo{author}{\bibfnamefont{D.}~\bibnamefont{Drechsel}},
  \bibinfo{journal}{Z. Phys.} \textbf{\bibinfo{volume}{A290}},
  \bibinfo{pages}{85} (\bibinfo{year}{1979}).

\bibitem[{\citenamefont{Girod and Reinhard}(1982)}]{Gir82a}
\bibinfo{author}{\bibfnamefont{M.}~\bibnamefont{Girod}} \bibnamefont{and}
  \bibinfo{author}{\bibfnamefont{P.-G.} \bibnamefont{Reinhard}},
  \bibinfo{journal}{Nucl. Phys.} \textbf{\bibinfo{volume}{A384}},
  \bibinfo{pages}{179} (\bibinfo{year}{1982}).

\bibitem[{\citenamefont{Bonche et~al.}(1991)\citenamefont{Bonche, Dobaczewski,
  Flocard, and Heenen}}]{Bon91a}
\bibinfo{author}{\bibfnamefont{P.}~\bibnamefont{Bonche}},
  \bibinfo{author}{\bibfnamefont{J.}~\bibnamefont{Dobaczewski}},
  \bibinfo{author}{\bibfnamefont{H.}~\bibnamefont{Flocard}}, \bibnamefont{and}
  \bibinfo{author}{\bibfnamefont{P.~H.} \bibnamefont{Heenen}},
  \bibinfo{journal}{Nucl. Phys.} \textbf{\bibinfo{volume}{A530}},
  \bibinfo{pages}{149} (\bibinfo{year}{1991}).

\bibitem[{\citenamefont{Heenen et~al.}(1993)\citenamefont{Heenen, Bonche,
  Dobaczewski, and Flocard}}]{Hee93a}
\bibinfo{author}{\bibfnamefont{P.~H.} \bibnamefont{Heenen}},
  \bibinfo{author}{\bibfnamefont{P.}~\bibnamefont{Bonche}},
  \bibinfo{author}{\bibfnamefont{J.}~\bibnamefont{Dobaczewski}},
  \bibnamefont{and} \bibinfo{author}{\bibfnamefont{H.}~\bibnamefont{Flocard}},
  \bibinfo{journal}{Nucl. Phys.} \textbf{\bibinfo{volume}{A561}},
  \bibinfo{pages}{367} (\bibinfo{year}{1993}).

\bibitem[{\citenamefont{Tajima et~al.}(1993)\citenamefont{Tajima, Bonche,
  Flocard, Heenen, and Weiss}}]{Taj93a}
\bibinfo{author}{\bibfnamefont{N.}~\bibnamefont{Tajima}},
  \bibinfo{author}{\bibfnamefont{P.}~\bibnamefont{Bonche}},
  \bibinfo{author}{\bibfnamefont{H.}~\bibnamefont{Flocard}},
  \bibinfo{author}{\bibfnamefont{P.~H.} \bibnamefont{Heenen}},
  \bibnamefont{and} \bibinfo{author}{\bibfnamefont{M.~S.} \bibnamefont{Weiss}},
  \bibinfo{journal}{Nucl. Phys.} \textbf{\bibinfo{volume}{A551}},
  \bibinfo{pages}{434} (\bibinfo{year}{1993}).

\bibitem[{\citenamefont{Fayans et~al.}(2000)\citenamefont{Fayans, Tolokonnikov,
  Trykov, and Zawischa}}]{Fay00a}
\bibinfo{author}{\bibfnamefont{S.~A.} \bibnamefont{Fayans}},
  \bibinfo{author}{\bibfnamefont{S.~V.} \bibnamefont{Tolokonnikov}},
  \bibinfo{author}{\bibfnamefont{E.~L.} \bibnamefont{Trykov}},
  \bibnamefont{and} \bibinfo{author}{\bibfnamefont{D.}~\bibnamefont{Zawischa}},
  \bibinfo{journal}{Nucl. Phys.} \textbf{\bibinfo{volume}{A676}},
  \bibinfo{pages}{49} (\bibinfo{year}{2000}).

\bibitem[{\citenamefont{Bertozzi et~al.}(1972)\citenamefont{Bertozzi, Friar,
  Heisenberg, and Negele}}]{Ber72a}
\bibinfo{author}{\bibfnamefont{W.}~\bibnamefont{Bertozzi}},
  \bibinfo{author}{\bibfnamefont{J.}~\bibnamefont{Friar}},
  \bibinfo{author}{\bibfnamefont{J.}~\bibnamefont{Heisenberg}},
  \bibnamefont{and} \bibinfo{author}{\bibfnamefont{J.~W.}
  \bibnamefont{Negele}}, \bibinfo{journal}{Phys. Lett.}
  \textbf{\bibinfo{volume}{B41}}, \bibinfo{pages}{408} (\bibinfo{year}{1972}).

\bibitem[{\citenamefont{Otten}(1989)}]{Otten}
\bibinfo{author}{\bibfnamefont{E.~W.} \bibnamefont{Otten}}, in
  \emph{\bibinfo{booktitle}{Treatise on Heavy-Ion Science}}, edited by
  \bibinfo{editor}{\bibfnamefont{A.~D.} \bibnamefont{Bromley}}
  (\bibinfo{publisher}{Plenum, New York}, \bibinfo{year}{1989}), vol.
  \bibinfo{volume}{8. Nuclei far from Stability}, pp.
  \bibinfo{pages}{517--638}.

\bibitem[{\citenamefont{Caurier et~al.}(2001)\citenamefont{Caurier, Langanke,
  Martinez-Pinedo, Nowacki, and Vogel}}]{Cau01a}
\bibinfo{author}{\bibfnamefont{E.}~\bibnamefont{Caurier}},
  \bibinfo{author}{\bibfnamefont{K.}~\bibnamefont{Langanke}},
  \bibinfo{author}{\bibfnamefont{G.}~\bibnamefont{Martinez-Pinedo}},
  \bibinfo{author}{\bibfnamefont{F.}~\bibnamefont{Nowacki}}, \bibnamefont{and}
  \bibinfo{author}{\bibfnamefont{P.}~\bibnamefont{Vogel}},
  \bibinfo{journal}{Phys. Lett.} \textbf{\bibinfo{volume}{B522}},
  \bibinfo{pages}{240} (\bibinfo{year}{2001}).

\bibitem[{\citenamefont{Angeli}(2004)}]{Ang04a}
\bibinfo{author}{\bibfnamefont{I.}~\bibnamefont{Angeli}},
  \bibinfo{journal}{Atom. Data Nucl. Data Tables}
  \textbf{\bibinfo{volume}{87}}, \bibinfo{pages}{185} (\bibinfo{year}{2004}).

\bibitem[{\citenamefont{Haensel and Jerzak}(1982)}]{Hae82a}
\bibinfo{author}{\bibfnamefont{P.}~\bibnamefont{Haensel}} \bibnamefont{and}
  \bibinfo{author}{\bibfnamefont{A.~J.} \bibnamefont{Jerzak}},
  \bibinfo{journal}{Phys. Lett.} \textbf{\bibinfo{volume}{B112}},
  \bibinfo{pages}{285} (\bibinfo{year}{1982}).

\end{thebibliography}



\end{document}